\documentclass[prb,a4paper,10pt,twocolumn,showpacs,floatfix,superscriptaddress,amsmath,amsfonts,amssymb,preprintnumbers,citeautoscript]{revtex4}
\pdfoutput=1
\usepackage[T1]{fontenc}\usepackage[latin1]{inputenc}
\usepackage{dcolumn,graphicx,color,booktabs,microtype,afterpage} \graphicspath{{Figures/}}
\usepackage[charter,greekuppercase=italicized]{mathdesign}\usepackage{sidecap}
\renewcommand{\figurename}{Fig.}
\renewcommand{\tablename}{Table}
\makeatletter\renewcommand{\fnum@figure}[1]{\figurename~\thefigure~(color online).}\makeatother
\makeatletter\renewcommand{\fnum@table}[1]{\tablename~\thetable.}\makeatother

\newcount\hh \newcount\mm
\hh=\time \divide\hh by 60
\mm=\hh \multiply\mm by 60 \mm=-\mm
\advance\mm by \time
\def\now{\number\hh:\ifnum\mm<10{}0\fi\number\mm}

\usepackage[colorlinks,plainpages=false,linkcolor=blue,urlcolor=blue,citecolor=blue,pdfpagemode=UseNone,pdfstartview=FitBH]{hyperref}

\newcommand{\half}{\frac{1}{\protect\raisebox{0.8pt}{\scriptsize 2}}}
\newcommand{\threehalf}{\frac{3}{\protect\raisebox{0.8pt}{\scriptsize 2}}}

\begin{document}

\makeatletter\renewcommand{\ps@plain}{%
\def\@evenhead{\hfill\itshape\rightmark}%
\def\@oddhead{\itshape\leftmark\hfill}%
\renewcommand{\@evenfoot}{\hfill\small{--~\thepage~--}\hfill}%
\renewcommand{\@oddfoot}{\hfill\small{--~\thepage~--}\hfill}%
}\makeatother\pagestyle{plain}


\title{~\vspace{-1ex}\\Possible realization of an antiferromagnetic Griffiths phase in Ba(Fe$_{1-x}$Mn$_x$)$_2$As$_2$}

\author{D.\,S.\,Inosov}\email[Corresponding author: \vspace{8pt}]{Dmytro.Inosov@tu-dresden.de}
\affiliation{Max-Planck-Institut für Festkörperforschung, Heisenbergstraße 1, D-70569 Stuttgart, Germany}
\affiliation{Institut für Festkörperphysik, TU Dresden, D-01069 Dresden, Germany}

\author{G.\,Friemel}
\affiliation{Max-Planck-Institut für Festkörperforschung, Heisenbergstraße 1, D-70569 Stuttgart, Germany}

\author{J.\,T.~Park}
\affiliation{Max-Planck-Institut für Festkörperforschung, Heisenbergstraße 1, D-70569 Stuttgart, Germany}
\affiliation{Forschungsneutronenquelle Heinz Maier-Leibnitz (FRM-II), TU München, D-85747 Garching, Germany}

\author{A.\,C.\,Walters}
\affiliation{Max-Planck-Institut für Festkörperforschung, Heisenbergstraße 1, D-70569 Stuttgart, Germany}

\author{Y.~Texier}
\affiliation{Laboratoire de Physique des Solides, Univ. Paris-Sud, UMR8502, CNRS\,--\,F-91405 Orsay Cedex, France}

\author{Y.~Laplace}
\affiliation{Laboratoire de Physique des Solides, Univ. Paris-Sud, UMR8502, CNRS\,--\,F-91405 Orsay Cedex, France}

\author{J.\,Bobroff}
\affiliation{Laboratoire de Physique des Solides, Univ. Paris-Sud, UMR8502, CNRS\,--\,F-91405 Orsay Cedex, France}

\author{V.~Hinkov}
\affiliation{Max-Planck-Institut für Festkörperforschung, Heisenbergstraße 1, D-70569 Stuttgart, Germany}
\affiliation{Physikalisches Institut, Julius-Maximilians-Universität Würzburg, Am Hubland, D-97074 Würzburg}

\author{D.\,L.\,Sun}
\affiliation{Max-Planck-Institut für Festkörperforschung, Heisenbergstraße 1, D-70569 Stuttgart, Germany}

\author{Y.~Liu}
\affiliation{Max-Planck-Institut für Festkörperforschung, Heisenbergstraße 1, D-70569 Stuttgart, Germany}

\author{R.\,Khasanov}
\affiliation{Laboratory for Muon Spin Spectroscopy, Paul Scherrer Institut, CH-5232 Villigen PSI, Switzerland}

\author{K.\,Sedlak}
\affiliation{Laboratory for Muon Spin Spectroscopy, Paul Scherrer Institut, CH-5232 Villigen PSI, Switzerland}

\author{Ph.\,Bourges}
\affiliation{Laboratoire L{\' e}on Brillouin, CEA-CNRS, CEA Saclay, F-91191 Gif-sur-Yvette Cedex, France}

\author{Y.\,Sidis}
\affiliation{Laboratoire L{\' e}on Brillouin, CEA-CNRS, CEA Saclay, F-91191 Gif-sur-Yvette Cedex, France}

\author{A.\,Ivanov}
\affiliation{Institut Laue-Langevin, 6 rue Jules Horowitz, F-38042 Grenoble Cedex 9, France}

\author{C.\,T.~Lin}
\affiliation{Max-Planck-Institut für Festkörperforschung, Heisenbergstraße 1, D-70569 Stuttgart, Germany}

\author{T.~Keller}
\affiliation{Max-Planck-Institut für Festkörperforschung, Heisenbergstraße 1, D-70569 Stuttgart, Germany}
\affiliation{Forschungsneutronenquelle Heinz Maier-Leibnitz (FRM-II), TU München, D-85747 Garching, Germany}

\author{B.\,Keimer}
\affiliation{Max-Planck-Institut für Festkörperforschung, Heisenbergstraße 1, D-70569 Stuttgart, Germany}

\begin{abstract}
\noindent We investigate magnetic ordering in metallic Ba(Fe$_{1-x}$Mn$_x$)$_2$As$_2$ and discuss the unusual magnetic phase, which was recently discovered for Mn concentrations $x>10$\%. We argue that it can be understood as a Griffiths-type phase that forms above the quantum critical point associated with the suppression of the stripe-antiferromagnetic spin-density-wave (SDW) order in BaFe$_2$As$_2$ by the randomly introduced localized Mn moments acting as strong magnetic impurities. While the SDW transition at $x=0$, 2.5\% and 5\% remains equally sharp, in the $x=12$\% sample we observe an abrupt smearing of the antiferromagnetic transition in temperature and a considerable suppression of the spin gap in the magnetic excitation spectrum. According to our muon-spin-relaxation, nuclear magnetic resonance and neutron-scattering data, antiferromagnetically ordered rare regions start forming in the $x=12$\% sample significantly above the N\'eel temperature of the parent compound. Upon cooling, their volume grows continuously, leading to an increase in the magnetic Bragg intensity and to the gradual opening of a partial spin gap in the magnetic excitation spectrum. Using neutron Larmor diffraction, we also demonstrate that the magnetically ordered volume is characterized by a finite orthorhombic distortion, which could not be resolved in previous diffraction studies most probably due to its coexistence with the tetragonal phase and a microstrain-induced broadening of the Bragg reflections. We argue that Ba(Fe$_{1-x}$Mn$_x$)$_2$As$_2$ could represent an interesting model spin-glass system, in which localized magnetic moments are randomly embedded into a SDW metal with Fermi surface nesting.
\end{abstract}

\keywords{Griffiths phase, spin waves, antiferromagnetism, anisotropy gap, inelastic neutron scattering, iron pnictide superconductors}
\pacs{75.50.Lk 75.50.Ee 74.70.Xa 75.30.Ds 76.75.+i 78.70.Nx}

\maketitle\enlargethispage{3pt}

\vspace{-5pt}\section{Introduction}\enlargethispage{8pt}

\vspace{-5pt}\subsection{Magnetic phase transitions in disordered systems}\vspace{-5pt}

It is well established that intrinsic randomness, often present in real condensed-matter systems in the form of quenched substitutional disorder, can exert a crucial influence on the behavior of the system's thermodynamic parameters close to a phase transition \cite{Vojta06, LohneysenRosch07}. Such effects have been studied in detail in several model systems, most notably in disordered Ising or Heisenberg ferro- and antiferromagnets \cite{Griffiths69, BallesterosFernandez98, BhattLee82, Sandvik02, VajkGreven02, SknepnekVojta04, VojtaSchmalian05}. It has been demonstrated that sufficiently strong disorder can alter the critical scaling behavior of a phase transition, or even lead to the appearance of qualitatively new electronic or magnetic states. In particular, quantum phase transitions can be smeared because of the coexistence of disordered (paramagnetic) regions and locally ordered clusters within the so-called Griffiths region of a phase diagram \cite{Griffiths69, TanaskovicMiranda04, LohneysenRosch07}, which has been observed experimentally in various real materials \cite{BinekKleemann98, SalamonLin02, SalamonChun03, WangSun07, GuoYoung08, GuoYoung10, KrivoruchkoMarchenko10, EreminaFazlizhanov11}.

The specifics of itinerant magnetic systems \cite{Vojta10, NozadzeVojta11, NozadzeVojta12}, which are of the most relevance to our present study, is determined by the presence of long-range Ruderman-Kittel-Kasuya-Yosida (RKKY) interactions \cite{RudermanKittel54, Kasuya56, Yosida57, FischerKlein75} between local magnetic moments that induce correlations between the magnetically ordered rare regions, leading to the formation of so-called cluster glass (CG) phases preceding uniform ordering \cite{CastroNetoJones00, DobrosavljevicMiranda05, WesterkampDeppe09, UbaidKassisVojta10}. At present, theoretical understanding of rare-region effects in itinerant systems still remains a topic of active research and is yet far from being complete \cite{Vojta10, NozadzeVojta11}. It has been also noted \cite{NozadzeVojta11} that most of the experimental reports of Griffiths-type phases in itinerant systems are concerned with ferromagnetic metals, while there are barely any clear-cut experimental observations of such phases in itinerant antiferromagnets. Metallic compounds with pronounced Fermi-surface nesting, which are close to a spin-density-wave (SDW) instability, are especially promising as model systems for demonstration of the above-mentioned effects, because the RKKY interaction is known to be enhanced at the nesting vector \cite{InosovEvtushinsky09}. Hence, if localized magnetic moments are randomly embedded into such a metal to form a so-called RKKY spin glass (SG) \cite{ShellCowen82, BinderYoung86, FischerHertz99}, the long-range superexchange between them \cite{BulaevskiiPanyukov86} is expected to support magnetic correlations between antiferromagnetic (AFM) rare regions with the same SDW wave vector. The RKKY interaction in layered metals with Fermi surface nesting has been considered theoretically, for example, in Refs.~\citenum{AristovMaleyev97, AkbariEremin11, AkbariThalmeier13}. However, thermodynamic properties of such strongly nested systems with randomly embedded local magnetic moments have not been investigated, to the best of our knowledge.

\vspace{-4pt}\subsection{Phase diagram of Ba(Fe$_{1-x}$Mn$_x$)$_2$As$_2$}\vspace{-5pt}

Layered iron pnictides \cite{RenZhao09} are among the most actively studied metallic materials, in which Fermi surface nesting is generally considered to be responsible for the formation of an AFM spin-density-wave state at low temperatures \cite{LumsdenChristianson10review}. They have attracted enormous attention in recent years mainly because of the high superconducting transition temperatures that can be induced in these systems by chemical substitution or pressure \cite{Chu09, PaglioneGreene10, Johnston10, Stewart11}. In particular, the so-called `122' compounds with the body-centered-tetragonal ThCr$_2$Si$_2$-type structure, such as $A$Fe$_2$As$_2$ ($A$\,=\,Ba, Sr or Ca), usually exhibit superconductivity upon transition-metal doping on the Fe site \cite{CanfieldBudko10}. Prominent exceptions are Mn- and Cr-substituted systems \cite{SefatSingh09, MartyChristianson11, PandeyAnand11, KimKhim10, ThalerHodovanets11, KimKreyssig10, KimPratt11}, which exhibit no superconductivity, but instead show unusual magnetic behavior that is not typical for their stoichiometric parent compounds. Moreover, it has been demonstrated that substituting Mn for Fe in a hole-doped Ba$_{1-x}$K$_x$Fe$_2$As$_2$ leads to a much more rapid suppression of the superconducting transition temperature, $T_\text{c}$, as compared to other transition-metal elements \cite{ChengShen10, LiGuo12}. Our recent nuclear magnetic resonance (NMR) measurements \cite{TexierLaplace12} indicate that this distinct behavior results from the localization of additional Mn holes, which prevents the change in the electron count within the conductance band, in contrast to Co or Ni dopants, but instead stabilizes local magnetic moments on the Mn sites. Their absolute value was initially assessed at 2.58\,$\mu_\text{B}$ from dc magnetization measurements \cite{ChengShen10}, yet this quantity is likely overestimated according to a more recent analysis \cite{Bobroff_private}. Such a localized magnetic behavior extends to the pure and doped BaMn$_2$As$_2$ compounds, in which large spin-5/2 local moments have also been reported \cite{SinghEllern09, SinghGreen09, JohnstonMcQueeney11, PandeyDhaka12, BaoJiang12, LamsalTucker13}.

The Ba(Fe$_{1-x}$Mn$_x$)$_2$As$_2$ (BFMA) system reportedly changes its ground-state structure from orthorhombic to tetragonal at a critical Mn concentration of $x_\text{c}\approx10$\%, while its $(\piup,0)$ magnetic ordering wave vector remains unchanged \cite{KimKreyssig10, KimPratt11}. This observation is surprising, because the anisotropic arrangement of magnetic moments in the stripe-AFM state, characterized by this propagation vector, is expected to break the tetragonal symmetry of the crystal and naturally lead to an orthorhombic distortion, as it happens in the BaFe$_2$As$_2$ and in many other iron pnictides. The SDW ordering temperature, $T_\text{N}$, is initially reduced upon Mn substitution, like in Sr(Fe$_{1-x}$Mn$_x$)$_2$As$_2$ \cite{KasinathanOrmeci09}, for $x<x_\text{c}$, but starts to increase again above this critical concentration. This is accompanied by a drastic broadening of the phase transition in temperature \cite{ThalerHodovanets11, KimKreyssig10}. So far, both the unusual suppression of the structural distortion and this nonmonotonic behavior of the ordering temperature remain unexplained. They appear to be unique to BFMA, as they are not observed in the very similar Ba(Fe$_{1-x}$Cr$_x$)$_2$As$_2$ system, which changes its ground state abruptly from the stripe-AFM SDW to a checkerboard (G-type) AFM order, typical for pure BaCr$_2$As$_2$ \cite{SinghSefat09}, at $\sim$\,30\% Cr concentration \cite{MartyChristianson11}.

\begin{figure}[t]
\includegraphics[width=\columnwidth]{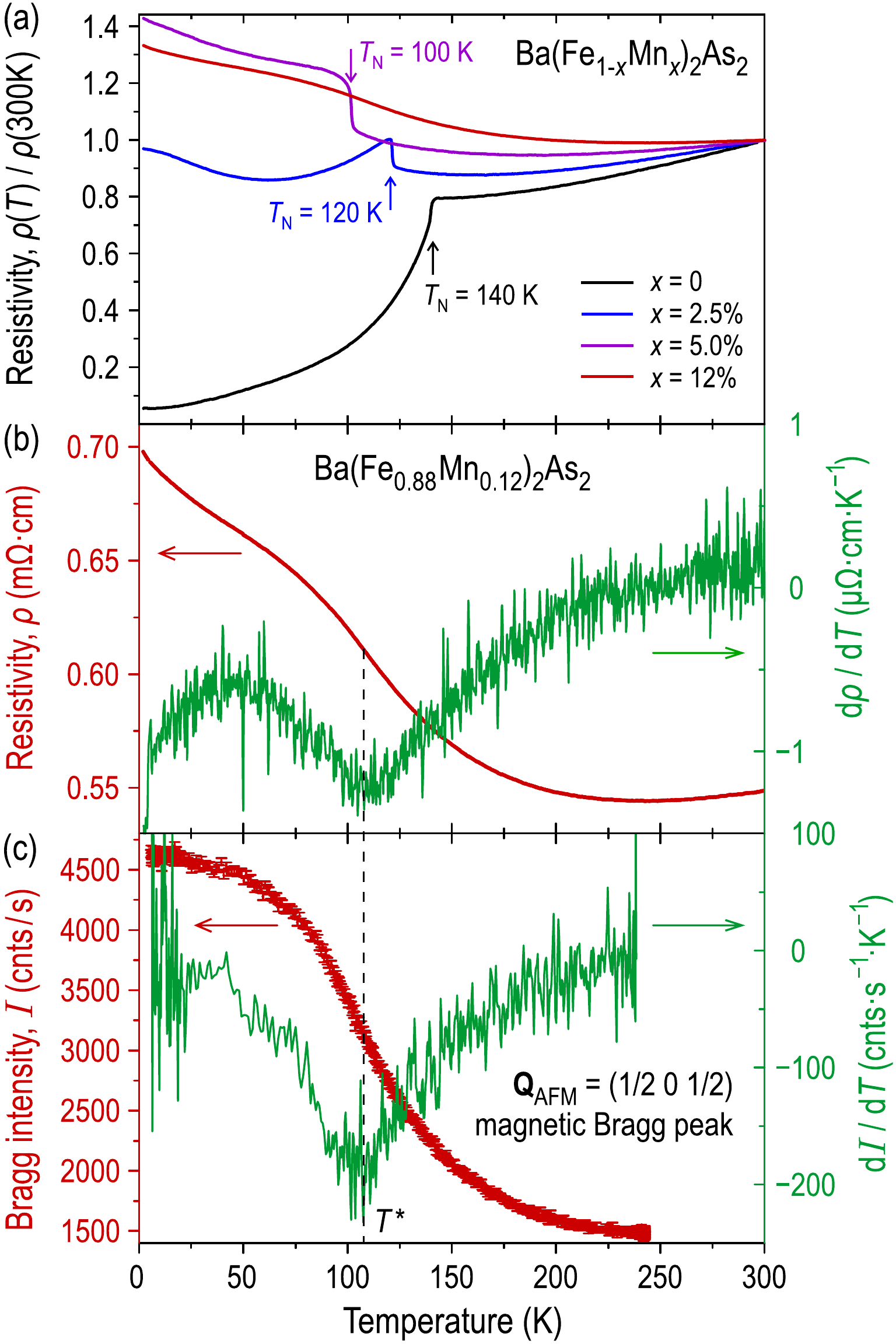}
\caption{Characterization of the magnetic transitions in Ba(Fe$_{1-x}$Mn$_x$)$_2$As$_2$. (a)~Temperature dependence of the normalized in-plane resistivity, $\rho(T)/\rho(\text{300\,K})$, for all samples used in the present study. (b)~In-plane resistivity (smooth curve) and its temperature derivative (noisy curve) for the $x=12$\% sample, exhibiting an inflection-point anomaly at $T^\ast\approx105$\,K. (c)~Temperature dependence of the elastic neutron scattering intensity at the $(\smash{\half}0\smash{\half})_{\text{Fe}_1}$ magnetic Bragg peak position (monotonic curve) and its temperature derivative \mbox{with a minimum at $T^\ast$.}}
\label{Fig:Transition}
\end{figure}

Finally, in a recent inelastic neutron scattering (INS) experiment on a BFMA sample with $x=7.5$\% ($x<x_\text{c}$, $T_\text{N}=80$\,K), the presence of an additional branch of short-range quasielastic spin fluctuations was demonstrated at the $(\piup,\piup)$ wave vector, corresponding to the checkerboard-type AFM order that is not observed in the parent compound \cite{TuckerPratt12}. This result indicates a tendency to the formation of antiferromagnetically polarized N\'eel regions around Mn local moments, which compete with the stripe SDW order of the parent compound and are likely responsible for the initial reduction of $T_\text{N}$ at low Mn concentrations ($x<x_\text{c}$).

\vspace{-5pt}\section{Sample preparation and characterization}\label{Sec:Characterization}

\vspace{-5pt}\subsection{Single crystals of Ba(Fe$_{1-x}$Mn$_x$)$_2$As$_2$}\vspace{-5pt}

For the present study, we used three single-crystalline BFMA samples with Mn concentrations of 2.5\%, 5.0\%, and 12\% and a reference sample of the pure parent BaFe$_2$As$_2$ compound. These samples are identical to those studied in Refs.\,\citenum{TexierLaplace12} and \citenum{ParkFriemel12}, respectively. All single crystals were grown from self-flux in zirconia crucibles sealed in quartz ampoules under argon atmosphere, as described elsewhere \cite{LiuSun10}. All four compositions have been characterized using dc resistivity, NMR, and muon spin relaxation ($\mu$SR) spectroscopy. INS experiments were performed only on the $x=0$ and $x=12$\% samples, which represented arrays of multiple single crystals with a total mass of the order of 1\,g, coaligned to a mosaicity of $\sim$\,2$^\circ$ using a real-time digital x-ray Laue backscattering camera. In addition, the $x=12$\% sample was investigated by neutron Larmor diffraction. The lattice parameters corresponding to this composition, as measured on a triple-axis neutron spectrometer during sample alignment at room temperature, were $a=b=3.97(4)$\,\AA\ (which is nearly the same as in BaFe$_2$As$_2$) and $c=13.44(5)$\,\AA\ (about 1\% larger than in BaFe$_2$As$_2$ \cite{RotterTegel08PRB}). These relative changes in the unit cell dimensions are similar to those reported for Sr(Fe$_{1-x}$Mn$_x$)$_2$As$_2$ in an earlier study \cite{KimKhim10}.

\vspace{-5pt}\subsection{Resistivity and elastic neutron scattering}\vspace{-5pt}

The temperature dependence of the in-plane resistivity, $\rho(T)$, for all four BFMA samples, normalized to its room-temperature values, is shown in Fig.\,\ref{Fig:Transition}\,(a). In agreement with Ref.\,\onlinecite{KimKreyssig10}, we observe sharp anomalies in $\rho(T)$ at the SDW transition for all samples with $x<x_\text{c}$, whereas for the $x=12$\% sample the resistivity curve is smooth. This observation is consistent with the absence of anomalies in the temperature dependence of the specific heat \cite{PopovichBoris10}. Only after differentiation [Fig.\,\ref{Fig:Transition}\,(b)], an inflection point is revealed near $T^\ast\approx105$\,K, somewhat above the SDW transition temperature of the $x=5.0$\% sample, in agreement with the increasing tendency for $T^\ast$ in this composition range that was reported in Ref.\,\onlinecite{KimKreyssig10}.

\begin{figure}[b]
\includegraphics[width=\columnwidth]{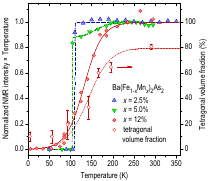}
\caption{Temperature dependence of the main $^{75}$As NMR line intensity (filled symbols) for samples with different Mn concentrations, normalized to the respective high-temperature saturation values. Empty circles show the volume fraction of the tetragonal phase in the $x=12$\% sample (right vertical axis), as measured by neutron Larmor diffraction (see text).\vspace{-4pt}}
\label{Fig:NMR_wipeout}
\end{figure}

To establish the origin of this $T^\ast$-anomaly in the resistivity, in Fig.\,\ref{Fig:Transition}\,(c) we compare it with the temperature dependence of the magnetic Bragg intensity (without background subtraction), measured on the same sample at the $(\half 0 \half)_{\text{Fe}_1}$ magnetic Bragg peak by means of elastic neutron scattering. Here and henceforth, the subscript ``Fe$_1$'' indicates that the reciprocal-lattice vector, $(H\,K\,L)$, is given in the unfolded notation corresponding to the Fe-sublattice (one Fe atom per unit cell) \cite{ParkInosov10}, and its coordinates are presented in reciprocal lattice units (r.l.u.), defined as 1\,$\text{r.l.u.} = 2\sqrt{2}\piup/a$ for the $H$ and $K$ directions and as 1\,$\text{r.l.u.} = 4\piup/c$ along the $L$ direction, where $a$ and $c$ are the lattice constants of the crystal in the tetragonal ($I4/mmm$) symmetry. First, we note that in contrast to the sharp order-parameter-like onset of the magnetic Bragg scattering at $T_\text{N}$ that is typical for most iron-arsenide parent compounds \cite{CruzHuang08, HuangQiu08, ZhaoRatcliff08, KanekoHoser08, GoldmanArgyriou08, ParkFriemel12}, here we see a smeared transition with a gradual onset around $\sim$\,240\,K, which lies approximately 100\,K above the ordering temperature of BaFe$_2$As$_2$. One possible explanation for this smearing, which we will later substantiate by direct measurements, is a disorder-induced separation of the sample into spacial regions with different local values of $T_\text{N}$ that leads to a gradual change of the magnetically ordered volume with temperature. However, the conventional random-$T_\text{N}$ type of disorder \cite{Vojta06} alone, which one would expect from a locally inhomogeneous distribution of the Mn atoms, can not explain the dramatic enhancement of the onset temperature. Indeed, at small Mn concentrations, $T_\text{N}$ is suppressed as a function of $x$ and therefore an inhomogeneous Mn distribution should result in the spread of local $T_\text{N}$ values between zero and at most 140\,K, i.e. we would normally expect it to be limited from above by the transition temperature of the parent compound. This conventional type of behavior is found, for instance, in Ba(Fe$_{0.99}$Ni$_{0.01}$)$_2$As$_2$, where despite the strong disorder the transition is merely suppressed by Ni substitution with no significant broadening, according to a recent $^{57}$Fe Mössbauer spectroscopy study \cite{OlariuBonville12}. In contrast, the behavior of magnetic Bragg intensity in Ba(Fe$_{0.88}$Mn$_{0.12}$)$_2$As$_2$ is qualitatively different, because at 140\,K it already reaches 27\% of its saturation value, suggesting that the local $T_\text{N}$ exceeds that of the pure BaFe$_2$As$_2$ in approximately 1/4 of the sample volume. Hence, we must conclude that although individual Mn impurities tend to suppress the ordering temperature, at sufficiently large concentrations (perhaps at $x\gtrsim x_\text{c}$) there exists an increasing probability of finding certain local configurations of Mn moments (rare regions) that stabilize the $(\piup,0)$ type of order sufficiently to reverse the downward trend in the onset temperature, as can be seen in the published phase diagram \cite{KimKreyssig10}. For this to happen, collective effects of several Mn moments (deviation from the dilute limit) must be at play. In Fig.\,\ref{Fig:Transition}\,(c), we also show the temperature derivative of the magnetic Bragg intensity, whose striking similarity with the $\mathrm{d}\rho(T)/\mathrm{d}T$ curve in Fig.\,\ref{Fig:Transition}\,(b) leaves no doubt about the magnetic origin of the $T^\ast$-anomaly.

\vspace{-5pt}\subsection{Nuclear magnetic resonance}\vspace{-5pt}

In Ref.\,\onlinecite{TexierLaplace12}, we already reported a detailed NMR study performed on the same set of BFMA samples. Without reiterating the results of that work, here we will only be interested in the $T$-dependence of the paramagnetic (PM) volume fraction, which can be directly measured by following the main $^{75}$As NMR line wipeout as a function of temperature. The NMR line intensity, multiplied by temperature, is plotted in Fig.\,\ref{Fig:NMR_wipeout} for samples with different Mn content. For the convenience of comparison, the high-temperature saturation values for every dataset were normalized to unity. The plotted quantity therefore serves as a direct gauge of the nonmagnetic fraction of the sample volume. For both $x=2.5$\% and $x=5.0$\%, the NMR line intensity sharply drops to zero at the SDW ordering temperature, indicating a transition to the magnetically ordered state in the whole volume of the sample: The freezing of the Fe moments results in a strong shift of the NMR line out of our limited observation window. In the $x=12$\% sample, however, a gradual intensity drop starts already near $\sim$\,240\,K, well above the ordering temperature of the parent compound, and progresses down to $\widetilde{T}_\text{N}\approx50$\,K, where the entire signal is lost. The shape of the transition curve is strikingly similar to that of the magnetic Bragg intensity in Fig.\,\ref{Fig:Transition}\,(c), which unequivocally confirms that the smearing of the magnetic transition occurs due to the gradual expansion of the regions with static magnetic moments and to the corresponding reduction in the PM volume upon cooling, most naturally explained by the broad distribution of the local ordering temperatures. We note that even in the $x=5.0$\% sample, a small, but similarly gradual wipeout of the NMR line can be seen below 200\,K, which leads to only a 10\% reduction of the PM volume upon reaching $T_\text{N}$.

\vspace{-5pt}\subsection{Neutron Larmor diffraction and orthorhombicity}\label{Sec:Larmor1}\vspace{-5pt}

Perhaps the most surprising property of the BFMA system, according to previous neutron and x-ray diffraction studies \cite{KimKreyssig10}, is the complete suppression of the tetragonal-to-orthorhombic structural phase transition for $x>x_\text{c}$, which reportedly holds down to the lowest temperatures despite the presence of the well established $(\pi,0,\pi)_{\text{Fe}_1}$ stripe-AFM order that appears to be identical to that in the parent compound. This observation is very difficult to explain, because the stripe-AFM order obviously breaks the $C_4$ rotational symmetry, and the corresponding orthorhombic distortion is anticipated due to the non-vanishing magnetoelastic coupling. In Ref.\,\onlinecite{KimKreyssig10}, the authors speculate that a new double-$\mathbf{Q}$ magnetic structure with an order parameter of the form $\Delta_1\mathrm{e}^{\mathrm{i}\,(\pi,0)\cdot\mathbf{R}}+\Delta_2\mathrm{e}^{\mathrm{i}\,(0,\pi)\cdot\mathbf{R}}$ (with both $\Delta_1\neq0$ and $\Delta_2\neq0$), theoretically suggested by Eremin and Chubukov \cite{EreminChubukov10}, could be reconciled with their experimental observations. We find this explanation theoretically compelling, yet unpersuasive, as it is hard to imagine that in the presence of very strong magnetic disorder and the dramatically broadened distribution of the local transition temperatures, the system could keep the delicate balance between the $\Delta_1$ and $\Delta_2$ order parameters over macroscopic volumes. Apparently, such an exotic order that has never been observed in any clean iron-pnictide compound requires precisely tuned conditions to be stabilized, which are unlikely in a magnetically inhomogeneous system with randomly embedded local moments.

\begin{figure}[t!]
\includegraphics[width=\columnwidth]{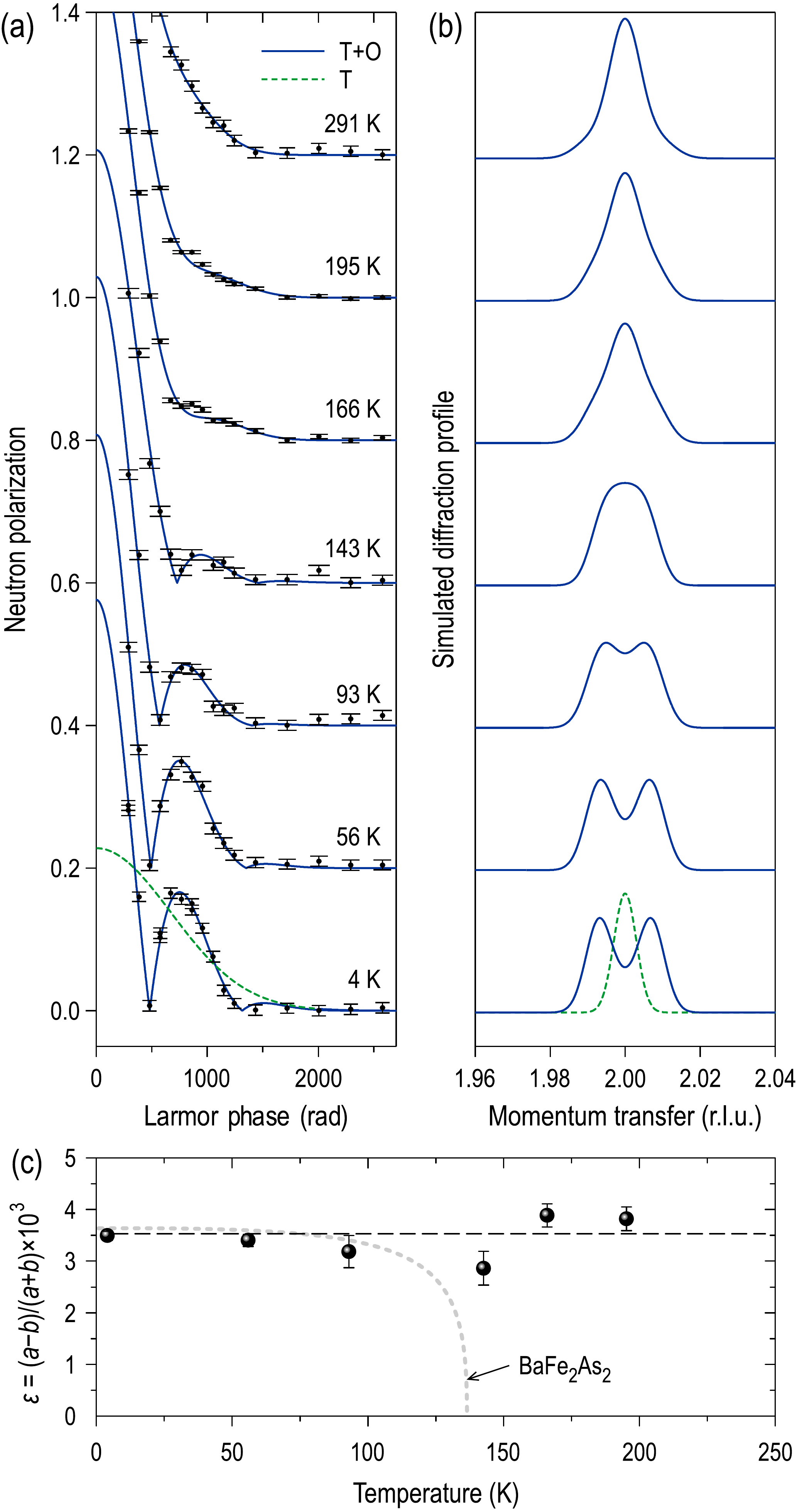}
\caption{Neutron Larmor-diffraction measurements of the orthorhombic splitting in Ba(Fe$_{0.88}$Mn$_{0.12}$)$_2$As$_2$. (a)~Experimental data for different temperatures (indicated above each curve), fitted to a model containing a mixture of the orthorhombic (O) and tetragonal (T) phases (solid lines). The dashed line shows a failed fit of the $T=4$\,K data assuming a single tetragonal phase  \cite{KimKreyssig10}. For clarity, each dataset is offset vertically by an increment of 0.2 units from the one below it. (b)~Modeled diffraction profiles, corresponding to every temperature in panel (a), as they would look like in an x-ray diffraction experiment with infinitesimally small resolution. These models account for the experimentally determined orthorhombic splitting, ratio of the tetragonal and orthorhombic phase volumes, and the peak broadening due to the finite width of the microstrain distribution, as extracted from the fits in panel (a). (c)~Temperature dependence of the orthorhombicity parameter, $\varepsilon=(a-b)/(a+b)$, extracted from the same fits. The dashed line is a temperature-independent fit. The grey dotted line is the corresponding dependence for the parent BaFe$_2$As$_2$, reproduced from Ref.\,\onlinecite{BlombergKreyssig12} for comparison.}
\label{Fig:Larmor}\vspace{-10em}
\end{figure}

In search for an alternative explanation for the missing orthorhombicity, we have performed neutron Larmor diffraction measurements on our $x=12$\% sample, which is very similar to the $x=11.8$\% sample from Ref.\,\onlinecite{KimKreyssig10}, if judged by the shape of the resistive transition, the temperature dependence of the magnetic Bragg intensity, and the value of $T^\ast$. Neutron Larmor diffraction \cite{Rekveldt00, RekveldtKeller01, RekveldtKraan02} is a polarized-neutron technique known to be extremely sensitive to minor structural distortions and the lattice-spacing spread, ${\scriptstyle\Delta}d/d$, with resolution better than $10^{-5}$, which does not depend on beam collimation and monochromaticity and is independent of the mosaic spread. The detailed principle of this technique is explained, for instance, in Ref.\,\onlinecite{KellerRekveldt02}.

Our measurements were done at the neutron resonant spin-echo triple-axis spectrometer TRISP installed at the FRM-II research reactor in Garching, Germany. The neutron polarization was measured as a function of the Larmor precession phase, controlled by the magnitude of the magnetic field that was applied in the same direction before and after the sample. To be sensitive to variations in the $d$-spacing of the $(200)_{\text{Fe}_1}$ Bragg reflection, the magnetic field boundaries were made parallel to the (200)$\mathrm{_{Fe_{1}}}$ Bragg planes. The results are shown in Fig.\,\ref{Fig:Larmor}\,(a). In Larmor diffraction, the measured polarized-neutron intensity is proportional to the Fourier transform of the $d$-spacing distribution \cite{RekveldtKraan02, KellerRekveldt02}. This means that for a single mean value of $d$, distributed with a certain full width at half maximum (FWHM), the measured signal would monotonically decrease with increasing magnetic field (increasing Larmor phase). However, for two closely spaced characteristic values of $d$, one will observe destructive and constructive interference in the measured neutron polarization. Larmor diffraction is therefore highly sensitive to orthorhombic distortions, as it can distinguish very clearly between a single Bragg peak in the case of a tetragonal crystal and a pair of peaks that are split due to an orthorhombic distortion, even if this splitting is too small to be resolved by conventional neutron or x-ray diffraction.

The appearance of a pronounced minimum in the low-temperature data measured on the $x=12$\% sample [Fig.\,\ref{Fig:Larmor}(a), bottom curve] is therefore definitive evidence $\hspace*{\columnwidth}\vspace*{6em}$ $\hspace*{\columnwidth}$ that the majority of the sample is orthorhombic. At higher temperatures, it proved impossible to fit the data under the assumption that the whole sample was either orthorhombic or tetragonal. However, by assuming a \emph{coexistence} of orthorhombic and tetragonal phases, the data could be fitted consistently at all temperatures, with all parameters nearly independent of temperature apart from the orthorhombic and tetragonal fractions of the sample volume. The latter fraction is plotted in Fig.\,\ref{Fig:NMR_wipeout} as a function of temperature (empty symbols), showing an increase upon warming that is consistent with that of the PM volume fraction measured on the same sample by NMR and exhibiting a similarly broadened transition with a comparable width and centered at approximately the same temperature. Note that at high temperatures, the fitting of the Larmor diffraction data systematically underestimates the tetragonal volume fraction by $\sim$\,20\%, which is most likely due to a deviation of the ${\scriptstyle\Delta}d/d$ distribution from a perfect Gaussian shape that cannot be trivially accounted for. In the experimental data, such a deviation is difficult to distinguish from a small admixture of the orthorhombic phase, which explains the 20\% reduction of the high-temperature saturation value in Fig.\,\ref{Fig:NMR_wipeout} from the expected 100\%. Otherwise, the similar shapes of the curves describing the evolution of the PM and tetragonal volume fractions let us conclude that only the PM part of the sample remains tetragonal, whereas the remaining magnetically ordered fraction is orthorhombic. The corresponding orthorhombicity parameter, obtained from the same fits and plotted in Fig.\,3\,(c), turns out to be nearly independent of temperature, with a mean magnitude of $(a-b)/(a+b) = 3.5(1) \times 10^{-3}$ that is almost identical to that found in the orthorhombic phase of the undoped BaFe$_2$As$_2$ (Ref.\,\onlinecite{HuangQiu08}).

\makeatletter\renewcommand{\fnum@figure}[1]{\figurename~\thefigure.}\makeatother
\begin{figure}[b]
\hspace*{-1ex}\includegraphics[width=0.85\columnwidth]{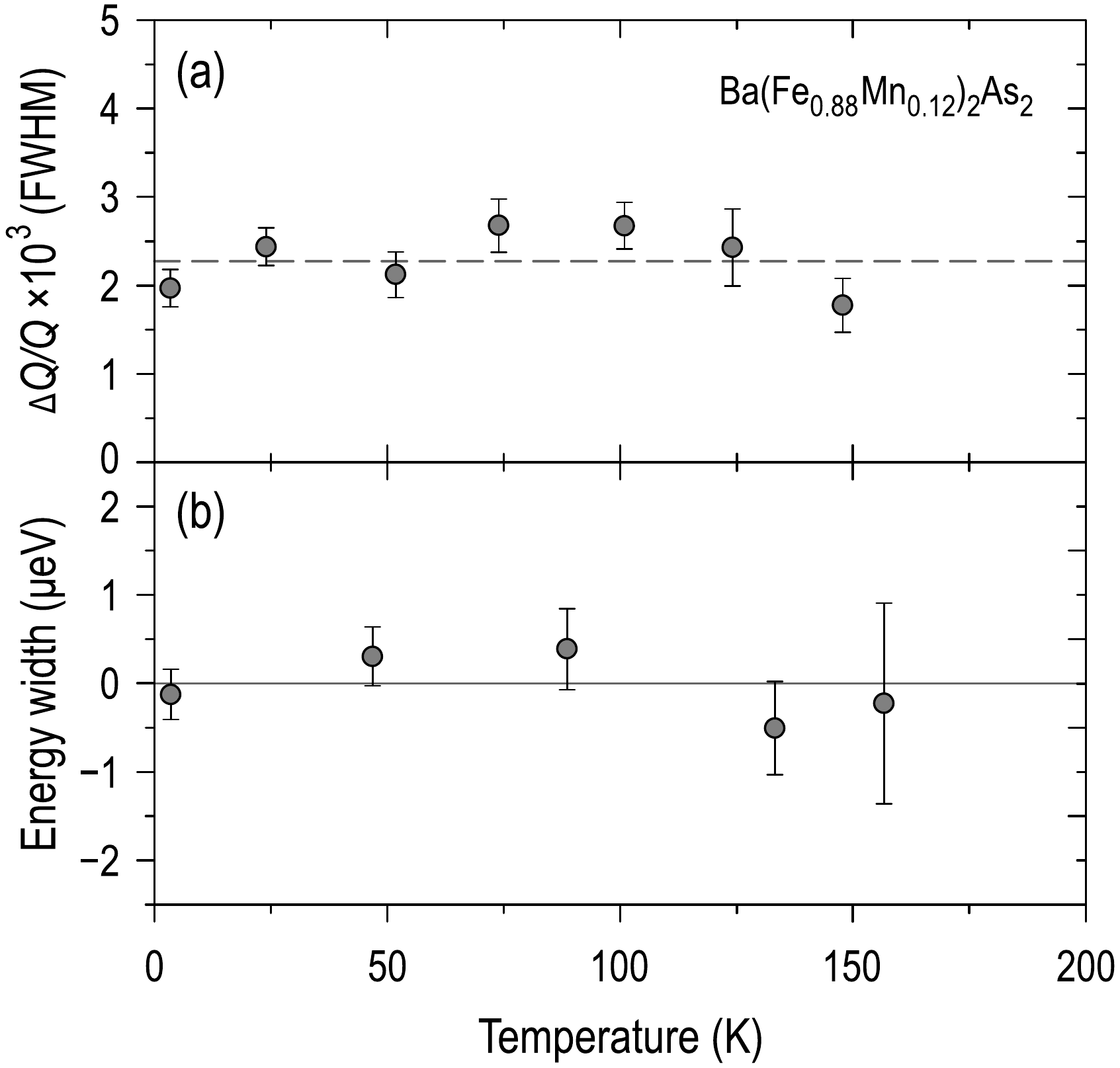}
\caption{(a)~Temperature dependence of the momentum width, ${\scriptstyle\Delta}Q/Q$, measured on the $(\half0\frac{7}{\protect\raisebox{0.8pt}{\scriptsize 2}})_{\text{Fe}_1}$ magnetic Bragg reflection. (b)~The energy width measured on the $(\half0\half)_{\text{Fe}_1}$ magnetic Bragg reflection vs. temperature. Measurements above 160\,K were unfeasible due to the dramatically reduced intensity of the signal.\vspace{-4pt}}
\label{Fig:Larmor_widths}
\end{figure}
\makeatletter\renewcommand{\fnum@figure}[1]{\figurename~\thefigure~(color online).}\makeatother

As another parameter of the fits in Fig.\,\ref{Fig:Larmor}\,(a), we have also obtained the FWHM of the microstrain distribution, which describes the lattice-spacing spread, ${\scriptstyle\Delta}d/d$, and the intrinsic width of the Bragg reflection that would be measured in a conventional diffraction experiment if both the diffractometer resolution and the sample mosaic were infinitesimally small. For the $(200)_{\text{Fe}_1}$ Bragg peak, this width is nearly independent of temperature and amounts to ${\scriptstyle\Delta}d/d = 4.6(1) \times 10^{-3}$, which is comparable to the orthorhombic distortion. For the out-of-plane $(002)_{\text{Fe}_1}$ reflection, ${\scriptstyle\Delta}d/d$ marginally increases from $1.37(1)\times 10^{-3}$ at room temperature to $1.44(1)\times 10^{-3}$ at $T=6$\,K. In Fig.\,3\,(b), we reconstruct the scattering function, $S(Q)$, from the parameters of the fits in Fig.\,3\,(a). These model curves correspond to the longitudinal Bragg-peak profiles that would be measured in a conventional x-ray or neutron diffraction experiment under the assumption of an infinitesimally small diffractometer reso\-lution. Even at the lowest temperature of 4\,K, we observe some intrinsic overlap of the two orthorhombic peaks due to the broad microstrain distribution, so there is no doubt that the sizeable intrinsic variation of the $d$-spacing would make it exceedingly difficult to observe the orthorhombic distortion directly using traditional diffraction methods. At higher temperatures, the splitting would be additionally masked by the coexisting tetragonal phase. This appears to be the most likely reason for the reported absence of orthorhombicity in a similar sample \cite{KimKreyssig10}.

\vspace{-5pt}\subsection{Intrinsic width of the magnetic Bragg peaks}\vspace{-5pt}

We now turn our attention to the evolution of the momentum and energy widths of the magnetic Bragg peaks with temperature in the $x=12$\% sample. The momentum width of the $(\half0\frac{7}{\protect\raisebox{0.8pt}{\scriptsize 2}})_{\text{Fe}_1}$ magnetic Bragg peak was measured using Larmor diffraction in the same experimental setup as described in section \ref{Sec:Larmor1}. We find no temperature dependence of this width up to 150\,K [Fig.\,\ref{Fig:Larmor_widths}\,(a)], with the mean value of the normalized FWHM ${\scriptstyle\Delta}Q/Q = 2.3(1) \times 10^{-3}$. In general, the momentum width of a commensurate magnetic Bragg peak is determined by both the structural microstrain ${\scriptstyle\Delta}d/d$ and the size of the ordered magnetic domains that could lead to an additional finite-size broadening. One might expect that since the magnetically ordered fraction of the sample becomes smaller with increasing temperature, the ordered magnetic domains would shrink upon warming, thereby increasing the momentum width. However, in our case we find the momentum width to be independent of temperature, which suggests that the magnetic ordering remains long range at least up to 150\,K. Under this assumption, the sole source of the broadening is the structural microstrain, which in the case of the $(\half0\frac{7}{\protect\raisebox{0.8pt}{\scriptsize 2}})_{\text{Fe}_1}$ magnetic Bragg peak lies between the values of ${\scriptstyle\Delta}d/d$ that were found in section \ref{Sec:Larmor1} for the $(200)_{\text{Fe}_1}$ and $(002)_{\text{Fe}_1}$ structural peaks. Such an anisotropy in the width of the microstrain distribution is typical for the iron pnictides and has been \mbox{already reported previously} \cite{InosovLeineweber09}.

The magnetic Bragg peak energy width was measured using the neutron resonance spin-echo (NRSE) technique at the TRISP spectrometer. In NRSE, the dependence of neutron polarization on the magnitude of the magnetic fields before and after the sample is proportional to the Fourier transform of the lineshape of magnetic fluctuations \cite{Keller03}. NRSE spectroscopy routinely provides accurate measurements of energy widths down to the $\mu$eV range at TRISP. In Fig.\,\ref{Fig:Larmor_widths}\,(b) we show the energy width of the $(\half 0 \half)_{\text{Fe}_1}$ magnetic Bragg peak in the $x=12$\% BFMA sample as a function of temperature. We find that the width is vanishingly small at all temperatures, meaning that the observed peak remains static and shows no quasielastic behavior up to at least 160\,K within our instrumental resolution. In other words, its
characteristic lifetime $\tau$ is longer than $\sim$\,1\,ns, which is the typical timescale over which the NRSE measurement was sensitive. We therefore conclude that the magnetic order in BFMA remains truly static and long range above the critical Mn concentration even at temperatures that are comparable with the $T_\text{N}$ of the parent compound.

\vspace{-5pt}\subsection{Thermal expansion coefficient}\vspace{-5pt}

\begin{figure}[t]
\includegraphics[width=\columnwidth]{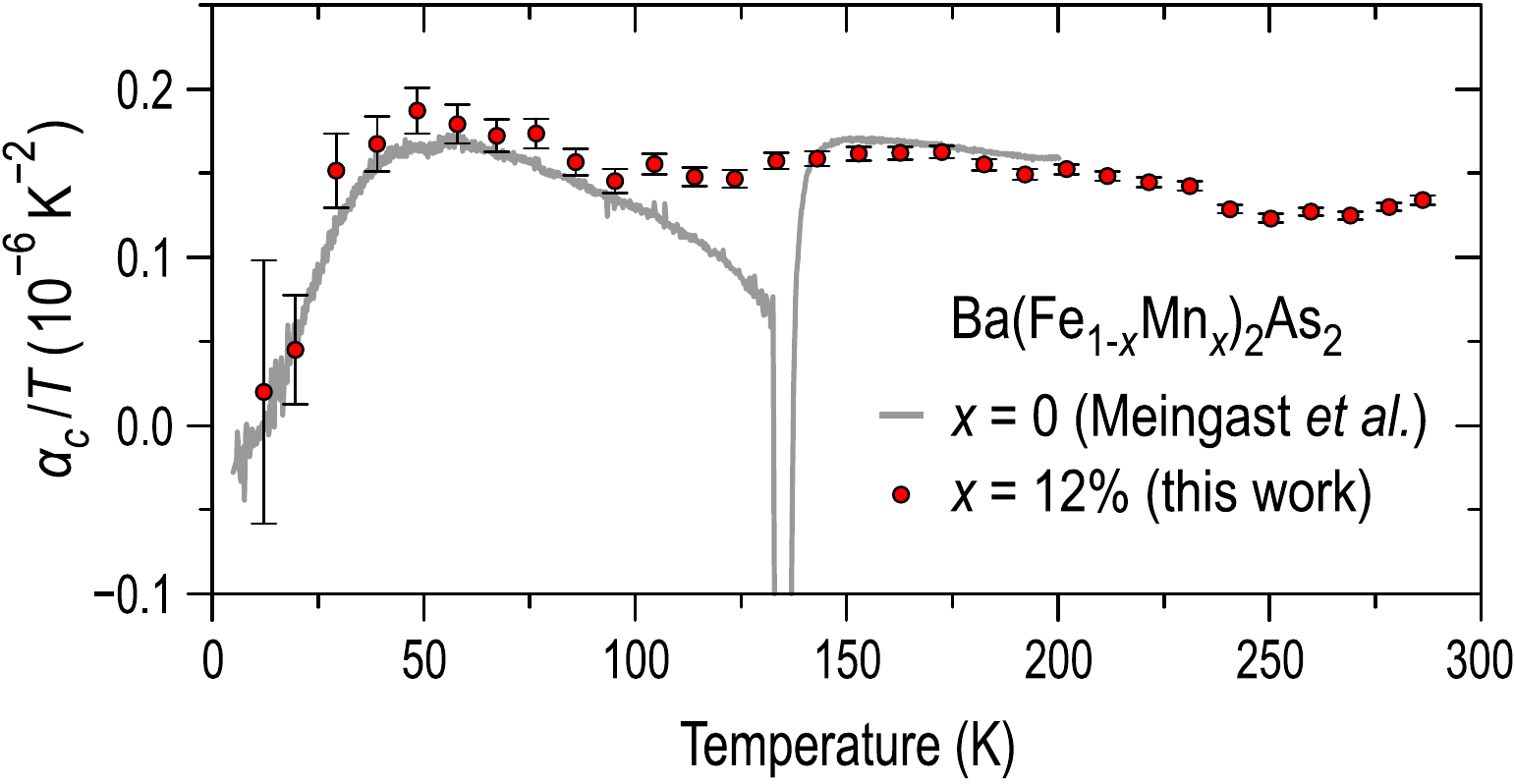}
\caption{Temperature dependence of the $c$-axis linear thermal expansion coefficient, $\alpha_c/T$, for Ba(Fe$_{0.88}$Mn$_{0.12}$)$_2$As$_2$ as measured by polarized-neutron Larmor diffraction (circles). The grey line shows the analogous dependence for the pure BaFe$_2$As$_2$, reproduced from Ref.\,\onlinecite{MeingastHardy12} for comparison.}
\label{Fig:ThermalExpansion}\vspace{-5pt}
\end{figure}

The magnetic and structural phase transitions in iron pnictides typically have a pronounced signature in the temperature dependence of the thermal expansion coefficients \cite{BudkoNi09, MeingastHardy12, BohmerBurger12}. Linear thermal expansion can be directly measured using neutron Larmor diffraction by following the shift of the total Larmor precession phase vs. temperature, even though the precision of this type of measurements is typically inferior to the state of the art capacitive dilatometry. To avoid the complications related to the coexistence of the tetragonal and orthorhombic phases and the resulting nontrivial structure of the in-plane Bragg reflections, here we will only concentrate on the $c$-axis isobaric linear thermal expansion coefficient,
\begin{equation}\alpha_c = \frac{1}{c}\frac{\partial\![c(T)-c(0)]}{\partial\!T},\end{equation}
measured on the $(004)_{\text{Fe}_1}$ structural Bragg reflection of the $x=12$\% BFMA sample. It is presented in Fig.\,\ref{Fig:ThermalExpansion} as the $\alpha_c/T$ ratio in order to emphasize the asymptotic behavior at $T\rightarrow0$. We compare it with the equivalent result of the BaFe$_2$As$_2$ dilatometry measurements from the literature \cite{MeingastHardy12}. No significant changes in the absolute values of the $\alpha_c/T$ coefficient upon Mn substitution can be observed either in the low- or high-temperatures regions, whereas in the immediate vicinity of the SDW transition the sharp anomaly at $T_\text{N}$ is replaced by a broad and shallow minimum near $T^\ast$, reminiscent of the one seen in the $T$-derivative of the resistivity [Fig.\,1(b)].

\vspace{-5pt}\section{$\mu$SR spectroscopy}\label{Sec:MuSR}

\begin{figure*}[t]\vspace{-0.8em}
\includegraphics[width=\textwidth]{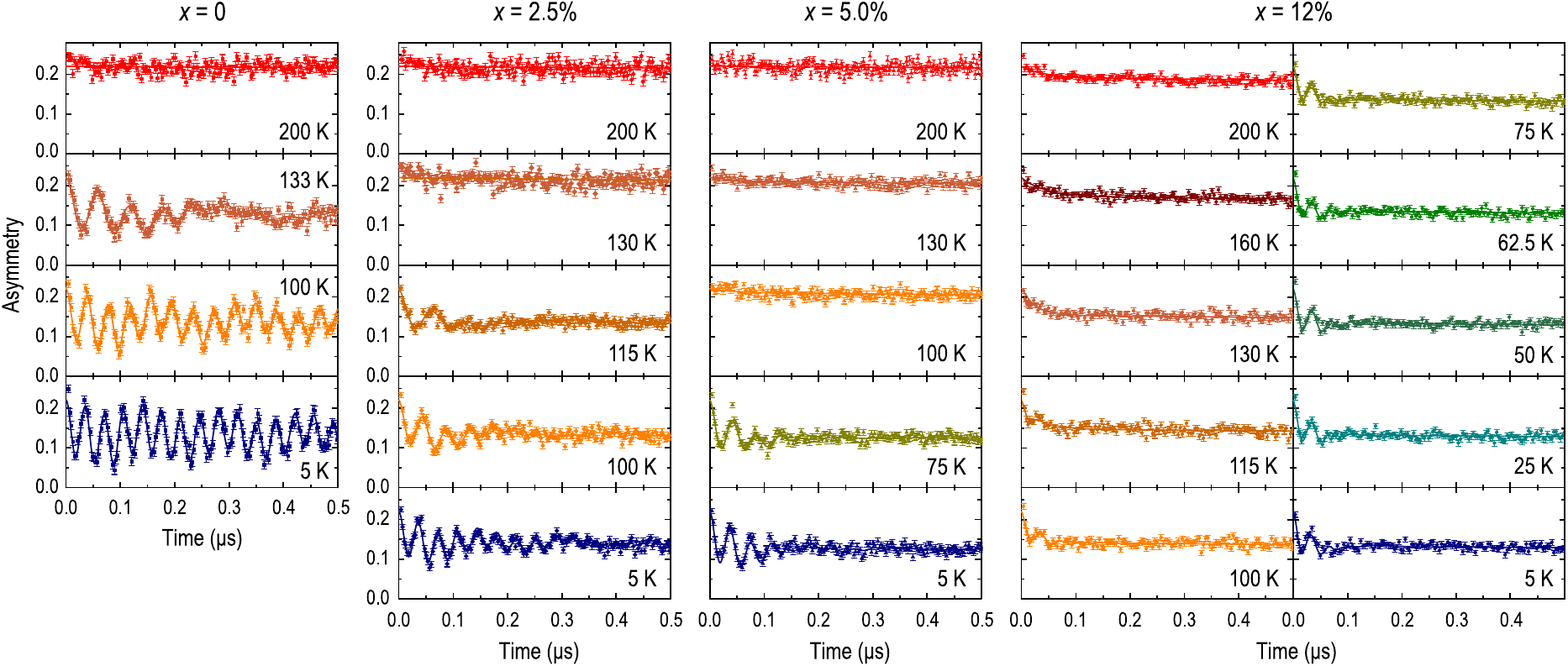}
\caption{Zero-field $\mu$SR data collected on the forward-backward pair of detectors at various temperatures (as indicated in the panels) on samples with four different Mn concentrations. The solid lines represent fits described in the text.}
\label{Fig:muSR-ZF}\vspace{-2pt}
\end{figure*}

\vspace{-5pt}\subsection{Experimental details}\vspace{-5pt}

Muon-spin-rotation spectroscopy \cite{Amato97, Blundell99} is a very powerful tool when it comes to studying magnetism in samples with several coexisting phases. As spin-polarized muons are implanted in the sample, the precession of their magnetic moment is determined by the value of the local magnetic field at the muon site. Therefore, this method is sensitive to the statistical distribution of the local magnetic environments in the sample in a very similar way to NMR. For a system that exhibits static magnetism, $\mu$SR can therefore offer valuable information about the degree of magnetic ordering (long- vs. short-range, commensurate vs. incommensurate, etc.), the value of the static magnetic moment, its homogeneity in the sample, and the magnetic volume fraction. By performing measurements in a weak transverse field, one can also accurately estimate the fraction of the sample volume with no static magnetism, i.e. PM or nonmagnetic. This is achieved by counting the fraction of muons that feel no internal magnetic field, so that their precession frequency matches the magnitude of the applied field. In particular, $\mu$SR spectroscopy has already accumulated a long track record of studying phase-separation phenomena in both iron-pnictide and iron-chalcogenide superconductors \cite{DrewNiedermayer09, ParkInosov09, GokoAczel09, TakeshitaKadono09, WiesenmayerLuetkens11, KhasanovSanna11, ShermadiniKrztonMaziopa11, CharnukhaCvitkovic12, ShermadiniLuetkens12, BernhardWang12}.

We performed our $\mu$SR measurements on BFMA single crystals with all four available compositions ($x = 0$, 2.5\%, 5.0\%, and 12\%) using the DOLLY instrument at the muon source of the Paul Scherrer Institute in Villigen, Switzerland. The incident muons were polarized parallel to the beam direction, and the samples were mounted with their $c$-axes turned by 45$^\circ$ in the horizontal plane with respect to the muon beam. Because the internal magnetic field at the muon site in the AFM phase is directed parallel to the crystallographic $c$-axis \cite{AczelBaggio08}, in this experimental geometry the signal could be counted both on the left-right and forward-backward pairs of positron detectors.

\vspace{-7pt}\subsection{Zero-field $\mu$SR (AFM phase)}\vspace{-6pt}

Figure~\ref{Fig:muSR-ZF} shows $\mu$SR data measured in zero magnetic field on samples with different Mn concentrations as a function of temperature. The parent compound (leftmost column), which we used here as a reference sample, showed pronounced oscillations in the time dependence of the muon asymmetry below $T_\text{N}$ with two characteristic frequencies, in agreement with Ref.\,\onlinecite{AczelBaggio08}. Upon increasing Mn concentration, we observed an increase in the depolarization rate of the oscillating signal, as can be seen from the comparison of the lowest-temperature ($T=5$\,K) datasets in Fig.\,\ref{Fig:muSR-ZF}. This trend is indicative of the increasing inhomogeneity in the system that leads to a broadening of the local-field distribution at the muon site. As a result, the $T=5$\,K dataset for the $x=5.0$\% sample looks qualitatively similar to the one measured on the parent compound at $T=133$\,K, immediately below the SDW transition.

\begin{figure}[b]
\hspace*{-0.01\columnwidth}\includegraphics[width=1.02\columnwidth]{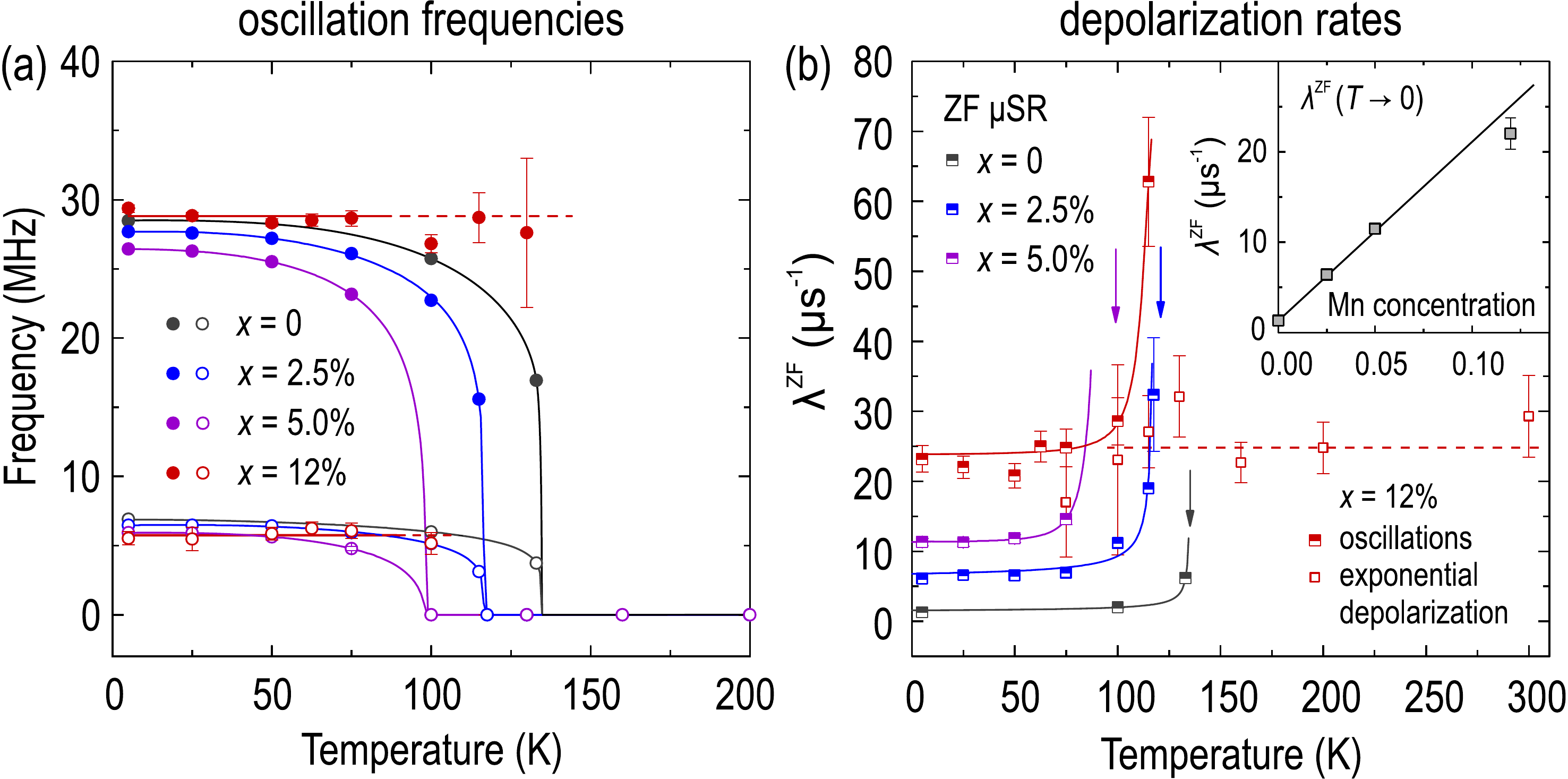}
\caption{Fitting parameters for the zero-field $\mu$SR data. (a)~Temperature dependencies of the muon oscillation frequencies. (b)~The same for the muon depolarization rate. Arrows indicate transition temperatures. For the $x=12$\% sample, depolarization rates for the exponentially decaying component of the $\mu$SR signal are additionally plotted with empty symbols. The lines are guides to the eyes. The inset shows the $T\rightarrow0$ limit of the depolarization rate, which is a measure of the degree of magnetic disorder in the ground state of the system, as a function of Mn concentration. The line is a \mbox{linear fit to these data}.}\vspace{-0.4em}
\label{Fig:muSR-Freq}
\end{figure}

At a temperature of 200\,K, which lies significantly above $T_\text{N}$, we observed no loss of the muon asymmetry either in the $x=2.5$\% or in the $x=5.0$\% sample. This proves that samples with $x<x_\text{c}$ remain fully PM at this temperature. However, the $x=12$\% sample shows a noticeable SG-like exponential depolarization of the $\mu$SR signal even at $T=200$\,K, which points at the nucleation of static magnetic islands in the small fraction of the sample volume. This signal persists down to $\sim$\,75\,K, where it coexists with the rapidly depolarizing oscillatory component. Knowing that the onset of the static $(\half 0 \half)_{\text{Fe}_1}$ magnetic Bragg peak can be observed in the same temperature range [Fig.\,\ref{Fig:Transition}(c)], we can associate these islands with AFM rare regions. The size of such static magnetic domains must be sufficiently small to explain the absence of clear oscillations in the muon asymmetry down to 130\,K in temperature. Therefore, to support the long-range AFM order that is evidenced by the sharp magnetic Bragg peaks (Fig.\,\ref{Fig:Larmor_widths}), long-range AFM correlations between these domains must be present, possibly mediated by the nesting-assisted RKKY exchange interaction \cite{AkbariEremin11, AkbariThalmeier13}. It is natural to associate this type of order with an RKKY SG or a CG phase \cite{ShellCowen82, BinderYoung86, FischerHertz99, Vojta10}.

\begin{figure*}[t]
\includegraphics[width=0.87\textwidth]{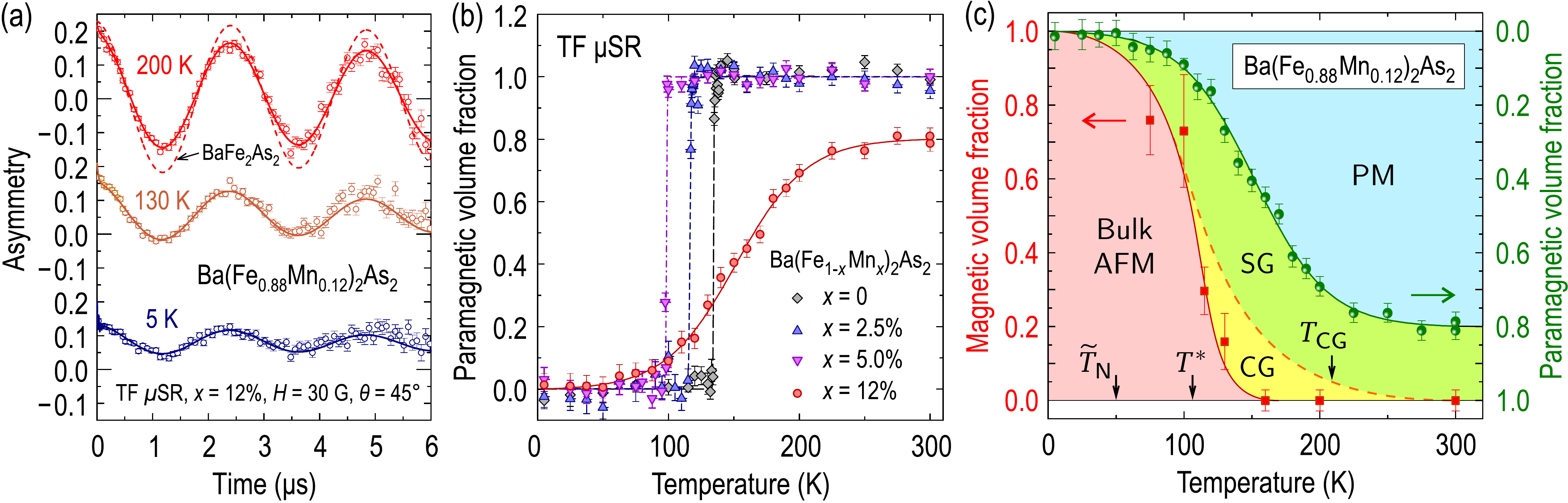}
\caption{(a)~Time dependence of the transverse-field $\mu$SR asymmetry at three selected temperatures for the $x=12$\% sample. Fitting results are shown with solid lines. The dashed line shows the respective fit for pure BaFe$_2$As$_2$ at $T=200$\,K for comparison. (b)~Temperature dependence of the PM volume fraction extracted from the transverse-field $\mu$SR data. (c)~The phase diagram of the $x=12$\% sample, summarizing the zero-field (squares) and transverse-field (circles) $\mu$SR data. The $T$-dependence of the AFM Bragg peak intensity from Fig.\,\ref{Fig:Transition}, rescaled to its maximum and minimum values, is shown by the dashed line. The plot shows volume fractions of the bulk ordered AFM phase (oscillating $\mu$SR signal in zero field), the CG phase (rapid exponential muon depolarization in zero field accompanied by a magnetic Bragg peak in neutron diffraction evidencing long-range magnetic correlations), the SG phase (muon depolarization in zero field without any long-range magnetic order), and the PM phase ($\mu$SR oscillations in the transverse field).}
\label{Fig:muSR-TF}
\end{figure*}

In order to extract quantitative information from the zero-field $\mu$SR data, we have fitted the time-dependence of the muon asymmetry with the following model: \begin{equation}A(t)=A_0\bigl[P_\text{osc}(t)+P_\text{SG}(t)+P_\text{PM}(t)\bigr],\end{equation}
where $A_0$ is the initial asymmetry, while the $P_\text{osc}(t)$, $P_\text{SG}(t)$, and $P_\text{PM}(t)$ terms represent the oscillating, exponentially depolarizing, and PM components of the $\mu$SR signal, respectively. These, in turn, can be described by
\begin{align}
&\!\!\!P_\text{osc}(t)=\frac{\upsilon_\text{osc}}{2}\Biggl[\sum_{i=1}^2 p_i\cos(2\piup\nu_it+\varphi)\,\mathrm{e}^{-\lambda^\text{ZF}_it}+\mathrm{e}^{-\lambda^\text{LO}t}\Biggr]\text{;}\\
&\!\!\!P_\text{SG}(t)\,=\frac{\upsilon_\text{SG}}{2}\,\bigl[\mathrm{e}^{-\lambda^\text{SG}t}+\mathrm{e}^{-\lambda^\text{LO}t}\bigr]\text{;}~~
P_\text{PM}(t) =\upsilon_\text{PM}\,\mathrm{e}^{-\lambda^\text{PM}t}\text{.}
\end{align}
Here $\upsilon_\text{osc}$, $\upsilon_\text{SG}$, and $\upsilon_\text{PM}$ stand for the volume fractions of the corresponding phases; $\nu_i$ are the two muon precession frequencies; $p_i$ are the fractions of the muons on the two muon stopping sites corresponding to these frequencies (such that $p_1+p_2=1$); $\varphi$ is the initial phase of the muon spin; $\lambda^\text{ZF}$ and $\lambda^\text{SG}$ are the depolarization rates for the oscillating and for the rapidly decaying SG-like parts of the zero-field $\mu$SR signal, respectively; $\lambda^\text{LO}$ describes the slow relaxation of the muon polarization component longitudinal to the local magnetic field, originating from the 45$^\circ$ rotation of the sample's $c$-axis with respect to the muon beam in our experimental geometry; $\lambda^\text{PM}$ represents the slow depolarization rate of the PM response. As we fitted the experimental data, we fixed $\upsilon_\text{PM}$ to the PM volume fraction determined from the transverse-field $\mu$SR, as described below. The $\upsilon_\text{SG}$ volume fraction was considered zero for all samples except for $x=12$\%, where it was treated as a free fitting parameter within the full width of the smeared phase transition.

Further insight is gained by directly plotting the temperature dependent fitting parameters of the zero-field $\mu$SR data, such as the oscillation frequencies and the depolarization rates (Fig.\,\ref{Fig:muSR-Freq}). A nonmonotonic dependence of the oscillation frequencies on Mn concentration is revealed by Fig.\,\ref{Fig:muSR-Freq}\,(a). Initially, for the $x=0$, $x=2.5$\%, and $x=5.0$\% samples, the oscillation frequency decreases with Mn substitution, whereas for the $x=12$\% sample it is remarkably restored to roughly the same value as in the parent compound. Moreover, the oscillation frequencies in the $x=12$\% sample no longer exhibit the order-parameter-like suppression as a function of temperature, which is typical for samples with sharp AFM transitions. Instead, they remain approximately constant in the whole range of temperatures where the frequency can be properly defined ($T \lesssim 130$\,K), possibly with a weak local minimum at $T^\ast$.

In Fig.\,\ref{Fig:muSR-Freq}\,(b), we also show the depolarization rate of the zero-field $\mu$SR signal, $\lambda^\text{ZF}(T)$. For $x=0$, $x=2.5$\%, and $x=5.0$\% samples, the depolarization rate is only defined for the oscillatory response below $T_\text{N}$, as shown by solid lines. For the $x=12$\% sample, we also plot in addition the depolarization rate for the SG-like phase that exhibits a rapid exponential depolarization without oscillations in a $T$-dependent fraction of the muons stopping in the sample, $\lambda^\text{SG}(T)$. This parameter, which turns out to be nearly constant within the accuracy of our fits, can only be measured at elevated temperatures ($T\gtrsim75$\,K) and is plotted in Fig.\,\ref{Fig:muSR-Freq}\,(b) with empty symbols (dashed line). To demonstrate that the actual amount of magnetic disorder introduced in the system with Mn substitution is indeed proportional to $x$, in the inset to Fig.\,\ref{Fig:muSR-Freq}\,(b) we plot the $x$-dependence of the depolarization rate in the zero-temperature limit, $\lambda^\text{ZF}(T\rightarrow 0)$, resulting from the empirical fits of $\lambda^\text{ZF}(T)$. This quantity is a good measure of the degree of magnetic disorder in the ground state of the system. As expected, it shows a nearly perfect linear increase with Mn concentration, which confirms that the nominal Mn content is statistically distributed within the crystals, and that the exceptional behavior of the $x=12$\% sample is not a consequence of macroscale Mn inhomogeneities at this particular composition. A qualitatively similar enhancement of the depolarization rate with increasing Mn concentration has been also reported recently in the LaFe$_{1-x}$Mn$_x$AsO series of samples \cite{FrankovskyLuetkens13}.\vspace*{5pt}

\vspace{-7pt}\subsection{Transverse-field $\mu$SR (paramagnetic phase)}\vspace{-6pt}\label{Sec:TF-MuSR}

To measure the temperature dependence of the PM volume fraction in our samples, we have applied a weak transverse field of 30\,G and measured the fraction of the muons that experienced slow precession in the external field, as shown in Fig.\,\ref{Fig:muSR-TF}\,(a). A constant part of the observed oscillation amplitude, which persists down to the base temperature (5\,K curve) and originates from muons stopping outside of the sample, has been subtracted during the fitting process. The remaining ($T$-dependent) amplitude of the oscillations, normalized to the maximum muon asymmetry, is plotted in Fig.\,\ref{Fig:muSR-TF}\,(b) vs. temperature for all the four sample compositions. In agreement with the corresponding NMR result (Fig.\,\ref{Fig:NMR_wipeout}), the $x=0$, $x=2.5$\%, and $x=5.0$\% samples exhibit sharp magnetic transitions in their full volume, whereas in the $x=12$\% sample the volume fraction of the PM phase changes gradually from 0 at low temperatures to $\sim$80\% at 300\,K. The remaining 20\% of the volume fraction at 300\,K can be naturally ascribed to the magnetic clusters that are responsible for the SG-like exponential depolarization of the muon asymmetry in zero field, which is observed in a comparable volume fraction of the sample. The width of the smeared transition is perfectly consistent with the results of NMR measurements discussed earlier. However, both in NMR and in $\mu$SR, the transition happens over a narrower range of temperatures than in elastic neutron scattering or in resistivity (Fig.\,\ref{Fig:Transition}). As a consequence, the midpoint of both NMR and $\mu$SR transitions is shifted to $\sim$\,150\,K, which is significantly higher than $T^\ast$.

\vspace{-6pt}\subsection{Phase diagram for $x=12$\%}\vspace{-6pt}

In Fig.\,\ref{Fig:muSR-TF}\,(c), we present a phase diagram that summarizes the results of both zero-field and transverse-field $\mu$SR measurements and elastic neutron scattering for the $x=12$\% composition. It shows the temperature evolution of the volume fractions corresponding to the bulk ordered AFM phase (oscillating $\mu$SR signal in zero field), the CG phase (rapid exponential muon depolarization in zero field accompanied by a magnetic Bragg peak in neutron diffraction evidencing long-range magnetic correlations), the SG phase (muon depolarization in zero field without any long-range magnetic order), and the PM phase ($\mu$SR oscillations in the transverse field). This lets us define several characteristic temperature scales for this particular sample composition. Below $\widetilde{T}_{\rm}\approx50$\,K, the sample exhibits bulk AFM order in its whole volume. This is consistent with the monotonic trend of N\'eel temperature suppression with Mn substitution, already established at lower concentrations.

At higher temperatures, the system enters the Griffiths regime of multiple coexisting phases. Above $\sim$150\,K, oscillations in the zero-field $\mu$SR signal can no longer be observed, which indicates the disappearance of the bulk AFM ordered phase. The CG phase, characterized by long-range AFM correlations between static magnetic clusters that are too small or too inhomogeneous to produce muon oscillations, persists to somewhat higher temperatures. We define the characteristic offset temperature of the CG phase, $T_\text{CG}\approx210$\,K, by the 95\% suppression of the magnetic Bragg intensity with respect to its low-$T$ value. A weak exponential depolarization of the muon asymmetry in $\sim$20\% of the sample volume persists up to the room temperature, but with no traces of long-range AFM correlations in the elastic neutron scattering, which is suggestive of fully magnetically disordered static clusters similar to a dilute SG \cite{Yamazaki82, UemuraYamazaki85, Fischer85}.

\begin{figure}[t!]\vspace{-0.5ex}
\includegraphics[width=\columnwidth]{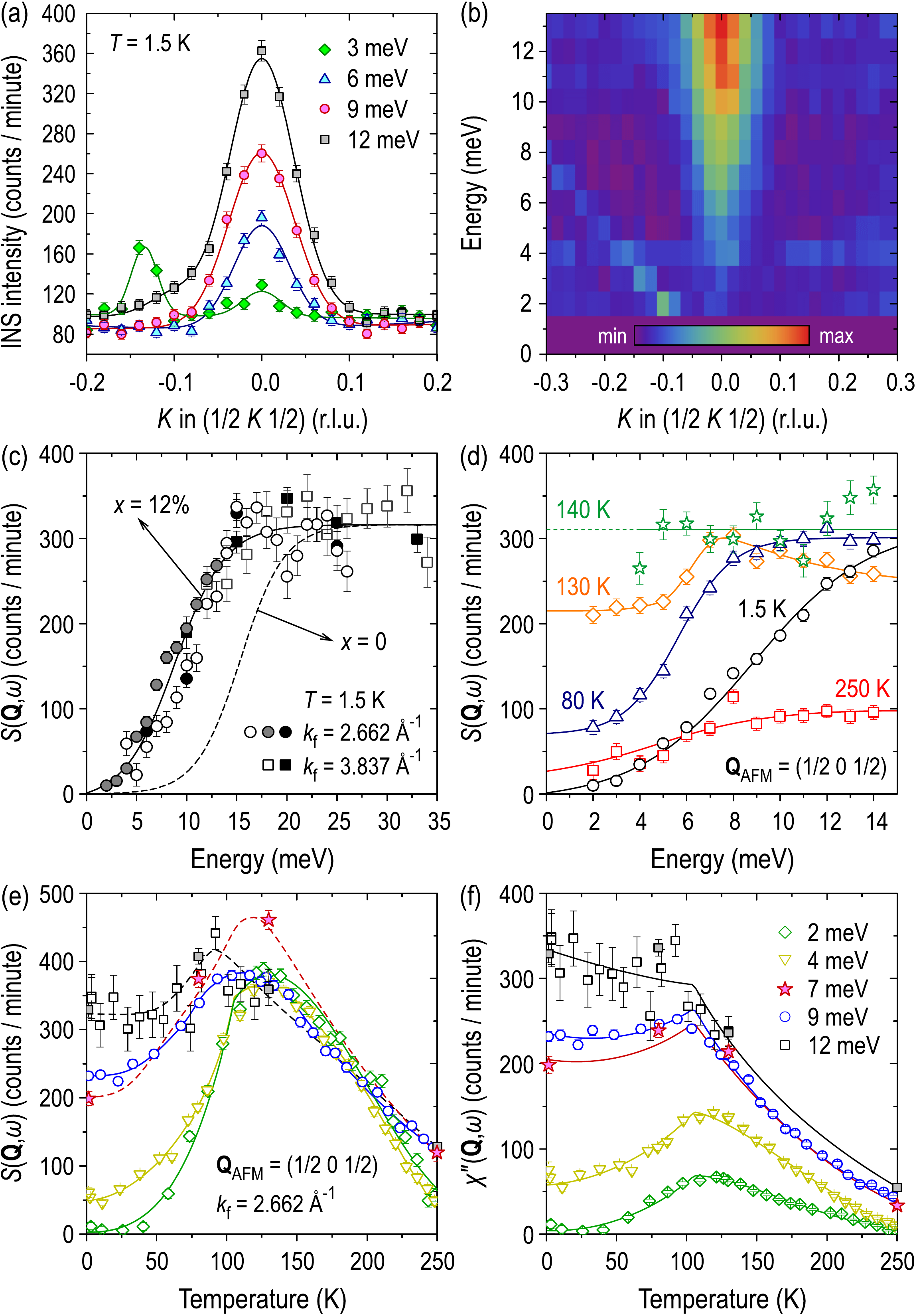}
\caption{INS data on the Ba(Fe$_{0.88}$Mn$_{0.12}$)$_2$As$_2$ sample at the magnetic ordering wave vector, $\mathbf{Q}_\text{AFM}$. (a)~Several representative unprocessed momentum scans, measured at $T=1.5$\,K along the $(\smash{\half}\kern.5pt K \smash{\half})_{{\rm Fe}_1}$ reciprocal-space direction with $k_\text{f}=2.662$\,\AA$\smash{^{-1}}$, centered at $\mathbf{Q}_\text{AFM}$. (b)~Color map of the low-energy INS intensity in the spin-gap region, compiled out of multiple low-temperature momentum scans such as those shown in panel (a). (c)~The background-subtracted scattering intensity, $S(\mathbf{Q},\omega)$, at $\mathbf{Q}_\text{AFM}=(\half 0\, L)_{{\rm Fe}_1}$, with $L=\half$ (grey points) or $L=\threehalf$ (all other points). The filled symbols were obtained from fits of the full momentum scans, such as those shown in panel (a), whereas empty symbols result from 3-point scans. The data taken with $k_\text{f}=2.662$\,\AA$^{-1}$ and 3.837\,\AA$^{-1}$ are shown with circles and squares, respectively. Datasets measured with different experimental conditions have been rescaled to match each other in the overlapping energy window. The solid curve is a guide to the eyes. The corresponding energy dependence for the BaFe$_2$As$_2$ parent compound from Ref.\,\onlinecite{ParkFriemel12} is shown for comparison as a dashed curve to emphasize the suppression of the spin gap by Mn substitution. (d)~Evolution of the low-energy part of $S(\mathbf{Q},\omega)$ with temperature, demonstrating a partial spin gap at intermediate temperatures with a magnitude that decreases upon heating. (e)~Temperature dependence of $S(\mathbf{Q},\omega)$ at various energies within the spin-gap region. (f)~The same for the dynamic spin susceptibility, $\chi''(\mathbf{Q},\omega)$, obtained from the data in panel (e) after Bose-factor normalization. The lines are guides to the eyes.}
\label{Fig:INS}\vspace{-1em}
\end{figure}

As one can see from Fig.\,\ref{Fig:muSR-TF}\,(c), the characteristic temperature $T^\ast$, defined in Ref.\,\onlinecite{KimKreyssig10} and in Fig.\,\ref{Fig:Transition} by the inflection point in the $T$-dependence of the resistivity, corresponds to the midpoint of the transition associated with the suppression of the bulk ordered AFM phase. This observation is not surprising, as one would expect the transport properties to be much stronger affected by the long-range static AFM order, leading to a Fermi surface reconstruction, than by dilute random inclusions of static magnetic clusters into the otherwise PM material. For a two-dimensional square lattice, the site percolation threshold amounts to 59.3\% \cite{VanDerMarck97}. Therefore, at 50\% filling of the sample volume by AFM ordered regions, the system is close to a percolative transition. In other words, at $T<T^\ast$ the AFM phase volume is mostly connected, whereas at $T>T^\ast$ it consists of disconnected clusters embedded in the magnetically disordered or PM matrix. Such a percolative crossover is the most likely reason for \mbox{the inflection point in the $T$-dependence of the resistivity}.

\vspace{-5pt}\section{Inelastic neutron scattering}

\vspace{-3pt}\subsection{Experimental details}\vspace{-5pt}

We have performed a series of INS measurements on the $x=12$\% BFMA compound using thermal-neutron triple-axis spectrometers IN8 (ILL, Grenoble, France), PUMA (FRM-II, Garching, Germany), and 1T (LLB, Saclay, France). All measurements were performed with the fixed final neutron wave vector, $k_\text{f}=2.662$\,\AA$^{-1}$ or 3.837\,\AA$^{-1}$. A~pyrolytic graphite filter was installed between the sample and the analyzer to eliminate the contamination from higher-order neutrons. The sample was mounted in one of the $(H\,K\,0)_{\text{Fe}_1}$, $(H\,0\,L)_{\text{Fe}_1}$, or $(H\,K\,H)_{\text{Fe}_1}$ scattering planes, depending on the particular goal of the experiment.

\vspace{-5pt}\subsection{Low-temperature spin gap}\vspace{-5pt}

\begin{figure}[t]\vspace{-0.5ex}
\includegraphics[width=\columnwidth]{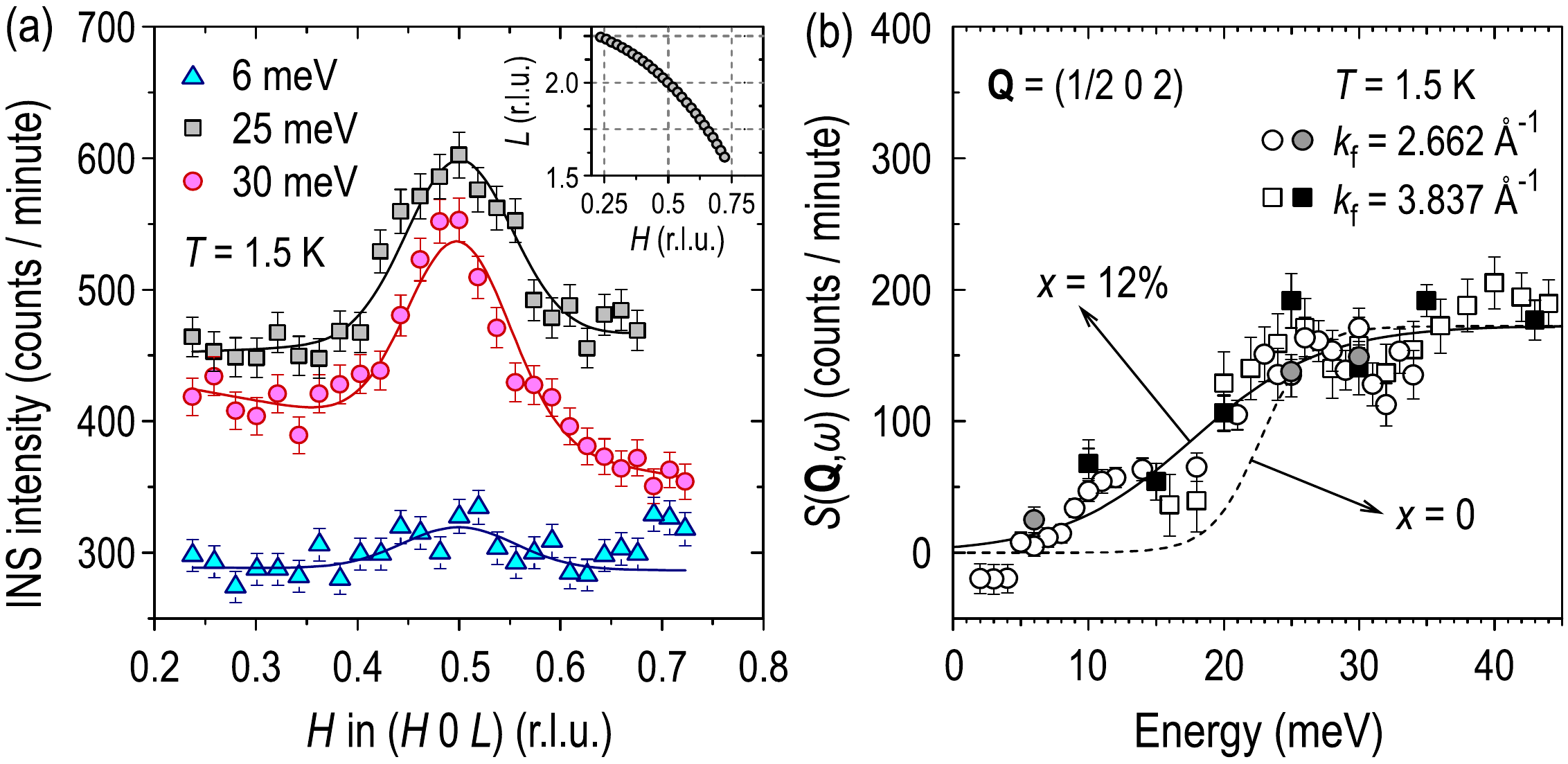}
\caption{INS data acquired on the $x=12$\% sample at the magnetic zone boundary ($L=2$). (a)~Three unprocessed momentum scans, measured at $T=1.5$\,K along the rocking trajectory in the $(H\,0\,L)_{{\rm Fe}_1}$ plane with $k_\text{f}=2.662$\,\AA$\smash{^{-1}}$. The inset shows the scan trajectory in the $(H,L)$ plane. (b)~The background-subtracted scattering intensity, $S(\mathbf{Q},\omega)$, taken at $\mathbf{Q}=(\half 0\, 2)_{{\rm Fe}_1}$. The filled symbols were obtained from fits of the full momentum scans, such as those shown in panel (a), whereas empty symbols result from 3-point scans. The data taken with $k_\text{f}=2.662$\,\AA$^{-1}$ and 3.837\,\AA$^{-1}$ are shown with circles and squares, respectively. The solid curve is a guide to the eyes. The corresponding energy dependence for the BaFe$_2$As$_2$ parent compound \cite{ParkFriemel12} is shown for comparison with the dashed curve.}
\label{Fig:INS_L2}
\end{figure}

We start our discussion of the INS data by presenting the low-energy spectrum of spin excitations in Ba(Fe$_{0.88}$Mn$_{0.12}$)$_2$As$_2$ at the magnetic ordering wave vector, $\mathbf{Q}_\text{AFM}$. In Fig.\,\ref{Fig:INS}\,(a), we show several representative low-temperature momentum scans along the Brillouin zone boundary, centered at $(\half0\half)_{\text{Fe}_1}$. A number of such scans is also summarized in Fig.\,\ref{Fig:INS}\,(b) as a color map. We observe a notable depletion of the scattering intensity at low energies, reminiscent of the spin anisotropy gap in the parent compound \cite{ParkFriemel12}. However, in contrast to BaFe$_2$As$_2$, where the intensity completely vanishes below $\sim$\,10\,meV in the AFM state, here the onset energy of magnetic fluctuations is strongly reduced, so that weak remnant spectral weight persists at least down to 2--3\,meV. This can be best seen in Fig.\,\ref{Fig:INS}\,(c), where we plot the scattering function, $S(\mathbf{Q},\omega)$, obtained by measuring the background-subtracted amplitude of the peak at various energies and by combining data from $L=\half$ and $L=\threehalf$ acquired with different $k_\text{f}$. Indeed, a comparison of our data with an equivalent result for BaFe$_2$As$_2$ from Ref.\,\onlinecite{ParkFriemel12} (dashed curve) shows a reduction of the spin-gap energy from $\sim$\,10\,meV in BaFe$_2$As$_2$ to $\sim$\,3\,meV in Ba(Fe$_{0.88}$Mn$_{0.12}$)$_2$As$_2$, with a weak intensity tail extending to even lower energies. Note that despite this dramatic spin-gap reduction, the characteristic ordering temperature ($T^\ast$) in BFMA is \mbox{only 25\% lower than in the parent compound}.

With increasing temperature, the spin gap in the $x=12$\% BFMA sample is suppressed as shown in Fig.\,\ref{Fig:INS}\,(d). Instead of a gradual order-parameter-like reduction of the gap energy, which one would expect for a SDW transition, here the gap energy remains nearly constant with temperature, whereas the magnetic intensity inside the gap is continuously increasing, so that the spin gap is completely filled in upon reaching $T\approx140$\,K, which coincides with the ordering temperature of the parent compound. This unusual behavior can be naturally explained in the framework of the phase-separation scenario, which we have already established in sections \ref{Sec:Characterization} and \ref{Sec:MuSR}. The spin-excitation spectrum should be considered as a sum of two components: gapless excitations originating from the PM phase and gapped spin-wave-like excitations from the magnetically ordered regions. As the PM volume of the sample increases upon warming at the expense of the AFM phase, the anisotropy gap appears to be filled in. At the same time, the characteristic energy scale of the residual partial gap in the low-energy magnetic spectrum is nearly unaffected, because it is mainly determined by the rare regions with relatively high local values of $T_\text{N}$.

Further insight is obtained by following the temperature dependence of the INS intensity at several fixed energies, shown in Fig.\,\ref{Fig:INS}\,(e). To account for the thermal population factor, in Fig.\,\ref{Fig:INS}\,(f) we have also plotted the imaginary part of the dynamical spin susceptibility, obtained from the same data after Bose-factor correction: $\chi''(\mathbf{Q},\omega)=(1-\mathrm{e}^{-\hslash\omega/k_\textup{B}T})\,S(\mathbf{Q},\omega)$. Remarkably, the anomalies related to the magnetic transition appear to be much sharper for the inelastic signal than for the magnetic Bragg peak in Fig.\,\ref{Fig:Transition}\,(c). This could be due to the fact that for a given energy transfer, $E$, only those magnetic regions whose spin gap is larger than this energy (i.e. those that are characterized by a sufficiently high local value of $T_\text{N}$) would yield an anomaly in the temperature dependence of the INS intensity. Therefore, this measurement effectively selects only a part of the magnetically ordered regions with $T_\text{N} \gtrsim E/k_\text{B}$, whereas the Bragg peak intensity in Fig.\,\ref{Fig:Transition}\,(c) originates from the whole magnetic volume of the sample independently of the local ordering temperature.

We have also studied the dispersion of the spin gap along the out-of-plane direction by measuring the spin-excitation spectrum at integer $L$, i.e. at the magnetic zone boundary. In Fig.\,\ref{Fig:INS_L2}\,(a), we show representative momentum scans through $(\half\,0\,2)$ measured at several energies along the rocking trajectory in the $(H\,0\,L)_{\text{Fe}_1}$ plane (see inset), whereas in Fig.\,\ref{Fig:INS_L2}\,(b) we plot the corresponding background-subtracted scattering function at the same wave vector, obtained in the same way as the similar spectrum in Fig.\,\ref{Fig:INS}\,(c). Again, the reference spectrum for the parent compound from Ref.\,\onlinecite{ParkFriemel12} is shown with the dashed curve for comparison. Here, the 20\,meV zone-boundary gap observed in BaFe$_2$As$_2$ is also strongly suppressed and smeared out upon Mn substitution, so that the gradual onset of magnetic fluctuations is found near $\sim$\,5\,meV, whereas the high-energy offset of the spin gap stays unchanged at $\sim$\,25\,meV. In the framework of a localized Heisenberg-type description of spin-wave excitations in iron pnictides \cite{HarrigerLuo11}, the spin-gap magnitude at integer $L$ is directly related to the value of the effective out-of-plane exchange constant, $J_\perp$. The observed smearing of this gap in BFMA is therefore indicative of a broad distribution of $J_\perp$ within the sample, whose maximal value coincides with that in the parent compound, whereas at the opposite extreme of this distribution a small fraction of the sample exhibits a quasi two-dimensional behavior with the much smaller zone-boundary gap of only 5\,meV.

\vspace{-5pt}\subsection{In-plane ellipticity and the absence of charge doping}\vspace{-5pt}

\begin{figure}[t]\vspace*{-3pt}
\includegraphics[width=0.85\columnwidth]{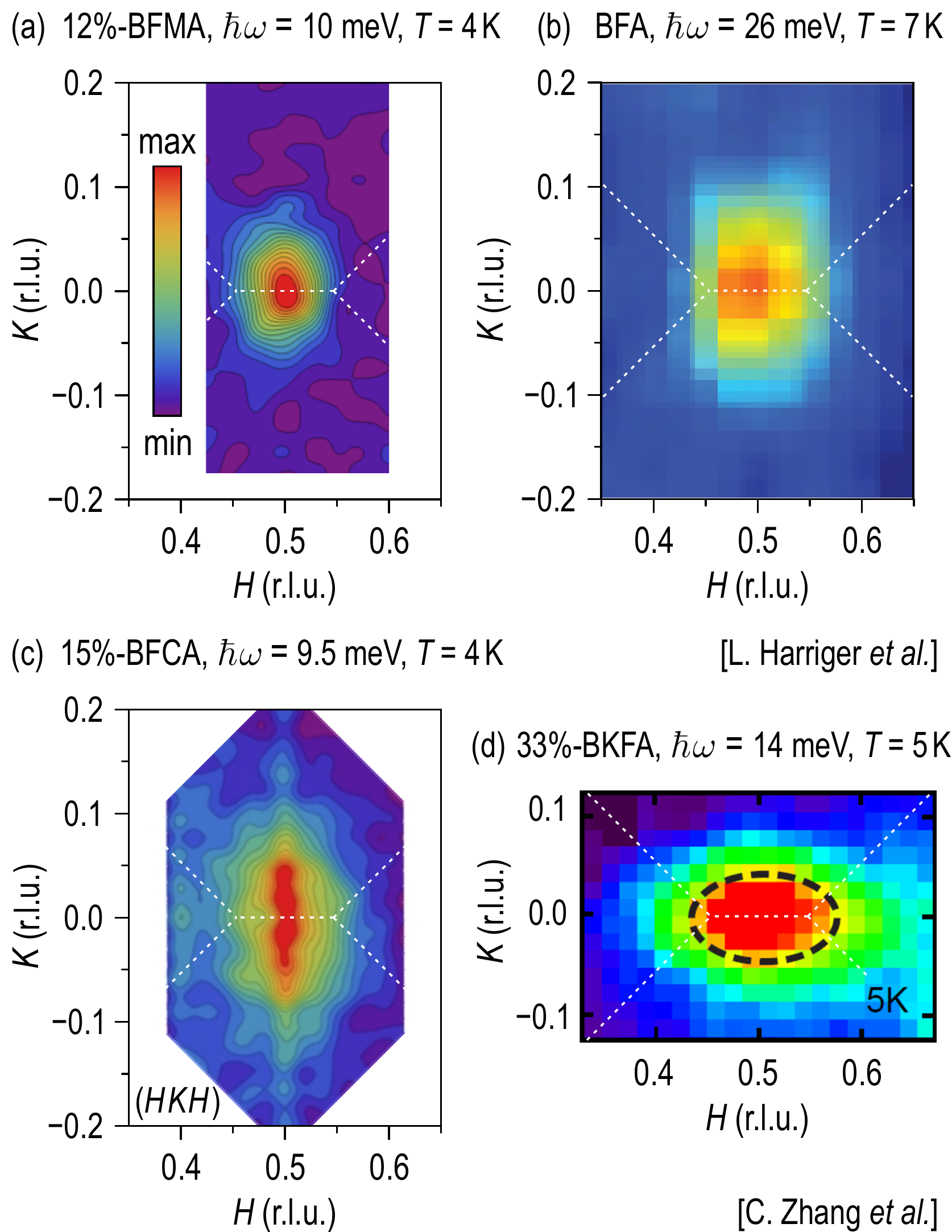}
\caption{Comparison of the elliptical cross-sections of the low-temperature INS intensity in the $Q_xQ_y$-plane projection for different compounds, measured at a constant energy, $\hslash\omega$, which is indicated above each panel. (a)~Ba(Fe$_{0.88}$Mn$_{0.12}$)$_2$As$_2$ sample from the present work. (b)~Parent BaFe$_2$As$_2$ compound from Ref.~\citenum{HarrigerLuo11}. (c)~Electron-doped Ba(Fe$_{0.85}$Co$_{0.15}$)$_2$As$_2$ sample from Ref.~\citenum{ParkInosov10}. (d)~Hole-doped Ba$_{0.67}$K$_{0.33}$Fe$_2$As$_2$ sample from Ref.\,\onlinecite{ZhangWang11}. The white dotted lines in all panels mark Brillouin-zone boundaries corresponding to the conventional body-centered tetragonal unit cell. Note that both the orientation and the momentum scales in all panels are identical.}
\label{Fig:Ellipticity}\vspace{-4pt}
\end{figure}

In iron arsenides, the ordering wave vector, $\mathbf{Q}_\text{AFM}=(\half 0 \half)_{\text{Fe}_1}$, lies on the axis of twofold rotational symmetry in the unfolded Brillouin zone, which determines its elliptical in-plane cross section. We demonstrated previously \cite{ParkInosov10} that the orientation of this ellipse and its aspect ratio can serve as an indirect measure of the doping level and can be well described by the band-structure theory. Indeed, in electron-doped Ba(Fe$_{1-x}$Co$_x$)$_2$As$_2$ (BFCA) the ellipse is strongly elongated along the transverse direction \cite{ParkInosov10, WangZhang13}, whereas in hole-doped Ba$_{0.67}$K$_{0.33}$Fe$_2$As$_2$ (BKFA) its longer axis flips to the longitudinal direction \cite{WangZhang13, ZhangWang11}. In comparison to the doped compounds, the cross section of magnetic excitations in BaFe$_2$As$_2$ is nearly isotropic, with only a weak transverse elongation \cite{HarrigerLuo11}. In Fig.\,\ref{Fig:Ellipticity}\,(a) we present a similar measurement of the in-plane cross section in Ba(Fe$_{0.88}$Mn$_{0.12}$)$_2$As$_2$, measured at $T=4$\,K at an energy transfer of 10\,meV. The color map represents an interpolation of several $K$-scans, measured with a regular step along the $H$ direction in the $(H\,K\,H)_{\text{Fe}_1}$ scattering plane. For comparison, we reproduce the corresponding maps for the pure BaFe$_2$As$_2$, electron-doped BFCA and hole-doped BKFA in Figs.\,\ref{Fig:Ellipticity}\,(b), (c) and (d), respectively. One can see that the $x=12$\% BFMA sample shows a nearly isotropic in-plane cross section of the INS intensity, which is characterized by the same aspect ratio and orientation as in the parent compound and is clearly different from the much more anisotropic response in the two superconducting samples. This indicates that the nesting properties and consequently the size of the Fermi surface sheets are not affected by the Mn substitution, in accordance with the absence of \mbox{either hole or electron doping demonstrated by NMR} \cite{TexierLaplace12}.

\vspace{-5pt}\subsection{Spin anisotropy of magnetic excitations}\vspace{-5pt}\enlargethispage{8pt}

Recent polarized-neutron scatting measurements \cite{QureshiSteffens12} revealed two components in the spin-wave spectrum of BaFe$_2$As$_2$, characterized by the out-of-plane and in-plane polarizations, with distinct zone-center spin gaps of 10\,meV and 16\,meV, respectively. This observation implies that the gradual onset of magnetic fluctuations, as measured by conventional unpolarized INS [e.g.~Fig.\,\ref{Fig:INS}\,(c) or Ref.\,\onlinecite{ParkFriemel12}], in fact represents a sum of two steplike functions with different onset energies, similar to those observed in copper oxides \cite{TranquadaShirane89, ShamotoSato93, BourgesSidis94, PetitgrandMaleyev99}. Usually, the onset of the in-plane scattering in iron pnictides can not be resolved as a separate step in the unpolarized data. As a result, one expects that the low-energy part of the spectrum between the spin-gap energy and the midpoint of the spin-gap edge has an out-of-plane polarization, in contrast to the higher-energy part of the spectrum that should be more isotropic. This gives us an opportunity to investigate the spin anisotropy of magnetic excitations in BFMA and to verify if they adhere to the same kind of behavior as in BaFe$_2$As$_2$ even without employing polarized neutrons.

\begin{figure}[t]\vspace*{-2pt}
\hspace*{-0.5em}\includegraphics[width=1.03\columnwidth]{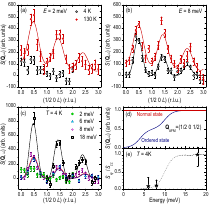}
\caption{(a,\,b)~$L$-dependence of the background-subtracted intensity in Ba(Fe$_{0.88}$Mn$_{0.12}$)$_2$As$_2$ in the low-temperature AFM state ($T=4$\,K) and in the PM state ($T=130$\,K), measured at an energy transfer of 2\,meV and 8\,meV, respectively. (c)\,Comparison of the low-temperature datasets ($T=4$\,K) at several energies. (d) Schematic representation of the scattering function in the normal and ordered states, the latter consisting of two broadened steplike functions corresponding to the magnetic scattering intensity with out-of-plane ($S_{zz}$) and in-plane ($S_{x{\kern-0.7pt}y}$) polarizations. (e) The $S_{x{\kern-0.7pt}y}/S_{zz}$ ratio extracted from the fits in panel (c). The expected energy dependence of this ratio is \mbox{schematically shown with the dotted line}.}
\label{Fig:INS_Ldep}
\end{figure}

For this purpose, we have investigated the $L$-depen\-dence of the scattering amplitude at the ordering wave vector in the $x=12$\% sample, as shown in Fig.\,\ref{Fig:INS_Ldep}. At the lowest energy of $\hslash\omega=2$\,meV, which lies well below the onset energy of the spin gap, no measurable signal was found in the magnetically ordered state at $T=4$\,K. At an elevated temperature of 130\,K, however, a periodic modulation of intensity with several maxima at half-integer $L$ values could be observed [Fig.\,\ref{Fig:INS_Ldep}\,(a)]. This behavior is typical for the PM state of the pure and lightly doped iron pnictides \cite{DialloPratt10, ParkInosov10}, indicating the three-dimensional nature of the isotropic paramagnon excitations above $T_\text{N}$ that ultimately gives rise to the $Q_z$-component of the magnetic propagation vector as the system enters the AFM state.

Above the spin-gap onset, a similar periodic modulation was observed both above and below $T_\text{N}$ [Fig.\,\ref{Fig:INS_Ldep}\,(b,\,c)]. At intermediate energies of $\hslash\omega=6$ and 8\,meV, the reduction of the scattering amplitude with increasing $L$ in the AFM state appears to be more rapid than expected for isotropic spin fluctuations following the Fe$^{2+}$ spin-only magnetic form factor \cite{ParkInosov10}. This behavior results from the out-of-plane polarization of the fluctuating moment, as the angle between the momentum transfer, $\mathbf{Q}$, and the $\mathbf{c}$-axis falls off with increasing $L$. By fitting the corresponding $L$ dependencies for various energy transfers, as shown in Fig.\,\ref{Fig:INS_Ldep}\,(c), we could extract the corresponding ratios of the magnetic scattering intensities with in-plane and out-of-plane polarizations, $S_{x{\kern-0.7pt}y}/S_{zz}$, which are presented in Fig.\,\ref{Fig:INS_Ldep}\,(e). These results are consistent with the presence of two spin gaps for different polarizations, as in the parent compound, though with reduced energy scales [see Fig.\,\ref{Fig:INS_Ldep}\,(d)]. We can therefore confirm that the low-energy onset of the magnetic signal, seen in Fig.\,\ref{Fig:INS}\,(c), originates predominantly from the out-of-plane polarized moments, whereas the spin-gap onset corresponding to the in-plane polarization is located above 8\,meV, according to Fig.\,\ref{Fig:INS_Ldep}\,(e).

\begin{figure}[t]\vspace{-3pt}
\hspace*{-1em}\includegraphics[width=0.9\columnwidth]{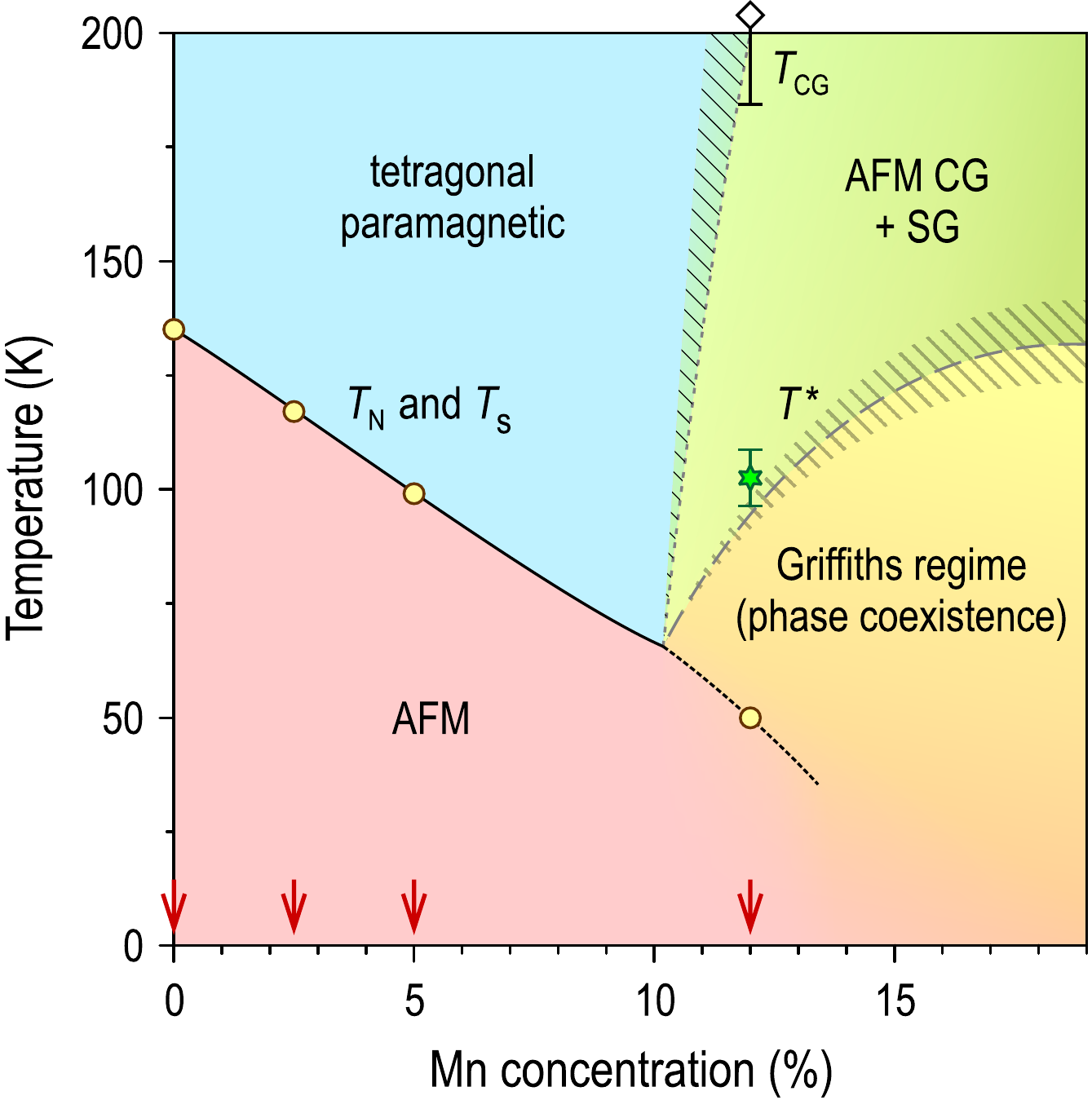}
\caption{Schematic phase diagram of BFMA after Refs.\,\citenum{KimKreyssig10},\,\citenum{KimPratt11} and the present work. Composition of the samples used in this study is indicated by arrows. The SDW transition temperatures ($T_\text{N}$ or $\widetilde{T}_\text{N}$), below which the whole volume of the samples remains fully magnetic, as determined by transverse-field $\mu$SR spectroscopy in section~\ref{Sec:TF-MuSR}, are marked by circles. The diamond symbol marks the onset of the elastic neutron-scattering intensity at the AFM wave vector, $T_\text{CG}$, which we define at 5\% of the magnetic Bragg peak's maximal intensity. It is associated with the formation of long-range magnetic correlations in the CG phase. The star symbol stands for $T^{\ast}$, defined by the position of the inflection point in the temperature dependence of the resistivity (Fig.\,\ref{Fig:Transition}) or by the 50\% reduction in the oscillating component of the muon asymmetry in zero field [Fig.\,\ref{Fig:muSR-TF}\,(c)].}
\label{Fig:PhaseDiagram}
\end{figure}

\vspace{-5pt}\section{Summary and discussion}

\vspace{-5pt}\subsection{The $x$-$T$ phase diagram of Ba(Fe$_{1-x}$Mn$_x$)$_2$As$_2$}\vspace{-5pt}

We summarize our results in a schematic phase diagram presented in Fig.\,\ref{Fig:PhaseDiagram}, where we plot various temperature scales characterizing magnetic order in BFMA vs. Mn concentration. Above the critical concentration of $x_\text{c}\approx10$\%, we distinguish three distinct crossover temperatures. Below $\widetilde{T}_\text{N}$ (circle), the sample orders antiferromagnetically in its whole volume, as determined by transverse-field $\mu$SR spectroscopy in section~\ref{Sec:TF-MuSR}. As the temperature is increased, the magnetically ordered volume fraction decreases, whereas the AFM order remains long-range, as evidenced by the persistence of the magnetic Bragg peak intensity and by its temperature-independent resolution-limited width. The inflection point observed in the resistivity at $T^\ast$ (star-shaped symbol in Fig.\,\ref{Fig:PhaseDiagram}) corresponds to the 50\% reduction in the magnetically ordered volume fraction (oscillating part of the muon asymmetry), i.e. to the midpoint of the smeared AFM transition. We also associate it with the percolation threshold of the magnetically ordered clusters, reminiscent of that found in Mn-substituted Sr$_3$Ru$_2$O$_7$ upon varying Mn content \cite{HossainBohnenbuck09, MesaYe12, HossainBohnenbuck12}.\enlargethispage{-5pt}

At $T>T^\ast$, the volume fraction of the AFM ordered clusters corresponding to the oscillatory component of the zero-field $\mu$SR signal rapidly vanishes. However, static magnetic moments still persist in most of the sample volume in the form of two distinct phases: (i) the CG phase, characterized by long-range AFM correlations responsible for the remnant magnetic Bragg peak intensity persisting up to 240\,K, and (ii) the SG phase that leads to a rapid exponential depolarization of the muon asymmetry without long-range AFM correlations. We define the CG onset temperature, $T_{\rm CG}$ (diamond symbol in Fig.\,\ref{Fig:PhaseDiagram}), at a point where the magnetic Bragg peak reaches 5\% of its maximal intensity. Above $T_{\rm CG}$, the PM volume fraction reaches its saturation value and becomes nearly temperature-independent, marking the upper boundary of the smeared AFM transition.

The region of phase coexistence, where magnetically ordered (CG-type) or spin-frozen (SG-type) clusters coexist with paramagnetic regions on the nanoscale within the sample, is in line with the Griffiths-phase concept \cite{Vojta06, Vojta10, NozadzeVojta11, NozadzeVojta12}. It is natural to associate the observed magnetic clusters with the AFM rare regions, which are pinned at the local statistical fluctuations of the Mn-ion distribution. As a result, the AFM quantum critical point that is typical for most families of iron-pnictide compounds is destroyed in the case of Mn substitution by the phase-transition smearing, giving way to a Griffiths-type behavior with the nanoscopic phase coexistence.

\vspace{-5pt}\subsection{Local moments in a metal with Fermi-surface nesting}\vspace{-5pt}

In the present study, we have uncovered the microscopic mechanisms that underlie the previously reported \cite{KimKreyssig10} smearing of the AFM phase transition in BFMA at high Mn concentrations. Most remarkably, we have demonstrated that long-range AFM correlations between the static magnetic clusters persist up to temperatures that are much higher than the $T_{\rm N}$ of the parent compound and exist well above the percolation threshold. Indeed, although nearly 80\% of the sample volume is paramagnetic at $T>T_{\rm CG}$, a clearly detectable magnetic Bragg peak persists in the $x=12$\% sample even above this temperature, at least up to 240\,K. Moreover, the absence of oscillations in the zero-field $\mu$SR response of the CG phase implies a nanoscopic size of the magnetic clusters, such that the muons locally implanted inside such clusters do not see them as a bulk ordered phase. They possibly represent individual Mn moments or small random configurations of such moments (rare regions) surrounded by the spin-polarization clouds of the neighboring Fe electrons.

These observations necessarily require the presence of some long-range magnetic interaction, acting between the small separated clusters through the PM volume in order to establish the long-range coherence of their magnetic moments. The most natural candidate for such an interaction is the RKKY exchange, which in the case of iron pnictides is known to be strongly affected by the nearly perfect nesting property of the Fermi surface \cite{AkbariEremin11, AkbariThalmeier13}. The BFMA compound therefore represents a model system, in which localized magnetic moments are randomly embedded into a SDW metal, providing an interesting playground for theorists to study the spin-glass behavior of magnetic impurities in metals with Fermi surface nesting.

So far, the influence of disorder on magnetic properties of iron pnictides has been mostly investigated only for the case of nonmagnetic impurities. For instance, in a recent theoretical study \cite{WeberMila12} is has been shown using Monte Carlo simulations that the introduction of non-magnetic impurity sites into the Fe sublattice can lead to the formation of anticollinear magnetic order, i.e. qualitatively alter the magnetic ground state of the material. There is also a persistent interest in understanding the influence of disorder on the superconducting properties of doped iron pnictides \cite{FernandesVavilov12, LiShen13}. Future theories extending these results to magnetic impurities, which have not been addressed in detail until now, should be informed by our present work. In particular, it would be desirable to explain theoretically the existence of a well defined critical concentration of Mn ions, $x_\text{c}$, below which no smearing of the AFM transition is observed. Understanding thermodynamical properties of a nesting-driven SDW metal with embedded local moments also represents a challenge that should be addressed in future studies.

~\vspace{2em}

\vspace{-5pt}\section*{Acknowledgments}\vspace{-5pt}

This work has been supported, in part, by the DFG within the priority program SPP\,1458, under Grants No.~\mbox{BO\,3537/1-1} and \mbox{IN\,209/1-2}, by the MPI\,--\,UBC Center for Quantum Materials, and by the ANR Pnictides. The authors are grateful to D.~Efremov, I.~Eremin, C.~Weber and A.\,Yaresko for stimulating discussions and encouragement.


\begin{thebibliography}{119}%
\makeatletter
\providecommand \@ifxundefined [1]{%
 \@ifx{#1\undefined}
}%
\providecommand \@ifnum [1]{%
 \ifnum #1\expandafter \@firstoftwo
 \else \expandafter \@secondoftwo
 \fi
}%
\providecommand \@ifx [1]{%
 \ifx #1\expandafter \@firstoftwo
 \else \expandafter \@secondoftwo
 \fi
}%
\providecommand \natexlab [1]{#1}%
\providecommand \enquote  [1]{``#1''}%
\providecommand \bibnamefont  [1]{#1}%
\providecommand \bibfnamefont [1]{#1}%
\providecommand \citenamefont [1]{#1}%
\providecommand \href@noop [0]{\@secondoftwo}%
\providecommand \href[0]{\begingroup \@sanitize@url \@href}%
\providecommand \@href[1]{\@@startlink{#1}\@@href}%
\providecommand \@@href[1]{\endgroup#1\@@endlink}%
\providecommand \@sanitize@url [0]{\catcode `\\12\catcode `\$12\catcode
  `\&12\catcode `\#12\catcode `\^12\catcode `\_12\catcode `\%12\relax}%
\providecommand \@@startlink[1]{}%
\providecommand \@@endlink[0]{}%
\providecommand \url  [0]{\begingroup\@sanitize@url \@url }%
\providecommand \@url [1]{\endgroup\@href {#1}{\urlprefix }}%
\providecommand \urlprefix  [0]{URL }%
\providecommand \Eprint [0]{\href}%
\providecommand \doibase [0]{http://dx.doi.org}%
\providecommand \selectlanguage [0]{\@gobble}%
\providecommand \bibinfo  [0]{\@secondoftwo}%
\providecommand \bibfield  [0]{\@secondoftwo}%
\providecommand \translation [1]{[#1]}%
\providecommand \BibitemOpen [0]{}%
\providecommand \bibitemStop [0]{}%
\providecommand \bibitemNoStop [0]{.\EOS\space}%
\providecommand \EOS [0]{\spacefactor3000\relax}%
\providecommand \BibitemShut  [1]{\csname bibitem#1\endcsname}%
\let\auto@bib@innerbib\@empty
\bibitem [{\citenamefont {Vojta}(2006)}]{Vojta06}%
  \BibitemOpen
  \bibfield  {author} {\bibinfo {author} {\bibfnamefont {T.}~\bibnamefont
  {Vojta}},\ }\href{http://iopscience.iop.org/0305-4470/39/22/R01/} {\bibfield
  {journal} {\bibinfo  {journal} {J.~Phys.~A: Math. Gen.}\ }\textbf {\bibinfo
  {volume} {39}},\ \bibinfo {pages} {R143} (\bibinfo {year}
  {2006})}\BibitemShut {NoStop}%
\bibitem [{\citenamefont {L\"ohneysen}\ \emph
  {\textit{et~al.}}(2007)\citenamefont {L\"ohneysen}, \citenamefont {Rosch},
  \citenamefont {Vojta},\ and\ \citenamefont {W\"olfle}}]{LohneysenRosch07}%
  \BibitemOpen
  \bibfield  {author} {\bibinfo {author} {\bibfnamefont {H.~v.}\ \bibnamefont
  {L\"ohneysen}}, \bibinfo {author} {\bibfnamefont {A.}~\bibnamefont {Rosch}},
  \bibinfo {author} {\bibfnamefont {M.}~\bibnamefont {Vojta}}, \ and\ \bibinfo
  {author} {\bibfnamefont {P.}~\bibnamefont {W\"olfle}},\
  }\href{http://link.aps.org/doi/10.1103/RevModPhys.79.1015} {\bibfield
  {journal} {\bibinfo  {journal} {Rev. Mod. Phys.}\ }\textbf {\bibinfo {volume}
  {79}},\ \bibinfo {pages} {1015} (\bibinfo {year} {2007})}\BibitemShut
  {NoStop}%
\bibitem [{\citenamefont {Griffiths}(1969)}]{Griffiths69}%
  \BibitemOpen
  \bibfield  {author} {\bibinfo {author} {\bibfnamefont {R.~B.}\ \bibnamefont
  {Griffiths}},\ }\href{http://link.aps.org/abstract/PRL/v23/p17} {\bibfield
  {journal} {\bibinfo  {journal} {Phys. Rev. Lett.}\ }\textbf {\bibinfo
  {volume} {23}},\ \bibinfo {pages} {17} (\bibinfo {year} {1969})}\BibitemShut
  {NoStop}%
\bibitem [{\citenamefont {Ballesteros}\ \emph
  {\textit{et~al.}}(1998)\citenamefont {Ballesteros}, \citenamefont
  {Fern\'andez}, \citenamefont {Mart\'in-Mayor}, \citenamefont {Mu\~noz
  Sudupe}, \citenamefont {Parisi},\ and\ \citenamefont
  {Ruiz-Lorenzo}}]{BallesterosFernandez98}%
  \BibitemOpen
  \bibfield  {author} {\bibinfo {author} {\bibfnamefont {H.~G.}\ \bibnamefont
  {Ballesteros}}, \bibinfo {author} {\bibfnamefont {L.~A.}\ \bibnamefont
  {Fern\'andez}}, \bibinfo {author} {\bibfnamefont {V.}~\bibnamefont
  {Mart\'in-Mayor}}, \bibinfo {author} {\bibfnamefont {A.}~\bibnamefont
  {Mu\~noz Sudupe}}, \bibinfo {author} {\bibfnamefont {G.}~\bibnamefont
  {Parisi}}, \ and\ \bibinfo {author} {\bibfnamefont {J.~J.}\ \bibnamefont
  {Ruiz-Lorenzo}},\ }\href{http://link.aps.org/doi/10.1103/PhysRevB.58.2740}
  {\bibfield  {journal} {\bibinfo  {journal} {Phys. Rev. B}\ }\textbf {\bibinfo
  {volume} {58}},\ \bibinfo {pages} {2740} (\bibinfo {year}
  {1998})}\BibitemShut {NoStop}%
\bibitem [{\citenamefont {Bhatt}\ and\ \citenamefont {Lee}(1982)}]{BhattLee82}%
  \BibitemOpen
  \bibfield  {author} {\bibinfo {author} {\bibfnamefont {R.~N.}\ \bibnamefont
  {Bhatt}}\ and\ \bibinfo {author} {\bibfnamefont {P.~A.}\ \bibnamefont
  {Lee}},\ }\href{http://link.aps.org/doi/10.1103/PhysRevLett.48.344}
  {\bibfield  {journal} {\bibinfo  {journal} {Phys. Rev. Lett.}\ }\textbf
  {\bibinfo {volume} {48}},\ \bibinfo {pages} {344} (\bibinfo {year}
  {1982})}\BibitemShut {NoStop}%
\bibitem [{\citenamefont {Sandvik}(2002)}]{Sandvik02}%
  \BibitemOpen
  \bibfield  {author} {\bibinfo {author} {\bibfnamefont {A.~W.}\ \bibnamefont
  {Sandvik}},\ }\href{http://link.aps.org/abstract/PRL/v89/e177201} {\bibfield
  {journal} {\bibinfo  {journal} {Phys. Rev. Lett.}\ }\textbf {\bibinfo
  {volume} {89}},\ \bibinfo {pages} {177201} (\bibinfo {year}
  {2002})}\BibitemShut {NoStop}%
\bibitem [{\citenamefont {Vajk}\ and\ \citenamefont
  {Greven}(2002)}]{VajkGreven02}%
  \BibitemOpen
  \bibfield  {author} {\bibinfo {author} {\bibfnamefont {O.~P.}\ \bibnamefont
  {Vajk}}\ and\ \bibinfo {author} {\bibfnamefont {M.}~\bibnamefont {Greven}},\
  }\href{http://link.aps.org/doi/10.1103/PhysRevLett.89.177202} {\bibfield
  {journal} {\bibinfo  {journal} {Phys. Rev. Lett.}\ }\textbf {\bibinfo
  {volume} {89}},\ \bibinfo {pages} {177202} (\bibinfo {year}
  {2002})}\BibitemShut {NoStop}%
\bibitem [{\citenamefont {Sknepnek}\ \emph {\textit{et~al.}}(2004)\citenamefont
  {Sknepnek}, \citenamefont {Vojta},\ and\ \citenamefont
  {Vojta}}]{SknepnekVojta04}%
  \BibitemOpen
  \bibfield  {author} {\bibinfo {author} {\bibfnamefont {R.}~\bibnamefont
  {Sknepnek}}, \bibinfo {author} {\bibfnamefont {T.}~\bibnamefont {Vojta}}, \
  and\ \bibinfo {author} {\bibfnamefont {M.}~\bibnamefont {Vojta}},\
  }\href{http://link.aps.org/doi/10.1103/PhysRevLett.93.097201} {\bibfield
  {journal} {\bibinfo  {journal} {Phys. Rev. Lett.}\ }\textbf {\bibinfo
  {volume} {93}},\ \bibinfo {pages} {097201} (\bibinfo {year}
  {2004})}\BibitemShut {NoStop}%
\bibitem [{\citenamefont {Vojta}\ and\ \citenamefont
  {Schmalian}(2005)}]{VojtaSchmalian05}%
  \BibitemOpen
  \bibfield  {author} {\bibinfo {author} {\bibfnamefont {T.}~\bibnamefont
  {Vojta}}\ and\ \bibinfo {author} {\bibfnamefont {J.}~\bibnamefont
  {Schmalian}},\ }\href{http://link.aps.org/doi/10.1103/PhysRevB.72.045438}
  {\bibfield  {journal} {\bibinfo  {journal} {Phys. Rev. B}\ }\textbf {\bibinfo
  {volume} {72}},\ \bibinfo {pages} {045438} (\bibinfo {year}
  {2005})}\BibitemShut {NoStop}%
\bibitem [{\citenamefont {Tanaskovi\ifmmode~\acute{c}\else \'{c}\fi{}}\ \emph
  {\textit{et~al.}}(2004)\citenamefont {Tanaskovi\ifmmode~\acute{c}\else
  \'{c}\fi{}}, \citenamefont {Miranda},\ and\ \citenamefont
  {Dobrosavljevi\ifmmode~\acute{c}\else \'{c}\fi{}}}]{TanaskovicMiranda04}%
  \BibitemOpen
  \bibfield  {author} {\bibinfo {author} {\bibfnamefont {D.}~\bibnamefont
  {Tanaskovi\ifmmode~\acute{c}\else \'{c}\fi{}}}, \bibinfo {author}
  {\bibfnamefont {E.}~\bibnamefont {Miranda}}, \ and\ \bibinfo {author}
  {\bibfnamefont {V.}~\bibnamefont {Dobrosavljevi\ifmmode~\acute{c}\else
  \'{c}\fi{}}},\ }\href{http://link.aps.org/doi/10.1103/PhysRevB.70.205108}
  {\bibfield  {journal} {\bibinfo  {journal} {Phys. Rev. B}\ }\textbf {\bibinfo
  {volume} {70}},\ \bibinfo {pages} {205108} (\bibinfo {year}
  {2004})}\BibitemShut {NoStop}%
\bibitem [{\citenamefont {Binek}\ \emph {\textit{et~al.}}(1998)\citenamefont
  {Binek}, \citenamefont {Kleemann},\ and\ \citenamefont
  {Belanger}}]{BinekKleemann98}%
  \BibitemOpen
  \bibfield  {author} {\bibinfo {author} {\bibfnamefont {C.}~\bibnamefont
  {Binek}}, \bibinfo {author} {\bibfnamefont {W.}~\bibnamefont {Kleemann}}, \
  and\ \bibinfo {author} {\bibfnamefont {D.~P.}\ \bibnamefont {Belanger}},\
  }\href{http://link.aps.org/doi/10.1103/PhysRevB.57.7791} {\bibfield
  {journal} {\bibinfo  {journal} {Phys. Rev. B}\ }\textbf {\bibinfo {volume}
  {57}},\ \bibinfo {pages} {7791} (\bibinfo {year} {1998})}\BibitemShut
  {NoStop}%
\bibitem [{\citenamefont {Salamon}\ \emph {\textit{et~al.}}(2002)\citenamefont
  {Salamon}, \citenamefont {Lin},\ and\ \citenamefont {Chun}}]{SalamonLin02}%
  \BibitemOpen
  \bibfield  {author} {\bibinfo {author} {\bibfnamefont {M.~B.}\ \bibnamefont
  {Salamon}}, \bibinfo {author} {\bibfnamefont {P.}~\bibnamefont {Lin}}, \ and\
  \bibinfo {author} {\bibfnamefont {S.~H.}\ \bibnamefont {Chun}},\
  }\href{http://link.aps.org/doi/10.1103/PhysRevLett.88.197203} {\bibfield
  {journal} {\bibinfo  {journal} {Phys. Rev. Lett.}\ }\textbf {\bibinfo
  {volume} {88}},\ \bibinfo {pages} {197203} (\bibinfo {year}
  {2002})}\BibitemShut {NoStop}%
\bibitem [{\citenamefont {Salamon}\ and\ \citenamefont
  {Chun}(2003)}]{SalamonChun03}%
  \BibitemOpen
  \bibfield  {author} {\bibinfo {author} {\bibfnamefont {M.~B.}\ \bibnamefont
  {Salamon}}\ and\ \bibinfo {author} {\bibfnamefont {S.~H.}\ \bibnamefont
  {Chun}},\ }\href{http://link.aps.org/doi/10.1103/PhysRevB.68.014411}
  {\bibfield  {journal} {\bibinfo  {journal} {Phys. Rev. B}\ }\textbf {\bibinfo
  {volume} {68}},\ \bibinfo {pages} {014411} (\bibinfo {year}
  {2003})}\BibitemShut {NoStop}%
\bibitem [{\citenamefont {Wang}\ \emph {\textit{et~al.}}(2007)\citenamefont
  {Wang}, \citenamefont {Sun}, \citenamefont {Liu}, \citenamefont {Xie},
  \citenamefont {Wang}, \citenamefont {Zhao}, \citenamefont {Shen},\ and\
  \citenamefont {Li}}]{WangSun07}%
  \BibitemOpen
  \bibfield  {author} {\bibinfo {author} {\bibfnamefont {J.~Z.}\ \bibnamefont
  {Wang}}, \bibinfo {author} {\bibfnamefont {J.~R.}\ \bibnamefont {Sun}},
  \bibinfo {author} {\bibfnamefont {G.~J.}\ \bibnamefont {Liu}}, \bibinfo
  {author} {\bibfnamefont {Y.~W.}\ \bibnamefont {Xie}}, \bibinfo {author}
  {\bibfnamefont {D.~J.}\ \bibnamefont {Wang}}, \bibinfo {author}
  {\bibfnamefont {T.~Y.}\ \bibnamefont {Zhao}}, \bibinfo {author}
  {\bibfnamefont {B.~G.}\ \bibnamefont {Shen}}, \ and\ \bibinfo {author}
  {\bibfnamefont {X.~G.}\ \bibnamefont {Li}},\
  }\href{http://link.aps.org/doi/10.1103/PhysRevB.76.104428} {\bibfield
  {journal} {\bibinfo  {journal} {Phys. Rev. B}\ }\textbf {\bibinfo {volume}
  {76}},\ \bibinfo {pages} {104428} (\bibinfo {year} {2007})}\BibitemShut
  {NoStop}%
\bibitem [{\citenamefont {Guo}\ \emph {\textit{et~al.}}(2008)\citenamefont
  {Guo}, \citenamefont {Young}, \citenamefont {Macaluso}, \citenamefont
  {Browne}, \citenamefont {Henderson}, \citenamefont {Chan}, \citenamefont
  {Henry},\ and\ \citenamefont {DiTusa}}]{GuoYoung08}%
  \BibitemOpen
  \bibfield  {author} {\bibinfo {author} {\bibfnamefont {S.}~\bibnamefont
  {Guo}}, \bibinfo {author} {\bibfnamefont {D.~P.}\ \bibnamefont {Young}},
  \bibinfo {author} {\bibfnamefont {R.~T.}\ \bibnamefont {Macaluso}}, \bibinfo
  {author} {\bibfnamefont {D.~A.}\ \bibnamefont {Browne}}, \bibinfo {author}
  {\bibfnamefont {N.~L.}\ \bibnamefont {Henderson}}, \bibinfo {author}
  {\bibfnamefont {J.~Y.}\ \bibnamefont {Chan}}, \bibinfo {author}
  {\bibfnamefont {L.~L.}\ \bibnamefont {Henry}}, \ and\ \bibinfo {author}
  {\bibfnamefont {J.~F.}\ \bibnamefont {DiTusa}},\
  }\href{http://link.aps.org/doi/10.1103/PhysRevLett.100.017209} {\bibfield
  {journal} {\bibinfo  {journal} {Phys. Rev. Lett.}\ }\textbf {\bibinfo
  {volume} {100}},\ \bibinfo {pages} {017209} (\bibinfo {year}
  {2008})}\BibitemShut {NoStop}%
\bibitem [{\citenamefont {Guo}\ \emph {\textit{et~al.}}(2010)\citenamefont
  {Guo}, \citenamefont {Young}, \citenamefont {Macaluso}, \citenamefont
  {Browne}, \citenamefont {Henderson}, \citenamefont {Chan}, \citenamefont
  {Henry},\ and\ \citenamefont {DiTusa}}]{GuoYoung10}%
  \BibitemOpen
  \bibfield  {author} {\bibinfo {author} {\bibfnamefont {S.}~\bibnamefont
  {Guo}}, \bibinfo {author} {\bibfnamefont {D.~P.}\ \bibnamefont {Young}},
  \bibinfo {author} {\bibfnamefont {R.~T.}\ \bibnamefont {Macaluso}}, \bibinfo
  {author} {\bibfnamefont {D.~A.}\ \bibnamefont {Browne}}, \bibinfo {author}
  {\bibfnamefont {N.~L.}\ \bibnamefont {Henderson}}, \bibinfo {author}
  {\bibfnamefont {J.~Y.}\ \bibnamefont {Chan}}, \bibinfo {author}
  {\bibfnamefont {L.~L.}\ \bibnamefont {Henry}}, \ and\ \bibinfo {author}
  {\bibfnamefont {J.~F.}\ \bibnamefont {DiTusa}},\
  }\href{http://link.aps.org/doi/10.1103/PhysRevB.81.144423} {\bibfield
  {journal} {\bibinfo  {journal} {Phys. Rev. B}\ }\textbf {\bibinfo {volume}
  {81}},\ \bibinfo {pages} {144423} (\bibinfo {year} {2010})}\BibitemShut
  {NoStop}%
\bibitem [{\citenamefont {Krivoruchko}\ \emph
  {\textit{et~al.}}(2010)\citenamefont {Krivoruchko}, \citenamefont
  {Marchenko},\ and\ \citenamefont {Melikhov}}]{KrivoruchkoMarchenko10}%
  \BibitemOpen
  \bibfield  {author} {\bibinfo {author} {\bibfnamefont {V.~N.}\ \bibnamefont
  {Krivoruchko}}, \bibinfo {author} {\bibfnamefont {M.~A.}\ \bibnamefont
  {Marchenko}}, \ and\ \bibinfo {author} {\bibfnamefont {Y.}~\bibnamefont
  {Melikhov}},\ }\href{http://link.aps.org/doi/10.1103/PhysRevB.82.064419}
  {\bibfield  {journal} {\bibinfo  {journal} {Phys. Rev. B}\ }\textbf {\bibinfo
  {volume} {82}},\ \bibinfo {pages} {064419} (\bibinfo {year}
  {2010})}\BibitemShut {NoStop}%
\bibitem [{\citenamefont {Eremina}\ \emph {\textit{et~al.}}(2011)\citenamefont
  {Eremina}, \citenamefont {Fazlizhanov}, \citenamefont {Yatsyk}, \citenamefont
  {Sharipov}, \citenamefont {Pyataev}, \citenamefont {Krug~von Nidda},
  \citenamefont {Pascher}, \citenamefont {Loidl}, \citenamefont {Glazyrin},\
  and\ \citenamefont {Mukovskii}}]{EreminaFazlizhanov11}%
  \BibitemOpen
  \bibfield  {author} {\bibinfo {author} {\bibfnamefont {R.~M.}\ \bibnamefont
  {Eremina}}, \bibinfo {author} {\bibfnamefont {I.~I.}\ \bibnamefont
  {Fazlizhanov}}, \bibinfo {author} {\bibfnamefont {I.~V.}\ \bibnamefont
  {Yatsyk}}, \bibinfo {author} {\bibfnamefont {K.~R.}\ \bibnamefont
  {Sharipov}}, \bibinfo {author} {\bibfnamefont {A.~V.}\ \bibnamefont
  {Pyataev}}, \bibinfo {author} {\bibfnamefont {H.-A.}\ \bibnamefont {Krug~von
  Nidda}}, \bibinfo {author} {\bibfnamefont {N.}~\bibnamefont {Pascher}},
  \bibinfo {author} {\bibfnamefont {A.}~\bibnamefont {Loidl}}, \bibinfo
  {author} {\bibfnamefont {K.~V.}\ \bibnamefont {Glazyrin}}, \ and\ \bibinfo
  {author} {\bibfnamefont {Y.~M.}\ \bibnamefont {Mukovskii}},\
  }\href{http://link.aps.org/doi/10.1103/PhysRevB.84.064410} {\bibfield
  {journal} {\bibinfo  {journal} {Phys. Rev. B}\ }\textbf {\bibinfo {volume}
  {84}},\ \bibinfo {pages} {064410} (\bibinfo {year} {2011})}\BibitemShut
  {NoStop}%
\bibitem [{\citenamefont {Vojta}(2010)}]{Vojta10}%
  \BibitemOpen
  \bibfield  {author} {\bibinfo {author} {\bibfnamefont {T.}~\bibnamefont
  {Vojta}},\ }\href{http://www.springerlink.com/content/mj7845322m07140p/}
  {\bibfield  {journal} {\bibinfo  {journal} {J.~Low Temp. Phys.}\ }\textbf
  {\bibinfo {volume} {161}},\ \bibinfo {pages} {299} (\bibinfo {year}
  {2010})}\BibitemShut {NoStop}%
\bibitem [{\citenamefont {Nozadze}\ and\ \citenamefont
  {Vojta}(2011)}]{NozadzeVojta11}%
  \BibitemOpen
  \bibfield  {author} {\bibinfo {author} {\bibfnamefont {D.}~\bibnamefont
  {Nozadze}}\ and\ \bibinfo {author} {\bibfnamefont {T.}~\bibnamefont
  {Vojta}},\
  }\href{http://epljournal.edpsciences.org/articles/epl/abs/2011/17/epl13781/epl13781.html}
  {\bibfield  {journal} {\bibinfo  {journal} {EPL}\ }\textbf {\bibinfo {volume}
  {95}},\ \bibinfo {pages} {57010} (\bibinfo {year} {2011})}\BibitemShut
  {NoStop}%
\bibitem [{\citenamefont {Nozadze}\ and\ \citenamefont
  {Vojta}(2012)}]{NozadzeVojta12}%
  \BibitemOpen
  \bibfield  {author} {\bibinfo {author} {\bibfnamefont {D.}~\bibnamefont
  {Nozadze}}\ and\ \bibinfo {author} {\bibfnamefont {T.}~\bibnamefont
  {Vojta}},\ }\href{http://link.aps.org/doi/10.1103/PhysRevB.85.174202}
  {\bibfield  {journal} {\bibinfo  {journal} {Phys. Rev. B}\ }\textbf {\bibinfo
  {volume} {85}},\ \bibinfo {pages} {174202} (\bibinfo {year}
  {2012})}\BibitemShut {NoStop}%
\bibitem [{\citenamefont {Ruderman}\ and\ \citenamefont
  {Kittel}(1954)}]{RudermanKittel54}%
  \BibitemOpen
  \bibfield  {author} {\bibinfo {author} {\bibfnamefont {M.~A.}\ \bibnamefont
  {Ruderman}}\ and\ \bibinfo {author} {\bibfnamefont {C.}~\bibnamefont
  {Kittel}},\ }\href{http://link.aps.org/doi/10.1103/PhysRev.96.99} {\bibfield
  {journal} {\bibinfo  {journal} {Phys. Rev.}\ }\textbf {\bibinfo {volume}
  {96}},\ \bibinfo {pages} {99} (\bibinfo {year} {1954})}\BibitemShut {NoStop}%
\bibitem [{\citenamefont {Kasuya}(1956)}]{Kasuya56}%
  \BibitemOpen
  \bibfield  {author} {\bibinfo {author} {\bibfnamefont {T.}~\bibnamefont
  {Kasuya}},\ }\href{http://ptp.ipap.jp/link?PTP/16/45/} {\bibfield  {journal}
  {\bibinfo  {journal} {Prog. Theor. Phys.}\ }\textbf {\bibinfo {volume}
  {16}},\ \bibinfo {pages} {45} (\bibinfo {year} {1956})}\BibitemShut {NoStop}%
\bibitem [{\citenamefont {Yosida}(1957)}]{Yosida57}%
  \BibitemOpen
  \bibfield  {author} {\bibinfo {author} {\bibfnamefont {K.}~\bibnamefont
  {Yosida}},\ }\href{http://link.aps.org/doi/10.1103/PhysRev.106.893}
  {\bibfield  {journal} {\bibinfo  {journal} {Phys. Rev.}\ }\textbf {\bibinfo
  {volume} {106}},\ \bibinfo {pages} {893} (\bibinfo {year}
  {1957})}\BibitemShut {NoStop}%
\bibitem [{\citenamefont {Fischer}\ and\ \citenamefont
  {Klein}(1975)}]{FischerKlein75}%
  \BibitemOpen
  \bibfield  {author} {\bibinfo {author} {\bibfnamefont {B.}~\bibnamefont
  {Fischer}}\ and\ \bibinfo {author} {\bibfnamefont {M.~W.}\ \bibnamefont
  {Klein}},\ }\href{http://link.aps.org/doi/10.1103/PhysRevB.11.2025}
  {\bibfield  {journal} {\bibinfo  {journal} {Phys. Rev.~B}\ }\textbf {\bibinfo
  {volume} {11}},\ \bibinfo {pages} {2025} (\bibinfo {year}
  {1975})}\BibitemShut {NoStop}%
\bibitem [{\citenamefont {Castro~Neto}\ and\ \citenamefont
  {Jones}(2000)}]{CastroNetoJones00}%
  \BibitemOpen
  \bibfield  {author} {\bibinfo {author} {\bibfnamefont {A.~H.}\ \bibnamefont
  {Castro~Neto}}\ and\ \bibinfo {author} {\bibfnamefont {B.~A.}\ \bibnamefont
  {Jones}},\ }\href{http://link.aps.org/doi/10.1103/PhysRevB.62.14975}
  {\bibfield  {journal} {\bibinfo  {journal} {Phys. Rev. B}\ }\textbf {\bibinfo
  {volume} {62}},\ \bibinfo {pages} {14975} (\bibinfo {year}
  {2000})}\BibitemShut {NoStop}%
\bibitem [{\citenamefont {Dobrosavljevi\ifmmode~\acute{c}\else \'{c}\fi{}}\
  and\ \citenamefont {Miranda}(2005)}]{DobrosavljevicMiranda05}%
  \BibitemOpen
  \bibfield  {author} {\bibinfo {author} {\bibfnamefont {V.}~\bibnamefont
  {Dobrosavljevi\ifmmode~\acute{c}\else \'{c}\fi{}}}\ and\ \bibinfo {author}
  {\bibfnamefont {E.}~\bibnamefont {Miranda}},\
  }\href{http://link.aps.org/doi/10.1103/PhysRevLett.94.187203} {\bibfield
  {journal} {\bibinfo  {journal} {Phys. Rev. Lett.}\ }\textbf {\bibinfo
  {volume} {94}},\ \bibinfo {pages} {187203} (\bibinfo {year}
  {2005})}\BibitemShut {NoStop}%
\bibitem [{\citenamefont {Westerkamp}\ \emph
  {\textit{et~al.}}(2009)\citenamefont {Westerkamp}, \citenamefont {Deppe},
  \citenamefont {K\"uchler}, \citenamefont {Brando}, \citenamefont {Geibel},
  \citenamefont {Gegenwart}, \citenamefont {Pikul},\ and\ \citenamefont
  {Steglich}}]{WesterkampDeppe09}%
  \BibitemOpen
  \bibfield  {author} {\bibinfo {author} {\bibfnamefont {T.}~\bibnamefont
  {Westerkamp}}, \bibinfo {author} {\bibfnamefont {M.}~\bibnamefont {Deppe}},
  \bibinfo {author} {\bibfnamefont {R.}~\bibnamefont {K\"uchler}}, \bibinfo
  {author} {\bibfnamefont {M.}~\bibnamefont {Brando}}, \bibinfo {author}
  {\bibfnamefont {C.}~\bibnamefont {Geibel}}, \bibinfo {author} {\bibfnamefont
  {P.}~\bibnamefont {Gegenwart}}, \bibinfo {author} {\bibfnamefont {A.~P.}\
  \bibnamefont {Pikul}}, \ and\ \bibinfo {author} {\bibfnamefont
  {F.}~\bibnamefont {Steglich}},\
  }\href{http://link.aps.org/doi/10.1103/PhysRevLett.102.206404} {\bibfield
  {journal} {\bibinfo  {journal} {Phys. Rev. Lett.}\ }\textbf {\bibinfo
  {volume} {102}},\ \bibinfo {pages} {206404} (\bibinfo {year}
  {2009})}\BibitemShut {NoStop}%
\bibitem [{\citenamefont {Ubaid-Kassis}\ \emph
  {\textit{et~al.}}(2010)\citenamefont {Ubaid-Kassis}, \citenamefont {Vojta},\
  and\ \citenamefont {Schroeder}}]{UbaidKassisVojta10}%
  \BibitemOpen
  \bibfield  {author} {\bibinfo {author} {\bibfnamefont {S.}~\bibnamefont
  {Ubaid-Kassis}}, \bibinfo {author} {\bibfnamefont {T.}~\bibnamefont {Vojta}},
  \ and\ \bibinfo {author} {\bibfnamefont {A.}~\bibnamefont {Schroeder}},\
  }\href{http://link.aps.org/doi/10.1103/PhysRevLett.104.066402} {\bibfield
  {journal} {\bibinfo  {journal} {Phys. Rev. Lett.}\ }\textbf {\bibinfo
  {volume} {104}},\ \bibinfo {pages} {066402} (\bibinfo {year}
  {2010})}\BibitemShut {NoStop}%
\bibitem [{\citenamefont {Inosov}\ \emph
  {\textit{et~al.}}(2009{\natexlab{a}})\citenamefont {Inosov}, \citenamefont
  {Evtushinsky}, \citenamefont {Koitzsch}, \citenamefont {Zabolotnyy},
  \citenamefont {Borisenko}, \citenamefont {Kordyuk}, \citenamefont {Frontzek},
  \citenamefont {Loewenhaupt}, \citenamefont {L\"oser}, \citenamefont {Mazilu},
  \citenamefont {Bitterlich}, \citenamefont {Behr}, \citenamefont {Hoffmann},
  \citenamefont {Follath},\ and\ \citenamefont
  {B\"uchner}}]{InosovEvtushinsky09}%
  \BibitemOpen
  \bibfield  {author} {\bibinfo {author} {\bibfnamefont {D.~S.}\ \bibnamefont
  {Inosov}}, \bibinfo {author} {\bibfnamefont {D.~V.}\ \bibnamefont
  {Evtushinsky}}, \bibinfo {author} {\bibfnamefont {A.}~\bibnamefont
  {Koitzsch}}, \bibinfo {author} {\bibfnamefont {V.~B.}\ \bibnamefont
  {Zabolotnyy}}, \bibinfo {author} {\bibfnamefont {S.~V.}\ \bibnamefont
  {Borisenko}}, \bibinfo {author} {\bibfnamefont {A.~A.}\ \bibnamefont
  {Kordyuk}}, \bibinfo {author} {\bibfnamefont {M.}~\bibnamefont {Frontzek}},
  \bibinfo {author} {\bibfnamefont {M.}~\bibnamefont {Loewenhaupt}}, \bibinfo
  {author} {\bibfnamefont {W.}~\bibnamefont {L\"oser}}, \bibinfo {author}
  {\bibfnamefont {I.}~\bibnamefont {Mazilu}}, \bibinfo {author} {\bibfnamefont
  {H.}~\bibnamefont {Bitterlich}}, \bibinfo {author} {\bibfnamefont
  {G.}~\bibnamefont {Behr}}, \bibinfo {author} {\bibfnamefont {J.-U.}\
  \bibnamefont {Hoffmann}}, \bibinfo {author} {\bibfnamefont {R.}~\bibnamefont
  {Follath}}, \ and\ \bibinfo {author} {\bibfnamefont {B.}~\bibnamefont
  {B\"uchner}},\ }\href{http://link.aps.org/doi/10.1103/PhysRevLett.102.046401}
  {\bibfield  {journal} {\bibinfo  {journal} {Phys. Rev. Lett.}\ }\textbf
  {\bibinfo {volume} {102}},\ \bibinfo {pages} {046401} (\bibinfo {year}
  {2009}{\natexlab{a}})}\BibitemShut {NoStop}%
\bibitem [{\citenamefont {Shell}\ \emph {\textit{et~al.}}(1982)\citenamefont
  {Shell}, \citenamefont {Cowen},\ and\ \citenamefont {Foiles}}]{ShellCowen82}%
  \BibitemOpen
  \bibfield  {author} {\bibinfo {author} {\bibfnamefont {J.}~\bibnamefont
  {Shell}}, \bibinfo {author} {\bibfnamefont {J.~A.}\ \bibnamefont {Cowen}}, \
  and\ \bibinfo {author} {\bibfnamefont {C.~L.}\ \bibnamefont {Foiles}},\
  }\href{http://link.aps.org/doi/10.1103/PhysRevB.25.6015} {\bibfield
  {journal} {\bibinfo  {journal} {Phys. Rev. B}\ }\textbf {\bibinfo {volume}
  {25}},\ \bibinfo {pages} {6015} (\bibinfo {year} {1982})}\BibitemShut
  {NoStop}%
\bibitem [{\citenamefont {Binder}\ and\ \citenamefont
  {Young}(1986)}]{BinderYoung86}%
  \BibitemOpen
  \bibfield  {author} {\bibinfo {author} {\bibfnamefont {K.}~\bibnamefont
  {Binder}}\ and\ \bibinfo {author} {\bibfnamefont {A.~P.}\ \bibnamefont
  {Young}},\ }\href{http://link.aps.org/doi/10.1103/RevModPhys.58.801}
  {\bibfield  {journal} {\bibinfo  {journal} {Rev. Mod. Phys.}\ }\textbf
  {\bibinfo {volume} {58}},\ \bibinfo {pages} {801} (\bibinfo {year}
  {1986})}\BibitemShut {NoStop}%
\bibitem [{\citenamefont {Fischer}\ and\ \citenamefont
  {Hertz}(1999)}]{FischerHertz99}%
  \BibitemOpen
  \bibfield  {author} {\bibinfo {author} {\bibfnamefont {K.~H.}\ \bibnamefont
  {Fischer}}\ and\ \bibinfo {author} {\bibfnamefont {J.~A.}\ \bibnamefont
  {Hertz}},\ }\href@noop {} {\emph {\bibinfo {title} {Spin glasses}}}\
  (\bibinfo  {publisher} {Cambridge Univ. Press},\ \bibinfo {year}
  {1999})\BibitemShut {NoStop}%
\bibitem [{\citenamefont {Bulaevskii}\ and\ \citenamefont
  {Panyukov}(1986)}]{BulaevskiiPanyukov86}%
  \BibitemOpen
  \bibfield  {author} {\bibinfo {author} {\bibfnamefont {L.~N.}\ \bibnamefont
  {Bulaevskii}}\ and\ \bibinfo {author} {\bibfnamefont {S.~V.}\ \bibnamefont
  {Panyukov}},\
  }\href{http://www.jetpletters.ac.ru/ps/1402/article_21282.shtml} {\bibfield
  {journal} {\bibinfo  {journal} {JETP Lett.}\ }\textbf {\bibinfo {volume}
  {43}},\ \bibinfo {pages} {240} (\bibinfo {year} {1986})}\BibitemShut
  {NoStop}%
\bibitem [{\citenamefont {Aristov}\ and\ \citenamefont
  {Maleyev}(1997)}]{AristovMaleyev97}%
  \BibitemOpen
  \bibfield  {author} {\bibinfo {author} {\bibfnamefont {D.~N.}\ \bibnamefont
  {Aristov}}\ and\ \bibinfo {author} {\bibfnamefont {S.~V.}\ \bibnamefont
  {Maleyev}},\ }\href{http://link.aps.org/doi/10.1103/PhysRevB.56.8841}
  {\bibfield  {journal} {\bibinfo  {journal} {Phys. Rev.~B}\ }\textbf {\bibinfo
  {volume} {56}},\ \bibinfo {pages} {8841} (\bibinfo {year}
  {1997})}\BibitemShut {NoStop}%
\bibitem [{\citenamefont {Akbari}\ \emph {\textit{et~al.}}(2011)\citenamefont
  {Akbari}, \citenamefont {Eremin},\ and\ \citenamefont
  {Thalmeier}}]{AkbariEremin11}%
  \BibitemOpen
  \bibfield  {author} {\bibinfo {author} {\bibfnamefont {A.}~\bibnamefont
  {Akbari}}, \bibinfo {author} {\bibfnamefont {I.}~\bibnamefont {Eremin}}, \
  and\ \bibinfo {author} {\bibfnamefont {P.}~\bibnamefont {Thalmeier}},\
  }\href{http://link.aps.org/doi/10.1103/PhysRevB.84.134513} {\bibfield
  {journal} {\bibinfo  {journal} {Phys. Rev.~B}\ }\textbf {\bibinfo {volume}
  {84}},\ \bibinfo {pages} {134513} (\bibinfo {year} {2011})}\BibitemShut
  {NoStop}%
\bibitem [{\citenamefont {Akbari}\ \emph {\textit{et~al.}}(2013)\citenamefont
  {Akbari}, \citenamefont {Thalmeier},\ and\ \citenamefont
  {Eremin}}]{AkbariThalmeier13}%
  \BibitemOpen
  \bibfield  {author} {\bibinfo {author} {\bibfnamefont {A.}~\bibnamefont
  {Akbari}}, \bibinfo {author} {\bibfnamefont {P.}~\bibnamefont {Thalmeier}}, \
  and\ \bibinfo {author} {\bibfnamefont {I.}~\bibnamefont {Eremin}},\
  }\href{http://iopscience.iop.org/1367-2630/15/3/033034/} {\bibfield
  {journal} {\bibinfo  {journal} {New J. Phys.}\ }\textbf {\bibinfo {volume}
  {15}},\ \bibinfo {pages} {033034} (\bibinfo {year} {2013})}\BibitemShut
  {NoStop}%
\bibitem [{\citenamefont {Ren}\ and\ \citenamefont {Zhao}(2009)}]{RenZhao09}%
  \BibitemOpen
  \bibfield  {author} {\bibinfo {author} {\bibfnamefont {Z.-A.}\ \bibnamefont
  {Ren}}\ and\ \bibinfo {author} {\bibfnamefont {Z.-X.}\ \bibnamefont {Zhao}},\
  }\href{http://onlinelibrary.wiley.com/doi/10.1002/adma.200901049/abstract}
  {\bibfield  {journal} {\bibinfo  {journal} {Adv. Mater.}\ }\textbf {\bibinfo
  {volume} {21}},\ \bibinfo {pages} {4584} (\bibinfo {year}
  {2009})}\BibitemShut {NoStop}%
\bibitem [{\citenamefont {Lumsden}\ and\ \citenamefont
  {Christianson}(2010)}]{LumsdenChristianson10review}%
  \BibitemOpen
  \bibfield  {author} {\bibinfo {author} {\bibfnamefont {M.~D.}\ \bibnamefont
  {Lumsden}}\ and\ \bibinfo {author} {\bibfnamefont {A.~D.}\ \bibnamefont
  {Christianson}},\ }\href{http://iopscience.iop.org/0953-8984/22/20/203203}
  {\bibfield  {journal} {\bibinfo  {journal} {J.~Phys.: Condens. Matter}\
  }\textbf {\bibinfo {volume} {22}},\ \bibinfo {pages} {203203} (\bibinfo
  {year} {2010})}\BibitemShut {NoStop}%
\bibitem [{\citenamefont {Chu}(2009)}]{Chu09}%
  \BibitemOpen
  \bibfield  {author} {\bibinfo {author} {\bibfnamefont {C.~W.}\ \bibnamefont
  {Chu}},\
  }\href{http://www.nature.com/nphys/journal/v5/n11/abs/nphys1449.html}
  {\bibfield  {journal} {\bibinfo  {journal} {Nature Phys.}\ }\textbf {\bibinfo
  {volume} {5}},\ \bibinfo {pages} {787} (\bibinfo {year} {2009})}\BibitemShut
  {NoStop}%
\bibitem [{\citenamefont {Paglione}\ and\ \citenamefont
  {Greene}(2010)}]{PaglioneGreene10}%
  \BibitemOpen
  \bibfield  {author} {\bibinfo {author} {\bibfnamefont {J.}~\bibnamefont
  {Paglione}}\ and\ \bibinfo {author} {\bibfnamefont {R.~L.}\ \bibnamefont
  {Greene}},\
  }\href{http://www.nature.com/nphys/journal/v6/n9/abs/nphys1759.html}
  {\bibfield  {journal} {\bibinfo  {journal} {Nature Phys.}\ }\textbf {\bibinfo
  {volume} {6}},\ \bibinfo {pages} {645} (\bibinfo {year} {2010})}\BibitemShut
  {NoStop}%
\bibitem [{\citenamefont {Johnston}(2010)}]{Johnston10}%
  \BibitemOpen
  \bibfield  {author} {\bibinfo {author} {\bibfnamefont {D.~C.}\ \bibnamefont
  {Johnston}},\
  }\href{http://www.tandfonline.com/doi/abs/10.1080/00018732.2010.513480}
  {\bibfield  {journal} {\bibinfo  {journal} {Adv. Phys.}\ }\textbf {\bibinfo
  {volume} {59}},\ \bibinfo {pages} {803} (\bibinfo {year} {2010})}\BibitemShut
  {NoStop}%
\bibitem [{\citenamefont {Stewart}(2011)}]{Stewart11}%
  \BibitemOpen
  \bibfield  {author} {\bibinfo {author} {\bibfnamefont {G.~R.}\ \bibnamefont
  {Stewart}},\ }\href{http://link.aps.org/doi/10.1103/RevModPhys.83.1589}
  {\bibfield  {journal} {\bibinfo  {journal} {Rev. Mod. Phys.}\ }\textbf
  {\bibinfo {volume} {83}},\ \bibinfo {pages} {1589} (\bibinfo {year}
  {2011})}\BibitemShut {NoStop}%
\bibitem [{\citenamefont {Canfield}\ and\ \citenamefont
  {Bud'ko}(2010)}]{CanfieldBudko10}%
  \BibitemOpen
  \bibfield  {author} {\bibinfo {author} {\bibfnamefont {P.~C.}\ \bibnamefont
  {Canfield}}\ and\ \bibinfo {author} {\bibfnamefont {S.~L.}\ \bibnamefont
  {Bud'ko}},\
  }\href{http://arjournals.annualreviews.org/doi/abs/10.1146/annurev-conmatphys-070909-104041}
  {\bibfield  {journal} {\bibinfo  {journal} {Annu. Rev. Condens. Matter
  Phys.}\ }\textbf {\bibinfo {volume} {1}},\ \bibinfo {pages} {11.1} (\bibinfo
  {year} {2010})}\BibitemShut {NoStop}%
\bibitem [{\citenamefont {Sefat}\ \emph {\textit{et~al.}}(2009)\citenamefont
  {Sefat}, \citenamefont {Singh}, \citenamefont {VanBebber}, \citenamefont
  {Mozharivskyj}, \citenamefont {McGuire}, \citenamefont {Jin}, \citenamefont
  {Sales}, \citenamefont {Keppens},\ and\ \citenamefont
  {Mandrus}}]{SefatSingh09}%
  \BibitemOpen
  \bibfield  {author} {\bibinfo {author} {\bibfnamefont {A.~S.}\ \bibnamefont
  {Sefat}}, \bibinfo {author} {\bibfnamefont {D.~J.}\ \bibnamefont {Singh}},
  \bibinfo {author} {\bibfnamefont {L.~H.}\ \bibnamefont {VanBebber}}, \bibinfo
  {author} {\bibfnamefont {Y.}~\bibnamefont {Mozharivskyj}}, \bibinfo {author}
  {\bibfnamefont {M.~A.}\ \bibnamefont {McGuire}}, \bibinfo {author}
  {\bibfnamefont {R.}~\bibnamefont {Jin}}, \bibinfo {author} {\bibfnamefont
  {B.~C.}\ \bibnamefont {Sales}}, \bibinfo {author} {\bibfnamefont
  {V.}~\bibnamefont {Keppens}}, \ and\ \bibinfo {author} {\bibfnamefont
  {D.}~\bibnamefont {Mandrus}},\
  }\href{http://link.aps.org/doi/10.1103/PhysRevB.79.224524} {\bibfield
  {journal} {\bibinfo  {journal} {Phys. Rev. B}\ }\textbf {\bibinfo {volume}
  {79}},\ \bibinfo {pages} {224524} (\bibinfo {year} {2009})}\BibitemShut
  {NoStop}%
\bibitem [{\citenamefont {Marty}\ \emph {\textit{et~al.}}(2011)\citenamefont
  {Marty}, \citenamefont {Christianson}, \citenamefont {Wang}, \citenamefont
  {Matsuda}, \citenamefont {Cao}, \citenamefont {VanBebber}, \citenamefont
  {Zarestky}, \citenamefont {Singh}, \citenamefont {Sefat},\ and\ \citenamefont
  {Lumsden}}]{MartyChristianson11}%
  \BibitemOpen
  \bibfield  {author} {\bibinfo {author} {\bibfnamefont {K.}~\bibnamefont
  {Marty}}, \bibinfo {author} {\bibfnamefont {A.~D.}\ \bibnamefont
  {Christianson}}, \bibinfo {author} {\bibfnamefont {C.~H.}\ \bibnamefont
  {Wang}}, \bibinfo {author} {\bibfnamefont {M.}~\bibnamefont {Matsuda}},
  \bibinfo {author} {\bibfnamefont {H.}~\bibnamefont {Cao}}, \bibinfo {author}
  {\bibfnamefont {L.~H.}\ \bibnamefont {VanBebber}}, \bibinfo {author}
  {\bibfnamefont {J.~L.}\ \bibnamefont {Zarestky}}, \bibinfo {author}
  {\bibfnamefont {D.~J.}\ \bibnamefont {Singh}}, \bibinfo {author}
  {\bibfnamefont {A.~S.}\ \bibnamefont {Sefat}}, \ and\ \bibinfo {author}
  {\bibfnamefont {M.~D.}\ \bibnamefont {Lumsden}},\
  }\href{http://link.aps.org/doi/10.1103/PhysRevB.83.060509} {\bibfield
  {journal} {\bibinfo  {journal} {Phys. Rev. B}\ }\textbf {\bibinfo {volume}
  {83}},\ \bibinfo {pages} {060509} (\bibinfo {year} {2011})}\BibitemShut
  {NoStop}%
\bibitem [{\citenamefont {Pandey}\ \emph {\textit{et~al.}}(2011)\citenamefont
  {Pandey}, \citenamefont {Anand},\ and\ \citenamefont
  {Johnston}}]{PandeyAnand11}%
  \BibitemOpen
  \bibfield  {author} {\bibinfo {author} {\bibfnamefont {A.}~\bibnamefont
  {Pandey}}, \bibinfo {author} {\bibfnamefont {V.~K.}\ \bibnamefont {Anand}}, \
  and\ \bibinfo {author} {\bibfnamefont {D.~C.}\ \bibnamefont {Johnston}},\
  }\href{http://link.aps.org/doi/10.1103/PhysRevB.84.014405} {\bibfield
  {journal} {\bibinfo  {journal} {Phys. Rev. B}\ }\textbf {\bibinfo {volume}
  {84}},\ \bibinfo {pages} {014405} (\bibinfo {year} {2011})}\BibitemShut
  {NoStop}%
\bibitem [{\citenamefont {Kim}\ \emph
  {\textit{et~al.}}(2010{\natexlab{a}})\citenamefont {Kim}, \citenamefont
  {Khim}, \citenamefont {Kim}, \citenamefont {Eom}, \citenamefont {Law},
  \citenamefont {Kremer}, \citenamefont {Shim},\ and\ \citenamefont
  {Kim}}]{KimKhim10}%
  \BibitemOpen
  \bibfield  {author} {\bibinfo {author} {\bibfnamefont {J.~S.}\ \bibnamefont
  {Kim}}, \bibinfo {author} {\bibfnamefont {S.}~\bibnamefont {Khim}}, \bibinfo
  {author} {\bibfnamefont {H.~J.}\ \bibnamefont {Kim}}, \bibinfo {author}
  {\bibfnamefont {M.~J.}\ \bibnamefont {Eom}}, \bibinfo {author} {\bibfnamefont
  {J.~M.}\ \bibnamefont {Law}}, \bibinfo {author} {\bibfnamefont {R.~K.}\
  \bibnamefont {Kremer}}, \bibinfo {author} {\bibfnamefont {J.~H.}\
  \bibnamefont {Shim}}, \ and\ \bibinfo {author} {\bibfnamefont {K.~H.}\
  \bibnamefont {Kim}},\
  }\href{http://link.aps.org/doi/10.1103/PhysRevB.82.024510} {\bibfield
  {journal} {\bibinfo  {journal} {Phys. Rev. B}\ }\textbf {\bibinfo {volume}
  {82}},\ \bibinfo {pages} {024510} (\bibinfo {year}
  {2010}{\natexlab{a}})}\BibitemShut {NoStop}%
\bibitem [{\citenamefont {Thaler}\ \emph {\textit{et~al.}}(2011)\citenamefont
  {Thaler}, \citenamefont {Hodovanets}, \citenamefont {Torikachvili},
  \citenamefont {Ran}, \citenamefont {Kracher}, \citenamefont {Straszheim},
  \citenamefont {Yan}, \citenamefont {Mun},\ and\ \citenamefont
  {Canfield}}]{ThalerHodovanets11}%
  \BibitemOpen
  \bibfield  {author} {\bibinfo {author} {\bibfnamefont {A.}~\bibnamefont
  {Thaler}}, \bibinfo {author} {\bibfnamefont {H.}~\bibnamefont {Hodovanets}},
  \bibinfo {author} {\bibfnamefont {M.~S.}\ \bibnamefont {Torikachvili}},
  \bibinfo {author} {\bibfnamefont {S.}~\bibnamefont {Ran}}, \bibinfo {author}
  {\bibfnamefont {A.}~\bibnamefont {Kracher}}, \bibinfo {author} {\bibfnamefont
  {W.}~\bibnamefont {Straszheim}}, \bibinfo {author} {\bibfnamefont {J.~Q.}\
  \bibnamefont {Yan}}, \bibinfo {author} {\bibfnamefont {E.}~\bibnamefont
  {Mun}}, \ and\ \bibinfo {author} {\bibfnamefont {P.~C.}\ \bibnamefont
  {Canfield}},\ }\href{http://link.aps.org/doi/10.1103/PhysRevB.84.144528}
  {\bibfield  {journal} {\bibinfo  {journal} {Phys. Rev. B}\ }\textbf {\bibinfo
  {volume} {84}},\ \bibinfo {pages} {144528} (\bibinfo {year}
  {2011})}\BibitemShut {NoStop}%
\bibitem [{\citenamefont {Kim}\ \emph
  {\textit{et~al.}}(2010{\natexlab{b}})\citenamefont {Kim}, \citenamefont
  {Kreyssig}, \citenamefont {Thaler}, \citenamefont {Pratt}, \citenamefont
  {Tian}, \citenamefont {Zarestky}, \citenamefont {Green}, \citenamefont
  {Bud'ko}, \citenamefont {Canfield}, \citenamefont {McQueeney},\ and\
  \citenamefont {Goldman}}]{KimKreyssig10}%
  \BibitemOpen
  \bibfield  {author} {\bibinfo {author} {\bibfnamefont {M.~G.}\ \bibnamefont
  {Kim}}, \bibinfo {author} {\bibfnamefont {A.}~\bibnamefont {Kreyssig}},
  \bibinfo {author} {\bibfnamefont {A.}~\bibnamefont {Thaler}}, \bibinfo
  {author} {\bibfnamefont {D.~K.}\ \bibnamefont {Pratt}}, \bibinfo {author}
  {\bibfnamefont {W.}~\bibnamefont {Tian}}, \bibinfo {author} {\bibfnamefont
  {J.~L.}\ \bibnamefont {Zarestky}}, \bibinfo {author} {\bibfnamefont {M.~A.}\
  \bibnamefont {Green}}, \bibinfo {author} {\bibfnamefont {S.~L.}\ \bibnamefont
  {Bud'ko}}, \bibinfo {author} {\bibfnamefont {P.~C.}\ \bibnamefont
  {Canfield}}, \bibinfo {author} {\bibfnamefont {R.~J.}\ \bibnamefont
  {McQueeney}}, \ and\ \bibinfo {author} {\bibfnamefont {A.~I.}\ \bibnamefont
  {Goldman}},\ }\href{http://link.aps.org/abstract/PRB/v82/p220503} {\bibfield
  {journal} {\bibinfo  {journal} {Phys. Rev.~B}\ }\textbf {\bibinfo {volume}
  {82}},\ \bibinfo {pages} {220503} (\bibinfo {year}
  {2010}{\natexlab{b}})}\BibitemShut {NoStop}%
\bibitem [{\citenamefont {Kim}\ \emph {\textit{et~al.}}(2011)\citenamefont
  {Kim}, \citenamefont {Pratt}, \citenamefont {Rustan}, \citenamefont {Tian},
  \citenamefont {Zarestky}, \citenamefont {Thaler}, \citenamefont {Bud'ko},
  \citenamefont {Canfield}, \citenamefont {McQueeney}, \citenamefont
  {Kreyssig},\ and\ \citenamefont {Goldman}}]{KimPratt11}%
  \BibitemOpen
  \bibfield  {author} {\bibinfo {author} {\bibfnamefont {M.~G.}\ \bibnamefont
  {Kim}}, \bibinfo {author} {\bibfnamefont {D.~K.}\ \bibnamefont {Pratt}},
  \bibinfo {author} {\bibfnamefont {G.~E.}\ \bibnamefont {Rustan}}, \bibinfo
  {author} {\bibfnamefont {W.}~\bibnamefont {Tian}}, \bibinfo {author}
  {\bibfnamefont {J.~L.}\ \bibnamefont {Zarestky}}, \bibinfo {author}
  {\bibfnamefont {A.}~\bibnamefont {Thaler}}, \bibinfo {author} {\bibfnamefont
  {S.~L.}\ \bibnamefont {Bud'ko}}, \bibinfo {author} {\bibfnamefont {P.~C.}\
  \bibnamefont {Canfield}}, \bibinfo {author} {\bibfnamefont {R.~J.}\
  \bibnamefont {McQueeney}}, \bibinfo {author} {\bibfnamefont {A.}~\bibnamefont
  {Kreyssig}}, \ and\ \bibinfo {author} {\bibfnamefont {A.~I.}\ \bibnamefont
  {Goldman}},\ }\href{http://link.aps.org/abstract/PRB/v83/p054514} {\bibfield
  {journal} {\bibinfo  {journal} {Phys. Rev.~B}\ }\textbf {\bibinfo {volume}
  {83}},\ \bibinfo {pages} {054514} (\bibinfo {year} {2011})}\BibitemShut
  {NoStop}%
\bibitem [{\citenamefont {Cheng}\ \emph {\textit{et~al.}}(2010)\citenamefont
  {Cheng}, \citenamefont {Shen}, \citenamefont {Hu},\ and\ \citenamefont
  {Wen}}]{ChengShen10}%
  \BibitemOpen
  \bibfield  {author} {\bibinfo {author} {\bibfnamefont {P.}~\bibnamefont
  {Cheng}}, \bibinfo {author} {\bibfnamefont {B.}~\bibnamefont {Shen}},
  \bibinfo {author} {\bibfnamefont {J.}~\bibnamefont {Hu}}, \ and\ \bibinfo
  {author} {\bibfnamefont {H.-H.}\ \bibnamefont {Wen}},\
  }\href{http://link.aps.org/doi/10.1103/PhysRevB.81.174529} {\bibfield
  {journal} {\bibinfo  {journal} {Phys. Rev.~B}\ }\textbf {\bibinfo {volume}
  {81}},\ \bibinfo {pages} {174529} (\bibinfo {year} {2010})}\BibitemShut
  {NoStop}%
\bibitem [{\citenamefont {Li}\ \emph {\textit{et~al.}}(2012)\citenamefont {Li},
  \citenamefont {Guo}, \citenamefont {Zhang}, \citenamefont {Yuan},
  \citenamefont {Tsujimoto}, \citenamefont {Wang}, \citenamefont {Sathish},
  \citenamefont {Sun}, \citenamefont {Yu}, \citenamefont {Yi}, \citenamefont
  {Yamaura}, \citenamefont {Takayama-Muromachiu}, \citenamefont {Shirako},
  \citenamefont {Akaogi},\ and\ \citenamefont {Kontani}}]{LiGuo12}%
  \BibitemOpen
  \bibfield  {author} {\bibinfo {author} {\bibfnamefont {J.}~\bibnamefont
  {Li}}, \bibinfo {author} {\bibfnamefont {Y.~F.}\ \bibnamefont {Guo}},
  \bibinfo {author} {\bibfnamefont {S.~B.}\ \bibnamefont {Zhang}}, \bibinfo
  {author} {\bibfnamefont {J.}~\bibnamefont {Yuan}}, \bibinfo {author}
  {\bibfnamefont {Y.}~\bibnamefont {Tsujimoto}}, \bibinfo {author}
  {\bibfnamefont {X.}~\bibnamefont {Wang}}, \bibinfo {author} {\bibfnamefont
  {C.~I.}\ \bibnamefont {Sathish}}, \bibinfo {author} {\bibfnamefont
  {Y.}~\bibnamefont {Sun}}, \bibinfo {author} {\bibfnamefont {S.}~\bibnamefont
  {Yu}}, \bibinfo {author} {\bibfnamefont {W.}~\bibnamefont {Yi}}, \bibinfo
  {author} {\bibfnamefont {K.}~\bibnamefont {Yamaura}}, \bibinfo {author}
  {\bibfnamefont {E.}~\bibnamefont {Takayama-Muromachiu}}, \bibinfo {author}
  {\bibfnamefont {Y.}~\bibnamefont {Shirako}}, \bibinfo {author} {\bibfnamefont
  {M.}~\bibnamefont {Akaogi}}, \ and\ \bibinfo {author} {\bibfnamefont
  {H.}~\bibnamefont {Kontani}},\
  }\href{http://link.aps.org/doi/10.1103/PhysRevB.85.214509} {\bibfield
  {journal} {\bibinfo  {journal} {Phys. Rev. B}\ }\textbf {\bibinfo {volume}
  {85}},\ \bibinfo {pages} {214509} (\bibinfo {year} {2012})}\BibitemShut
  {NoStop}%
\bibitem [{\citenamefont {Texier}\ \emph {\textit{et~al.}}(2012)\citenamefont
  {Texier}, \citenamefont {Laplace}, \citenamefont {Mendels}, \citenamefont
  {Park}, \citenamefont {Friemel}, \citenamefont {Sun}, \citenamefont {Inosov},
  \citenamefont {Lin},\ and\ \citenamefont {Bobroff}}]{TexierLaplace12}%
  \BibitemOpen
  \bibfield  {author} {\bibinfo {author} {\bibfnamefont {Y.}~\bibnamefont
  {Texier}}, \bibinfo {author} {\bibfnamefont {Y.}~\bibnamefont {Laplace}},
  \bibinfo {author} {\bibfnamefont {P.}~\bibnamefont {Mendels}}, \bibinfo
  {author} {\bibfnamefont {J.~T.}\ \bibnamefont {Park}}, \bibinfo {author}
  {\bibfnamefont {G.}~\bibnamefont {Friemel}}, \bibinfo {author} {\bibfnamefont
  {D.~L.}\ \bibnamefont {Sun}}, \bibinfo {author} {\bibfnamefont {D.~S.}\
  \bibnamefont {Inosov}}, \bibinfo {author} {\bibfnamefont {C.~T.}\
  \bibnamefont {Lin}}, \ and\ \bibinfo {author} {\bibfnamefont
  {J.}~\bibnamefont {Bobroff}},\
  }\href{http://dx.doi.org/10.1209/0295-5075/99/17002} {\bibfield  {journal}
  {\bibinfo  {journal} {EPL}\ }\textbf {\bibinfo {volume} {99}},\ \bibinfo
  {pages} {17002} (\bibinfo {year} {2012})}\BibitemShut {NoStop}%
\bibitem [{\citenamefont {Bobroff}()}]{Bobroff_private}%
  \BibitemOpen
  \bibfield  {author} {\bibinfo {author} {\bibfnamefont {J.}~\bibnamefont
  {Bobroff}},\ }\bibinfo {howpublished} {private communication}\BibitemShut
  {NoStop}%
\bibitem [{\citenamefont {Singh}\ \emph
  {\textit{et~al.}}(2009{\natexlab{a}})\citenamefont {Singh}, \citenamefont
  {Ellern},\ and\ \citenamefont {Johnston}}]{SinghEllern09}%
  \BibitemOpen
  \bibfield  {author} {\bibinfo {author} {\bibfnamefont {Y.}~\bibnamefont
  {Singh}}, \bibinfo {author} {\bibfnamefont {A.}~\bibnamefont {Ellern}}, \
  and\ \bibinfo {author} {\bibfnamefont {D.~C.}\ \bibnamefont {Johnston}},\
  }\href{http://link.aps.org/doi/10.1103/PhysRevB.79.094519} {\bibfield
  {journal} {\bibinfo  {journal} {Phys. Rev. B}\ }\textbf {\bibinfo {volume}
  {79}},\ \bibinfo {pages} {094519} (\bibinfo {year}
  {2009}{\natexlab{a}})}\BibitemShut {NoStop}%
\bibitem [{\citenamefont {Singh}\ \emph
  {\textit{et~al.}}(2009{\natexlab{b}})\citenamefont {Singh}, \citenamefont
  {Green}, \citenamefont {Huang}, \citenamefont {Kreyssig}, \citenamefont
  {McQueeney}, \citenamefont {Johnston},\ and\ \citenamefont
  {Goldman}}]{SinghGreen09}%
  \BibitemOpen
  \bibfield  {author} {\bibinfo {author} {\bibfnamefont {Y.}~\bibnamefont
  {Singh}}, \bibinfo {author} {\bibfnamefont {M.~A.}\ \bibnamefont {Green}},
  \bibinfo {author} {\bibfnamefont {Q.}~\bibnamefont {Huang}}, \bibinfo
  {author} {\bibfnamefont {A.}~\bibnamefont {Kreyssig}}, \bibinfo {author}
  {\bibfnamefont {R.~J.}\ \bibnamefont {McQueeney}}, \bibinfo {author}
  {\bibfnamefont {D.~C.}\ \bibnamefont {Johnston}}, \ and\ \bibinfo {author}
  {\bibfnamefont {A.~I.}\ \bibnamefont {Goldman}},\
  }\href{http://link.aps.org/doi/10.1103/PhysRevB.80.100403} {\bibfield
  {journal} {\bibinfo  {journal} {Phys. Rev. B}\ }\textbf {\bibinfo {volume}
  {80}},\ \bibinfo {pages} {100403(R)} (\bibinfo {year}
  {2009}{\natexlab{b}})}\BibitemShut {NoStop}%
\bibitem [{\citenamefont {Johnston}\ \emph {\textit{et~al.}}(2011)\citenamefont
  {Johnston}, \citenamefont {McQueeney}, \citenamefont {Lake}, \citenamefont
  {Honecker}, \citenamefont {Zhitomirsky}, \citenamefont {Nath}, \citenamefont
  {Furukawa}, \citenamefont {Antropov},\ and\ \citenamefont
  {Singh}}]{JohnstonMcQueeney11}%
  \BibitemOpen
  \bibfield  {author} {\bibinfo {author} {\bibfnamefont {D.~C.}\ \bibnamefont
  {Johnston}}, \bibinfo {author} {\bibfnamefont {R.~J.}\ \bibnamefont
  {McQueeney}}, \bibinfo {author} {\bibfnamefont {B.}~\bibnamefont {Lake}},
  \bibinfo {author} {\bibfnamefont {A.}~\bibnamefont {Honecker}}, \bibinfo
  {author} {\bibfnamefont {M.~E.}\ \bibnamefont {Zhitomirsky}}, \bibinfo
  {author} {\bibfnamefont {R.}~\bibnamefont {Nath}}, \bibinfo {author}
  {\bibfnamefont {Y.}~\bibnamefont {Furukawa}}, \bibinfo {author}
  {\bibfnamefont {V.~P.}\ \bibnamefont {Antropov}}, \ and\ \bibinfo {author}
  {\bibfnamefont {Y.}~\bibnamefont {Singh}},\
  }\href{http://link.aps.org/doi/10.1103/PhysRevB.84.094445} {\bibfield
  {journal} {\bibinfo  {journal} {Phys. Rev. B}\ }\textbf {\bibinfo {volume}
  {84}},\ \bibinfo {pages} {094445} (\bibinfo {year} {2011})}\BibitemShut
  {NoStop}%
\bibitem [{\citenamefont {Pandey}\ \emph {\textit{et~al.}}(2012)\citenamefont
  {Pandey}, \citenamefont {Dhaka}, \citenamefont {Lamsal}, \citenamefont {Lee},
  \citenamefont {Anand}, \citenamefont {Kreyssig}, \citenamefont {Heitmann},
  \citenamefont {McQueeney}, \citenamefont {Goldman}, \citenamefont {Harmon},
  \citenamefont {Kaminski},\ and\ \citenamefont {Johnston}}]{PandeyDhaka12}%
  \BibitemOpen
  \bibfield  {author} {\bibinfo {author} {\bibfnamefont {A.}~\bibnamefont
  {Pandey}}, \bibinfo {author} {\bibfnamefont {R.~S.}\ \bibnamefont {Dhaka}},
  \bibinfo {author} {\bibfnamefont {J.}~\bibnamefont {Lamsal}}, \bibinfo
  {author} {\bibfnamefont {Y.}~\bibnamefont {Lee}}, \bibinfo {author}
  {\bibfnamefont {V.~K.}\ \bibnamefont {Anand}}, \bibinfo {author}
  {\bibfnamefont {A.}~\bibnamefont {Kreyssig}}, \bibinfo {author}
  {\bibfnamefont {T.~W.}\ \bibnamefont {Heitmann}}, \bibinfo {author}
  {\bibfnamefont {R.~J.}\ \bibnamefont {McQueeney}}, \bibinfo {author}
  {\bibfnamefont {A.~I.}\ \bibnamefont {Goldman}}, \bibinfo {author}
  {\bibfnamefont {B.~N.}\ \bibnamefont {Harmon}}, \bibinfo {author}
  {\bibfnamefont {A.}~\bibnamefont {Kaminski}}, \ and\ \bibinfo {author}
  {\bibfnamefont {D.~C.}\ \bibnamefont {Johnston}},\
  }\href{http://link.aps.org/doi/10.1103/PhysRevLett.108.087005} {\bibfield
  {journal} {\bibinfo  {journal} {Phys. Rev. Lett.}\ }\textbf {\bibinfo
  {volume} {108}},\ \bibinfo {pages} {087005} (\bibinfo {year}
  {2012})}\BibitemShut {NoStop}%
\bibitem [{\citenamefont {Bao}\ \emph {\textit{et~al.}}(2012)\citenamefont
  {Bao}, \citenamefont {Jiang}, \citenamefont {Sun}, \citenamefont {Jiao},
  \citenamefont {Shen}, \citenamefont {Guo}, \citenamefont {Chen},
  \citenamefont {Feng}, \citenamefont {Yuan}, \citenamefont {Xu}, \citenamefont
  {Cao}, \citenamefont {Sasaki}, \citenamefont {Tanaka}, \citenamefont
  {Matsubayashi},\ and\ \citenamefont {Uwatoko}}]{BaoJiang12}%
  \BibitemOpen
  \bibfield  {author} {\bibinfo {author} {\bibfnamefont {J.-K.}\ \bibnamefont
  {Bao}}, \bibinfo {author} {\bibfnamefont {H.}~\bibnamefont {Jiang}}, \bibinfo
  {author} {\bibfnamefont {Y.-L.}\ \bibnamefont {Sun}}, \bibinfo {author}
  {\bibfnamefont {W.-H.}\ \bibnamefont {Jiao}}, \bibinfo {author}
  {\bibfnamefont {C.-Y.}\ \bibnamefont {Shen}}, \bibinfo {author}
  {\bibfnamefont {H.-J.}\ \bibnamefont {Guo}}, \bibinfo {author} {\bibfnamefont
  {Y.}~\bibnamefont {Chen}}, \bibinfo {author} {\bibfnamefont {C.-M.}\
  \bibnamefont {Feng}}, \bibinfo {author} {\bibfnamefont {H.-Q.}\ \bibnamefont
  {Yuan}}, \bibinfo {author} {\bibfnamefont {Z.-A.}\ \bibnamefont {Xu}},
  \bibinfo {author} {\bibfnamefont {G.-H.}\ \bibnamefont {Cao}}, \bibinfo
  {author} {\bibfnamefont {R.}~\bibnamefont {Sasaki}}, \bibinfo {author}
  {\bibfnamefont {T.}~\bibnamefont {Tanaka}}, \bibinfo {author} {\bibfnamefont
  {K.}~\bibnamefont {Matsubayashi}}, \ and\ \bibinfo {author} {\bibfnamefont
  {Y.}~\bibnamefont {Uwatoko}},\
  }\href{http://link.aps.org/doi/10.1103/PhysRevB.85.144523} {\bibfield
  {journal} {\bibinfo  {journal} {Phys. Rev.~B}\ }\textbf {\bibinfo {volume}
  {85}},\ \bibinfo {pages} {144523} (\bibinfo {year} {2012})}\BibitemShut
  {NoStop}%
\bibitem [{\citenamefont {Lamsal}\ \emph {\textit{et~al.}}(2013)\citenamefont
  {Lamsal}, \citenamefont {Tucker}, \citenamefont {Heitmann}, \citenamefont
  {Kreyssig}, \citenamefont {Jesche}, \citenamefont {Pandey}, \citenamefont
  {Tian}, \citenamefont {McQueeney}, \citenamefont {Johnston},\ and\
  \citenamefont {Goldman}}]{LamsalTucker13}%
  \BibitemOpen
  \bibfield  {author} {\bibinfo {author} {\bibfnamefont {J.}~\bibnamefont
  {Lamsal}}, \bibinfo {author} {\bibfnamefont {G.~S.}\ \bibnamefont {Tucker}},
  \bibinfo {author} {\bibfnamefont {T.~W.}\ \bibnamefont {Heitmann}}, \bibinfo
  {author} {\bibfnamefont {A.}~\bibnamefont {Kreyssig}}, \bibinfo {author}
  {\bibfnamefont {A.}~\bibnamefont {Jesche}}, \bibinfo {author} {\bibfnamefont
  {A.}~\bibnamefont {Pandey}}, \bibinfo {author} {\bibfnamefont
  {W.}~\bibnamefont {Tian}}, \bibinfo {author} {\bibfnamefont {R.~J.}\
  \bibnamefont {McQueeney}}, \bibinfo {author} {\bibfnamefont {D.~C.}\
  \bibnamefont {Johnston}}, \ and\ \bibinfo {author} {\bibfnamefont {A.~I.}\
  \bibnamefont {Goldman}},\
  }\href{http://link.aps.org/doi/10.1103/PhysRevB.87.144418} {\bibfield
  {journal} {\bibinfo  {journal} {Phys. Rev.~B}\ }\textbf {\bibinfo {volume}
  {87}},\ \bibinfo {pages} {144418} (\bibinfo {year} {2013})}\BibitemShut
  {NoStop}%
\bibitem [{\citenamefont {Kasinathan}\ \emph
  {\textit{et~al.}}(2009)\citenamefont {Kasinathan}, \citenamefont {Ormeci},
  \citenamefont {Koch}, \citenamefont {Burkhardt}, \citenamefont {Schnelle},
  \citenamefont {Leithe-Jasper},\ and\ \citenamefont
  {Rosner}}]{KasinathanOrmeci09}%
  \BibitemOpen
  \bibfield  {author} {\bibinfo {author} {\bibfnamefont {D.}~\bibnamefont
  {Kasinathan}}, \bibinfo {author} {\bibfnamefont {A.}~\bibnamefont {Ormeci}},
  \bibinfo {author} {\bibfnamefont {K.}~\bibnamefont {Koch}}, \bibinfo {author}
  {\bibfnamefont {U.}~\bibnamefont {Burkhardt}}, \bibinfo {author}
  {\bibfnamefont {W.}~\bibnamefont {Schnelle}}, \bibinfo {author}
  {\bibfnamefont {A.}~\bibnamefont {Leithe-Jasper}}, \ and\ \bibinfo {author}
  {\bibfnamefont {H.}~\bibnamefont {Rosner}},\
  }\href{http://iopscience.iop.org/1367-2630/11/2/025023} {\bibfield  {journal}
  {\bibinfo  {journal} {New J. Phys.}\ }\textbf {\bibinfo {volume} {11}},\
  \bibinfo {pages} {025023} (\bibinfo {year} {2009})}\BibitemShut {NoStop}%
\bibitem [{\citenamefont {Singh}\ \emph
  {\textit{et~al.}}(2009{\natexlab{c}})\citenamefont {Singh}, \citenamefont
  {Sefat}, \citenamefont {McGuire}, \citenamefont {Sales}, \citenamefont
  {Mandrus}, \citenamefont {VanBebber},\ and\ \citenamefont
  {Keppens}}]{SinghSefat09}%
  \BibitemOpen
  \bibfield  {author} {\bibinfo {author} {\bibfnamefont {D.~J.}\ \bibnamefont
  {Singh}}, \bibinfo {author} {\bibfnamefont {A.~S.}\ \bibnamefont {Sefat}},
  \bibinfo {author} {\bibfnamefont {M.~A.}\ \bibnamefont {McGuire}}, \bibinfo
  {author} {\bibfnamefont {B.~C.}\ \bibnamefont {Sales}}, \bibinfo {author}
  {\bibfnamefont {D.}~\bibnamefont {Mandrus}}, \bibinfo {author} {\bibfnamefont
  {L.~H.}\ \bibnamefont {VanBebber}}, \ and\ \bibinfo {author} {\bibfnamefont
  {V.}~\bibnamefont {Keppens}},\
  }\href{http://link.aps.org/doi/10.1103/PhysRevB.79.094429} {\bibfield
  {journal} {\bibinfo  {journal} {Phys. Rev. B}\ }\textbf {\bibinfo {volume}
  {79}},\ \bibinfo {pages} {094429} (\bibinfo {year}
  {2009}{\natexlab{c}})}\BibitemShut {NoStop}%
\bibitem [{\citenamefont {Tucker}\ \emph {\textit{et~al.}}(2012)\citenamefont
  {Tucker}, \citenamefont {Pratt}, \citenamefont {Kim}, \citenamefont {Ran},
  \citenamefont {Thaler}, \citenamefont {Granroth}, \citenamefont {Marty},
  \citenamefont {Tian}, \citenamefont {Zarestky}, \citenamefont {Lumsden},
  \citenamefont {Bud'ko}, \citenamefont {Canfield}, \citenamefont {Kreyssig},
  \citenamefont {Goldman},\ and\ \citenamefont {McQueeney}}]{TuckerPratt12}%
  \BibitemOpen
  \bibfield  {author} {\bibinfo {author} {\bibfnamefont {G.~S.}\ \bibnamefont
  {Tucker}}, \bibinfo {author} {\bibfnamefont {D.~K.}\ \bibnamefont {Pratt}},
  \bibinfo {author} {\bibfnamefont {M.~G.}\ \bibnamefont {Kim}}, \bibinfo
  {author} {\bibfnamefont {S.}~\bibnamefont {Ran}}, \bibinfo {author}
  {\bibfnamefont {A.}~\bibnamefont {Thaler}}, \bibinfo {author} {\bibfnamefont
  {G.~E.}\ \bibnamefont {Granroth}}, \bibinfo {author} {\bibfnamefont
  {K.}~\bibnamefont {Marty}}, \bibinfo {author} {\bibfnamefont
  {W.}~\bibnamefont {Tian}}, \bibinfo {author} {\bibfnamefont {J.~L.}\
  \bibnamefont {Zarestky}}, \bibinfo {author} {\bibfnamefont {M.~D.}\
  \bibnamefont {Lumsden}}, \bibinfo {author} {\bibfnamefont {S.~L.}\
  \bibnamefont {Bud'ko}}, \bibinfo {author} {\bibfnamefont {P.~C.}\
  \bibnamefont {Canfield}}, \bibinfo {author} {\bibfnamefont {A.}~\bibnamefont
  {Kreyssig}}, \bibinfo {author} {\bibfnamefont {A.~I.}\ \bibnamefont
  {Goldman}}, \ and\ \bibinfo {author} {\bibfnamefont {R.~J.}\ \bibnamefont
  {McQueeney}},\ }\href{http://link.aps.org/doi/10.1103/PhysRevB.86.020503}
  {\bibfield  {journal} {\bibinfo  {journal} {Phys. Rev.~B}\ }\textbf {\bibinfo
  {volume} {86}},\ \bibinfo {pages} {020503} (\bibinfo {year}
  {2012})}\BibitemShut {NoStop}%
\bibitem [{\citenamefont {Park}\ \emph {\textit{et~al.}}(2012)\citenamefont
  {Park}, \citenamefont {Friemel}, \citenamefont {Loew}, \citenamefont
  {Hinkov}, \citenamefont {Li}, \citenamefont {Min}, \citenamefont {Sun},
  \citenamefont {Ivanov}, \citenamefont {Piovano}, \citenamefont {Lin},
  \citenamefont {Keimer}, \citenamefont {Kwon},\ and\ \citenamefont
  {Inosov}}]{ParkFriemel12}%
  \BibitemOpen
  \bibfield  {author} {\bibinfo {author} {\bibfnamefont {J.~T.}\ \bibnamefont
  {Park}}, \bibinfo {author} {\bibfnamefont {G.}~\bibnamefont {Friemel}},
  \bibinfo {author} {\bibfnamefont {T.}~\bibnamefont {Loew}}, \bibinfo {author}
  {\bibfnamefont {V.}~\bibnamefont {Hinkov}}, \bibinfo {author} {\bibfnamefont
  {Y.}~\bibnamefont {Li}}, \bibinfo {author} {\bibfnamefont {B.~H.}\
  \bibnamefont {Min}}, \bibinfo {author} {\bibfnamefont {D.~L.}\ \bibnamefont
  {Sun}}, \bibinfo {author} {\bibfnamefont {A.}~\bibnamefont {Ivanov}},
  \bibinfo {author} {\bibfnamefont {A.}~\bibnamefont {Piovano}}, \bibinfo
  {author} {\bibfnamefont {C.~T.}\ \bibnamefont {Lin}}, \bibinfo {author}
  {\bibfnamefont {B.}~\bibnamefont {Keimer}}, \bibinfo {author} {\bibfnamefont
  {Y.~S.}\ \bibnamefont {Kwon}}, \ and\ \bibinfo {author} {\bibfnamefont
  {D.~S.}\ \bibnamefont {Inosov}},\
  }\href{http://link.aps.org/doi/10.1103/PhysRevB.86.024437} {\bibfield
  {journal} {\bibinfo  {journal} {Phys. Rev.~B}\ }\textbf {\bibinfo {volume}
  {86}},\ \bibinfo {pages} {024437} (\bibinfo {year} {2012})}\BibitemShut
  {NoStop}%
\bibitem [{\citenamefont {Liu}\ \emph {\textit{et~al.}}(2010)\citenamefont
  {Liu}, \citenamefont {Sun}, \citenamefont {Park},\ and\ \citenamefont
  {Lin}}]{LiuSun10}%
  \BibitemOpen
  \bibfield  {author} {\bibinfo {author} {\bibfnamefont {Y.}~\bibnamefont
  {Liu}}, \bibinfo {author} {\bibfnamefont {D.~L.}\ \bibnamefont {Sun}},
  \bibinfo {author} {\bibfnamefont {J.~T.}\ \bibnamefont {Park}}, \ and\
  \bibinfo {author} {\bibfnamefont {C.~T.}\ \bibnamefont {Lin}},\
  }\href{http://www.sciencedirect.com/science/article/pii/S0921453409007369}
  {\bibfield  {journal} {\bibinfo  {journal} {Physica~C: Superconductivity}\
  }\textbf {\bibinfo {volume} {470}},\ \bibinfo {pages} {S513} (\bibinfo {year}
  {2010})}\BibitemShut {NoStop}%
\bibitem [{\citenamefont {Rotter}\ \emph {\textit{et~al.}}(2008)\citenamefont
  {Rotter}, \citenamefont {Tegel}, \citenamefont {Johrendt}, \citenamefont
  {Schellenberg}, \citenamefont {Hermes},\ and\ \citenamefont
  {P\"ottgen}}]{RotterTegel08PRB}%
  \BibitemOpen
  \bibfield  {author} {\bibinfo {author} {\bibfnamefont {M.}~\bibnamefont
  {Rotter}}, \bibinfo {author} {\bibfnamefont {M.}~\bibnamefont {Tegel}},
  \bibinfo {author} {\bibfnamefont {D.}~\bibnamefont {Johrendt}}, \bibinfo
  {author} {\bibfnamefont {I.}~\bibnamefont {Schellenberg}}, \bibinfo {author}
  {\bibfnamefont {W.}~\bibnamefont {Hermes}}, \ and\ \bibinfo {author}
  {\bibfnamefont {R.}~\bibnamefont {P\"ottgen}},\
  }\href{http://link.aps.org/abstract/PRB/v78/e020503} {\bibfield  {journal}
  {\bibinfo  {journal} {Phys. Rev.~B}\ }\textbf {\bibinfo {volume} {78}},\
  \bibinfo {pages} {020503(R)} (\bibinfo {year} {2008})}\BibitemShut {NoStop}%
\bibitem [{\citenamefont {Popovich}\ \emph {\textit{et~al.}}(2010)\citenamefont
  {Popovich}, \citenamefont {Boris}, \citenamefont {Dolgov}, \citenamefont
  {Golubov}, \citenamefont {Sun}, \citenamefont {Lin}, \citenamefont {Kremer},\
  and\ \citenamefont {Keimer}}]{PopovichBoris10}%
  \BibitemOpen
  \bibfield  {author} {\bibinfo {author} {\bibfnamefont {P.}~\bibnamefont
  {Popovich}}, \bibinfo {author} {\bibfnamefont {A.~V.}\ \bibnamefont {Boris}},
  \bibinfo {author} {\bibfnamefont {O.~V.}\ \bibnamefont {Dolgov}}, \bibinfo
  {author} {\bibfnamefont {A.~A.}\ \bibnamefont {Golubov}}, \bibinfo {author}
  {\bibfnamefont {D.~L.}\ \bibnamefont {Sun}}, \bibinfo {author} {\bibfnamefont
  {C.~T.}\ \bibnamefont {Lin}}, \bibinfo {author} {\bibfnamefont {R.~K.}\
  \bibnamefont {Kremer}}, \ and\ \bibinfo {author} {\bibfnamefont
  {B.}~\bibnamefont {Keimer}},\
  }\href{http://link.aps.org/abstract/PRL/v105/e027003} {\bibfield  {journal}
  {\bibinfo  {journal} {Phys. Rev. Lett.}\ }\textbf {\bibinfo {volume} {105}},\
  \bibinfo {pages} {027003} (\bibinfo {year} {2010})}\BibitemShut {NoStop}%
\bibitem [{\citenamefont {Park}\ \emph {\textit{et~al.}}(2010)\citenamefont
  {Park}, \citenamefont {Inosov}, \citenamefont {Yaresko}, \citenamefont
  {Graser}, \citenamefont {Sun}, \citenamefont {Bourges}, \citenamefont
  {Sidis}, \citenamefont {Li}, \citenamefont {Kim}, \citenamefont {Haug},
  \citenamefont {Ivanov}, \citenamefont {Hradil}, \citenamefont {Schneidewind},
  \citenamefont {Link}, \citenamefont {Faulhaber}, \citenamefont {Glavatskyy},
  \citenamefont {Lin}, \citenamefont {Keimer},\ and\ \citenamefont
  {Hinkov}}]{ParkInosov10}%
  \BibitemOpen
  \bibfield  {author} {\bibinfo {author} {\bibfnamefont {J.~T.}\ \bibnamefont
  {Park}}, \bibinfo {author} {\bibfnamefont {D.~S.}\ \bibnamefont {Inosov}},
  \bibinfo {author} {\bibfnamefont {A.}~\bibnamefont {Yaresko}}, \bibinfo
  {author} {\bibfnamefont {S.}~\bibnamefont {Graser}}, \bibinfo {author}
  {\bibfnamefont {D.~L.}\ \bibnamefont {Sun}}, \bibinfo {author} {\bibfnamefont
  {P.}~\bibnamefont {Bourges}}, \bibinfo {author} {\bibfnamefont
  {Y.}~\bibnamefont {Sidis}}, \bibinfo {author} {\bibfnamefont
  {Y.}~\bibnamefont {Li}}, \bibinfo {author} {\bibfnamefont {J.-H.}\
  \bibnamefont {Kim}}, \bibinfo {author} {\bibfnamefont {D.}~\bibnamefont
  {Haug}}, \bibinfo {author} {\bibfnamefont {A.}~\bibnamefont {Ivanov}},
  \bibinfo {author} {\bibfnamefont {K.}~\bibnamefont {Hradil}}, \bibinfo
  {author} {\bibfnamefont {A.}~\bibnamefont {Schneidewind}}, \bibinfo {author}
  {\bibfnamefont {P.}~\bibnamefont {Link}}, \bibinfo {author} {\bibfnamefont
  {E.}~\bibnamefont {Faulhaber}}, \bibinfo {author} {\bibfnamefont
  {I.}~\bibnamefont {Glavatskyy}}, \bibinfo {author} {\bibfnamefont {C.~T.}\
  \bibnamefont {Lin}}, \bibinfo {author} {\bibfnamefont {B.}~\bibnamefont
  {Keimer}}, \ and\ \bibinfo {author} {\bibfnamefont {V.}~\bibnamefont
  {Hinkov}},\ }\href{http://link.aps.org/abstract/PRB/v82/e134503} {\bibfield
  {journal} {\bibinfo  {journal} {Phys. Rev.~B}\ }\textbf {\bibinfo {volume}
  {82}},\ \bibinfo {pages} {134503} (\bibinfo {year} {2010})}\BibitemShut
  {NoStop}%
\bibitem [{\citenamefont {{C. de la Cruz}}\ \emph
  {\textit{et~al.}}(2008)\citenamefont {{C. de la Cruz}}, \citenamefont
  {Huang}, \citenamefont {Lynn}, \citenamefont {Li}, \citenamefont {{Ratcliff
  II}}, \citenamefont {Zarestky}, \citenamefont {Mook}, \citenamefont {Chen},
  \citenamefont {Luo}, \citenamefont {Wang},\ and\ \citenamefont
  {Dai}}]{CruzHuang08}%
  \BibitemOpen
  \bibfield  {author} {\bibinfo {author} {\bibnamefont {{C. de la Cruz}}},
  \bibinfo {author} {\bibfnamefont {Q.}~\bibnamefont {Huang}}, \bibinfo
  {author} {\bibfnamefont {J.~W.}\ \bibnamefont {Lynn}}, \bibinfo {author}
  {\bibfnamefont {J.}~\bibnamefont {Li}}, \bibinfo {author} {\bibfnamefont
  {W.}~\bibnamefont {{Ratcliff II}}}, \bibinfo {author} {\bibfnamefont {J.~L.}\
  \bibnamefont {Zarestky}}, \bibinfo {author} {\bibfnamefont {H.~A.}\
  \bibnamefont {Mook}}, \bibinfo {author} {\bibfnamefont {G.~F.}\ \bibnamefont
  {Chen}}, \bibinfo {author} {\bibfnamefont {J.~L.}\ \bibnamefont {Luo}},
  \bibinfo {author} {\bibfnamefont {N.~L.}\ \bibnamefont {Wang}}, \ and\
  \bibinfo {author} {\bibfnamefont {P.}~\bibnamefont {Dai}},\
  }\href{http://www.nature.com/nature/journal/v453/n7197/abs/nature07057.html}
  {\bibfield  {journal} {\bibinfo  {journal} {Nature {\rm (London)}}\ }\textbf
  {\bibinfo {volume} {453}},\ \bibinfo {pages} {899} (\bibinfo {year}
  {2008})}\BibitemShut {NoStop}%
\bibitem [{\citenamefont {Huang}\ \emph {\textit{et~al.}}(2008)\citenamefont
  {Huang}, \citenamefont {Qiu}, \citenamefont {Bao}, \citenamefont {Green},
  \citenamefont {Lynn}, \citenamefont {Gasparovic}, \citenamefont {Wu},
  \citenamefont {Wu},\ and\ \citenamefont {Chen}}]{HuangQiu08}%
  \BibitemOpen
  \bibfield  {author} {\bibinfo {author} {\bibfnamefont {Q.}~\bibnamefont
  {Huang}}, \bibinfo {author} {\bibfnamefont {Y.}~\bibnamefont {Qiu}}, \bibinfo
  {author} {\bibfnamefont {W.}~\bibnamefont {Bao}}, \bibinfo {author}
  {\bibfnamefont {M.~A.}\ \bibnamefont {Green}}, \bibinfo {author}
  {\bibfnamefont {J.~W.}\ \bibnamefont {Lynn}}, \bibinfo {author}
  {\bibfnamefont {Y.~C.}\ \bibnamefont {Gasparovic}}, \bibinfo {author}
  {\bibfnamefont {T.}~\bibnamefont {Wu}}, \bibinfo {author} {\bibfnamefont
  {G.}~\bibnamefont {Wu}}, \ and\ \bibinfo {author} {\bibfnamefont {X.~H.}\
  \bibnamefont {Chen}},\
  }\href{http://link.aps.org/doi/10.1103/PhysRevLett.101.257003} {\bibfield
  {journal} {\bibinfo  {journal} {Phys. Rev. Lett.}\ }\textbf {\bibinfo
  {volume} {101}},\ \bibinfo {pages} {257003} (\bibinfo {year}
  {2008})}\BibitemShut {NoStop}%
\bibitem [{\citenamefont {Zhao}\ \emph {\textit{et~al.}}(2008)\citenamefont
  {Zhao}, \citenamefont {Ratcliff}, \citenamefont {Lynn}, \citenamefont {Chen},
  \citenamefont {Luo}, \citenamefont {Wang}, \citenamefont {Hu},\ and\
  \citenamefont {Dai}}]{ZhaoRatcliff08}%
  \BibitemOpen
  \bibfield  {author} {\bibinfo {author} {\bibfnamefont {J.}~\bibnamefont
  {Zhao}}, \bibinfo {author} {\bibfnamefont {W.}~\bibnamefont {Ratcliff}},
  \bibinfo {author} {\bibfnamefont {J.~W.}\ \bibnamefont {Lynn}}, \bibinfo
  {author} {\bibfnamefont {G.~F.}\ \bibnamefont {Chen}}, \bibinfo {author}
  {\bibfnamefont {J.~L.}\ \bibnamefont {Luo}}, \bibinfo {author} {\bibfnamefont
  {N.~L.}\ \bibnamefont {Wang}}, \bibinfo {author} {\bibfnamefont
  {J.}~\bibnamefont {Hu}}, \ and\ \bibinfo {author} {\bibfnamefont
  {P.}~\bibnamefont {Dai}},\
  }\href{http://link.aps.org/doi/10.1103/PhysRevB.78.140504} {\bibfield
  {journal} {\bibinfo  {journal} {Phys. Rev.~B}\ }\textbf {\bibinfo {volume}
  {78}},\ \bibinfo {pages} {140504} (\bibinfo {year} {2008})}\BibitemShut
  {NoStop}%
\bibitem [{\citenamefont {Kaneko}\ \emph {\textit{et~al.}}(2008)\citenamefont
  {Kaneko}, \citenamefont {Hoser}, \citenamefont {Caroca-Canales},
  \citenamefont {Jesche}, \citenamefont {Krellner}, \citenamefont {Stockert},\
  and\ \citenamefont {Geibel}}]{KanekoHoser08}%
  \BibitemOpen
  \bibfield  {author} {\bibinfo {author} {\bibfnamefont {K.}~\bibnamefont
  {Kaneko}}, \bibinfo {author} {\bibfnamefont {A.}~\bibnamefont {Hoser}},
  \bibinfo {author} {\bibfnamefont {N.}~\bibnamefont {Caroca-Canales}},
  \bibinfo {author} {\bibfnamefont {A.}~\bibnamefont {Jesche}}, \bibinfo
  {author} {\bibfnamefont {C.}~\bibnamefont {Krellner}}, \bibinfo {author}
  {\bibfnamefont {O.}~\bibnamefont {Stockert}}, \ and\ \bibinfo {author}
  {\bibfnamefont {C.}~\bibnamefont {Geibel}},\
  }\href{http://link.aps.org/doi/10.1103/PhysRevB.78.212502} {\bibfield
  {journal} {\bibinfo  {journal} {Phys. Rev.~B}\ }\textbf {\bibinfo {volume}
  {78}},\ \bibinfo {pages} {212502} (\bibinfo {year} {2008})}\BibitemShut
  {NoStop}%
\bibitem [{\citenamefont {Goldman}\ \emph {\textit{et~al.}}(2008)\citenamefont
  {Goldman}, \citenamefont {Argyriou}, \citenamefont {Ouladdiaf}, \citenamefont
  {Chatterji}, \citenamefont {Kreyssig}, \citenamefont {Nandi}, \citenamefont
  {Ni}, \citenamefont {Bud'ko}, \citenamefont {Canfield},\ and\ \citenamefont
  {McQueeney}}]{GoldmanArgyriou08}%
  \BibitemOpen
  \bibfield  {author} {\bibinfo {author} {\bibfnamefont {A.~I.}\ \bibnamefont
  {Goldman}}, \bibinfo {author} {\bibfnamefont {D.~N.}\ \bibnamefont
  {Argyriou}}, \bibinfo {author} {\bibfnamefont {B.}~\bibnamefont {Ouladdiaf}},
  \bibinfo {author} {\bibfnamefont {T.}~\bibnamefont {Chatterji}}, \bibinfo
  {author} {\bibfnamefont {A.}~\bibnamefont {Kreyssig}}, \bibinfo {author}
  {\bibfnamefont {S.}~\bibnamefont {Nandi}}, \bibinfo {author} {\bibfnamefont
  {N.}~\bibnamefont {Ni}}, \bibinfo {author} {\bibfnamefont {S.~L.}\
  \bibnamefont {Bud'ko}}, \bibinfo {author} {\bibfnamefont {P.~C.}\
  \bibnamefont {Canfield}}, \ and\ \bibinfo {author} {\bibfnamefont {R.~J.}\
  \bibnamefont {McQueeney}},\
  }\href{http://link.aps.org/abstract/PRB/v78/e100506} {\bibfield  {journal}
  {\bibinfo  {journal} {Phys. Rev.~B}\ }\textbf {\bibinfo {volume} {78}},\
  \bibinfo {eid} {100506} (\bibinfo {year} {2008})}\BibitemShut {NoStop}%
\bibitem [{\citenamefont {Olariu}\ \emph {\textit{et~al.}}(2012)\citenamefont
  {Olariu}, \citenamefont {Bonville}, \citenamefont {Rullier-Albenque},
  \citenamefont {Colson},\ and\ \citenamefont {Forget}}]{OlariuBonville12}%
  \BibitemOpen
  \bibfield  {author} {\bibinfo {author} {\bibfnamefont {A.}~\bibnamefont
  {Olariu}}, \bibinfo {author} {\bibfnamefont {P.}~\bibnamefont {Bonville}},
  \bibinfo {author} {\bibfnamefont {F.}~\bibnamefont {Rullier-Albenque}},
  \bibinfo {author} {\bibfnamefont {D.}~\bibnamefont {Colson}}, \ and\ \bibinfo
  {author} {\bibfnamefont {A.}~\bibnamefont {Forget}},\
  }\href{http://iopscience.iop.org/1367-2630/14/5/053044/} {\bibfield
  {journal} {\bibinfo  {journal} {New J. Phys.}\ }\textbf {\bibinfo {volume}
  {14}},\ \bibinfo {pages} {053044} (\bibinfo {year} {2012})}\BibitemShut
  {NoStop}%
\bibitem [{\citenamefont {Eremin}\ and\ \citenamefont
  {Chubukov}(2010)}]{EreminChubukov10}%
  \BibitemOpen
  \bibfield  {author} {\bibinfo {author} {\bibfnamefont {I.}~\bibnamefont
  {Eremin}}\ and\ \bibinfo {author} {\bibfnamefont {A.~V.}\ \bibnamefont
  {Chubukov}},\ }\href{http://link.aps.org/abstract/PRB/v81/e024511} {\bibfield
   {journal} {\bibinfo  {journal} {Phys. Rev.~B}\ }\textbf {\bibinfo {volume}
  {81}},\ \bibinfo {pages} {024511} (\bibinfo {year} {2010})}\BibitemShut
  {NoStop}%
\bibitem [{\citenamefont {Blomberg}\ \emph {\textit{et~al.}}(2012)\citenamefont
  {Blomberg}, \citenamefont {Kreyssig}, \citenamefont {Tanatar}, \citenamefont
  {Fernandes}, \citenamefont {Kim}, \citenamefont {Thaler}, \citenamefont
  {Schmalian}, \citenamefont {Bud'ko}, \citenamefont {Canfield}, \citenamefont
  {Goldman},\ and\ \citenamefont {Prozorov}}]{BlombergKreyssig12}%
  \BibitemOpen
  \bibfield  {author} {\bibinfo {author} {\bibfnamefont {E.~C.}\ \bibnamefont
  {Blomberg}}, \bibinfo {author} {\bibfnamefont {A.}~\bibnamefont {Kreyssig}},
  \bibinfo {author} {\bibfnamefont {M.~A.}\ \bibnamefont {Tanatar}}, \bibinfo
  {author} {\bibfnamefont {R.~M.}\ \bibnamefont {Fernandes}}, \bibinfo {author}
  {\bibfnamefont {M.~G.}\ \bibnamefont {Kim}}, \bibinfo {author} {\bibfnamefont
  {A.}~\bibnamefont {Thaler}}, \bibinfo {author} {\bibfnamefont
  {J.}~\bibnamefont {Schmalian}}, \bibinfo {author} {\bibfnamefont {S.~L.}\
  \bibnamefont {Bud'ko}}, \bibinfo {author} {\bibfnamefont {P.~C.}\
  \bibnamefont {Canfield}}, \bibinfo {author} {\bibfnamefont {A.~I.}\
  \bibnamefont {Goldman}}, \ and\ \bibinfo {author} {\bibfnamefont
  {R.}~\bibnamefont {Prozorov}},\
  }\href{http://link.aps.org/doi/10.1103/PhysRevB.85.144509} {\bibfield
  {journal} {\bibinfo  {journal} {Phys. Rev.~B}\ }\textbf {\bibinfo {volume}
  {85}},\ \bibinfo {pages} {144509} (\bibinfo {year} {2012})}\BibitemShut
  {NoStop}%
\bibitem [{\citenamefont {Rekveldt}(2000)}]{Rekveldt00}%
  \BibitemOpen
  \bibfield  {author} {\bibinfo {author} {\bibfnamefont {M.~T.}\ \bibnamefont
  {Rekveldt}},\ }\href{http://www.scientific.net/MSF.321-324.258} {\bibfield
  {journal} {\bibinfo  {journal} {Mater. Sci. Forum}\ }\textbf {\bibinfo
  {volume} {321--324}},\ \bibinfo {pages} {258} (\bibinfo {year}
  {2000})}\BibitemShut {NoStop}%
\bibitem [{\citenamefont {Rekveldt}\ \emph {\textit{et~al.}}(2001)\citenamefont
  {Rekveldt}, \citenamefont {Keller},\ and\ \citenamefont
  {Golub}}]{RekveldtKeller01}%
  \BibitemOpen
  \bibfield  {author} {\bibinfo {author} {\bibfnamefont {M.~T.}\ \bibnamefont
  {Rekveldt}}, \bibinfo {author} {\bibfnamefont {T.}~\bibnamefont {Keller}}, \
  and\ \bibinfo {author} {\bibfnamefont {R.}~\bibnamefont {Golub}},\
  }\href{http://iopscience.iop.org/0295-5075/54/3/342/} {\bibfield  {journal}
  {\bibinfo  {journal} {Europhys. Lett.}\ }\textbf {\bibinfo {volume} {54}},\
  \bibinfo {pages} {342} (\bibinfo {year} {2001})}\BibitemShut {NoStop}%
\bibitem [{\citenamefont {Rekveldt}\ \emph {\textit{et~al.}}(2002)\citenamefont
  {Rekveldt}, \citenamefont {Kraan},\ and\ \citenamefont
  {Keller}}]{RekveldtKraan02}%
  \BibitemOpen
  \bibfield  {author} {\bibinfo {author} {\bibfnamefont {M.~T.}\ \bibnamefont
  {Rekveldt}}, \bibinfo {author} {\bibfnamefont {W.}~\bibnamefont {Kraan}}, \
  and\ \bibinfo {author} {\bibfnamefont {T.}~\bibnamefont {Keller}},\
  }\href{http://scripts.iucr.org/cgi-bin/paper?S0021889801017812} {\bibfield
  {journal} {\bibinfo  {journal} {J. Appl. Cryst.}\ }\textbf {\bibinfo {volume}
  {35}},\ \bibinfo {pages} {28} (\bibinfo {year} {2002})}\BibitemShut {NoStop}%
\bibitem [{\citenamefont {Keller}\ \emph {\textit{et~al.}}(2002)\citenamefont
  {Keller}, \citenamefont {Rekveldt},\ and\ \citenamefont
  {Habicht}}]{KellerRekveldt02}%
  \BibitemOpen
  \bibfield  {author} {\bibinfo {author} {\bibfnamefont {T.}~\bibnamefont
  {Keller}}, \bibinfo {author} {\bibfnamefont {M.}~\bibnamefont {Rekveldt}}, \
  and\ \bibinfo {author} {\bibfnamefont {K.}~\bibnamefont {Habicht}},\
  }\href{http://www.springerlink.com/content/952xnm2b6j4cxl9b/} {\bibfield
  {journal} {\bibinfo  {journal} {Appl. Phys.~A: Mater. Sci. Process.}\
  }\textbf {\bibinfo {volume} {74}},\ \bibinfo {pages} {S127} (\bibinfo {year}
  {2002})}\BibitemShut {NoStop}%
\bibitem [{\citenamefont {Inosov}\ \emph
  {\textit{et~al.}}(2009{\natexlab{b}})\citenamefont {Inosov}, \citenamefont
  {Leineweber}, \citenamefont {Yang}, \citenamefont {Park}, \citenamefont
  {Christensen}, \citenamefont {Dinnebier}, \citenamefont {Sun}, \citenamefont
  {Niedermayer}, \citenamefont {Haug}, \citenamefont {Stephens}, \citenamefont
  {Stahn}, \citenamefont {Khvostikova}, \citenamefont {Lin}, \citenamefont
  {Andersen}, \citenamefont {Keimer},\ and\ \citenamefont
  {Hinkov}}]{InosovLeineweber09}%
  \BibitemOpen
  \bibfield  {author} {\bibinfo {author} {\bibfnamefont {D.~S.}\ \bibnamefont
  {Inosov}}, \bibinfo {author} {\bibfnamefont {A.}~\bibnamefont {Leineweber}},
  \bibinfo {author} {\bibfnamefont {X.}~\bibnamefont {Yang}}, \bibinfo {author}
  {\bibfnamefont {J.~T.}\ \bibnamefont {Park}}, \bibinfo {author}
  {\bibfnamefont {N.~B.}\ \bibnamefont {Christensen}}, \bibinfo {author}
  {\bibfnamefont {R.}~\bibnamefont {Dinnebier}}, \bibinfo {author}
  {\bibfnamefont {G.~L.}\ \bibnamefont {Sun}}, \bibinfo {author} {\bibfnamefont
  {C.}~\bibnamefont {Niedermayer}}, \bibinfo {author} {\bibfnamefont
  {D.}~\bibnamefont {Haug}}, \bibinfo {author} {\bibfnamefont {P.~W.}\
  \bibnamefont {Stephens}}, \bibinfo {author} {\bibfnamefont {J.}~\bibnamefont
  {Stahn}}, \bibinfo {author} {\bibfnamefont {O.}~\bibnamefont {Khvostikova}},
  \bibinfo {author} {\bibfnamefont {C.~T.}\ \bibnamefont {Lin}}, \bibinfo
  {author} {\bibfnamefont {O.~K.}\ \bibnamefont {Andersen}}, \bibinfo {author}
  {\bibfnamefont {B.}~\bibnamefont {Keimer}}, \ and\ \bibinfo {author}
  {\bibfnamefont {V.}~\bibnamefont {Hinkov}},\
  }\href{http://link.aps.org/abstract/PRB/v79/e224503} {\bibfield  {journal}
  {\bibinfo  {journal} {Phys. Rev.~B}\ }\textbf {\bibinfo {volume} {79}},\
  \bibinfo {eid} {224503} (\bibinfo {year} {2009}{\natexlab{b}})}\BibitemShut
  {NoStop}%
\bibitem [{\citenamefont {Keller}\ \emph {\textit{et~al.}}(2003)\citenamefont
  {Keller}, \citenamefont {Keimer}, \citenamefont {Habicht}, \citenamefont
  {Golub},\ and\ \citenamefont {Mezei}}]{Keller03}%
  \BibitemOpen
  \bibfield  {author} {\bibinfo {author} {\bibfnamefont {T.}~\bibnamefont
  {Keller}}, \bibinfo {author} {\bibfnamefont {B.}~\bibnamefont {Keimer}},
  \bibinfo {author} {\bibfnamefont {K.}~\bibnamefont {Habicht}}, \bibinfo
  {author} {\bibfnamefont {R.}~\bibnamefont {Golub}}, \ and\ \bibinfo {author}
  {\bibfnamefont {F.}~\bibnamefont {Mezei}},\ }\href@noop {} {\emph {\bibinfo
  {title} {Neutron Resonance Spin Echo\,--\,Triple Axis Spectrometry, Lecture
  Notes in Physics {\bf 601}}}}\ (\bibinfo  {publisher} {Springer, Berlin,
  Heidelberg},\ \bibinfo {year} {2003})\ p.~\bibinfo {pages} {74}\BibitemShut
  {NoStop}%
\bibitem [{\citenamefont {Bud'ko}\ \emph {\textit{et~al.}}(2009)\citenamefont
  {Bud'ko}, \citenamefont {Ni}, \citenamefont {Nandi}, \citenamefont
  {Schmiedeshoff},\ and\ \citenamefont {Canfield}}]{BudkoNi09}%
  \BibitemOpen
  \bibfield  {author} {\bibinfo {author} {\bibfnamefont {S.~L.}\ \bibnamefont
  {Bud'ko}}, \bibinfo {author} {\bibfnamefont {N.}~\bibnamefont {Ni}}, \bibinfo
  {author} {\bibfnamefont {S.}~\bibnamefont {Nandi}}, \bibinfo {author}
  {\bibfnamefont {G.~M.}\ \bibnamefont {Schmiedeshoff}}, \ and\ \bibinfo
  {author} {\bibfnamefont {P.~C.}\ \bibnamefont {Canfield}},\
  }\href{http://link.aps.org/doi/10.1103/PhysRevB.79.054525} {\bibfield
  {journal} {\bibinfo  {journal} {Phys. Rev.~B}\ }\textbf {\bibinfo {volume}
  {79}},\ \bibinfo {pages} {054525} (\bibinfo {year} {2009})}\BibitemShut
  {NoStop}%
\bibitem [{\citenamefont {Meingast}\ \emph {\textit{et~al.}}(2012)\citenamefont
  {Meingast}, \citenamefont {Hardy}, \citenamefont {Heid}, \citenamefont
  {Adelmann}, \citenamefont {B\"ohmer}, \citenamefont {Burger}, \citenamefont
  {Ernst}, \citenamefont {Fromknecht}, \citenamefont {Schweiss},\ and\
  \citenamefont {Wolf}}]{MeingastHardy12}%
  \BibitemOpen
  \bibfield  {author} {\bibinfo {author} {\bibfnamefont {C.}~\bibnamefont
  {Meingast}}, \bibinfo {author} {\bibfnamefont {F.}~\bibnamefont {Hardy}},
  \bibinfo {author} {\bibfnamefont {R.}~\bibnamefont {Heid}}, \bibinfo {author}
  {\bibfnamefont {P.}~\bibnamefont {Adelmann}}, \bibinfo {author}
  {\bibfnamefont {A.}~\bibnamefont {B\"ohmer}}, \bibinfo {author}
  {\bibfnamefont {P.}~\bibnamefont {Burger}}, \bibinfo {author} {\bibfnamefont
  {D.}~\bibnamefont {Ernst}}, \bibinfo {author} {\bibfnamefont
  {R.}~\bibnamefont {Fromknecht}}, \bibinfo {author} {\bibfnamefont
  {P.}~\bibnamefont {Schweiss}}, \ and\ \bibinfo {author} {\bibfnamefont
  {T.}~\bibnamefont {Wolf}},\
  }\href{http://link.aps.org/doi/10.1103/PhysRevLett.108.177004} {\bibfield
  {journal} {\bibinfo  {journal} {Phys. Rev. Lett.}\ }\textbf {\bibinfo
  {volume} {108}},\ \bibinfo {pages} {177004} (\bibinfo {year}
  {2012})}\BibitemShut {NoStop}%
\bibitem [{\citenamefont {B\"ohmer}\ \emph {\textit{et~al.}}(2012)\citenamefont
  {B\"ohmer}, \citenamefont {Burger}, \citenamefont {Hardy}, \citenamefont
  {Wolf}, \citenamefont {Schweiss}, \citenamefont {Fromknecht}, \citenamefont
  {v.~L\"ohneysen}, \citenamefont {Meingast}, \citenamefont {Mak},
  \citenamefont {Lortz}, \citenamefont {Kasahara}, \citenamefont {Terashima},
  \citenamefont {Shibauchi},\ and\ \citenamefont {Matsuda}}]{BohmerBurger12}%
  \BibitemOpen
  \bibfield  {author} {\bibinfo {author} {\bibfnamefont {A.~E.}\ \bibnamefont
  {B\"ohmer}}, \bibinfo {author} {\bibfnamefont {P.}~\bibnamefont {Burger}},
  \bibinfo {author} {\bibfnamefont {F.}~\bibnamefont {Hardy}}, \bibinfo
  {author} {\bibfnamefont {T.}~\bibnamefont {Wolf}}, \bibinfo {author}
  {\bibfnamefont {P.}~\bibnamefont {Schweiss}}, \bibinfo {author}
  {\bibfnamefont {R.}~\bibnamefont {Fromknecht}}, \bibinfo {author}
  {\bibfnamefont {H.}~\bibnamefont {v.~L\"ohneysen}}, \bibinfo {author}
  {\bibfnamefont {C.}~\bibnamefont {Meingast}}, \bibinfo {author}
  {\bibfnamefont {H.~K.}\ \bibnamefont {Mak}}, \bibinfo {author} {\bibfnamefont
  {R.}~\bibnamefont {Lortz}}, \bibinfo {author} {\bibfnamefont
  {S.}~\bibnamefont {Kasahara}}, \bibinfo {author} {\bibfnamefont
  {T.}~\bibnamefont {Terashima}}, \bibinfo {author} {\bibfnamefont
  {T.}~\bibnamefont {Shibauchi}}, \ and\ \bibinfo {author} {\bibfnamefont
  {Y.}~\bibnamefont {Matsuda}},\
  }\href{http://link.aps.org/doi/10.1103/PhysRevB.86.094521} {\bibfield
  {journal} {\bibinfo  {journal} {Phys. Rev.~B}\ }\textbf {\bibinfo {volume}
  {86}},\ \bibinfo {pages} {094521} (\bibinfo {year} {2012})}\BibitemShut
  {NoStop}%
\bibitem [{\citenamefont {Amato}(1997)}]{Amato97}%
  \BibitemOpen
  \bibfield  {author} {\bibinfo {author} {\bibfnamefont {A.}~\bibnamefont
  {Amato}},\ }\href{http://link.aps.org/doi/10.1103/RevModPhys.69.1119}
  {\bibfield  {journal} {\bibinfo  {journal} {Rev. Mod. Phys.}\ }\textbf
  {\bibinfo {volume} {69}},\ \bibinfo {pages} {1119} (\bibinfo {year}
  {1997})}\BibitemShut {NoStop}%
\bibitem [{\citenamefont {Blundell}(1999)}]{Blundell99}%
  \BibitemOpen
  \bibfield  {author} {\bibinfo {author} {\bibfnamefont {S.~J.}\ \bibnamefont
  {Blundell}},\
  }\href{http://www.tandfonline.com/doi/abs/10.1080/001075199181521} {\bibfield
   {journal} {\bibinfo  {journal} {Contemporary Physics}\ }\textbf {\bibinfo
  {volume} {40}},\ \bibinfo {pages} {175} (\bibinfo {year} {1999})}\BibitemShut
  {NoStop}%
\bibitem [{\citenamefont {Drew}\ \emph {\textit{et~al.}}(2009)\citenamefont
  {Drew}, \citenamefont {Niedermayer}, \citenamefont {Baker}, \citenamefont
  {Pratt}, \citenamefont {Blundell}, \citenamefont {Lancaster}, \citenamefont
  {Liu}, \citenamefont {Wu}, \citenamefont {Chen}, \citenamefont {Watanabe},
  \citenamefont {Malik}, \citenamefont {Dubroka}, \citenamefont {Roessle},
  \citenamefont {Kim}, \citenamefont {Baines},\ and\ \citenamefont
  {Bernhard}}]{DrewNiedermayer09}%
  \BibitemOpen
  \bibfield  {author} {\bibinfo {author} {\bibfnamefont {A.~J.}\ \bibnamefont
  {Drew}}, \bibinfo {author} {\bibfnamefont {C.}~\bibnamefont {Niedermayer}},
  \bibinfo {author} {\bibfnamefont {P.~J.}\ \bibnamefont {Baker}}, \bibinfo
  {author} {\bibfnamefont {F.~L.}\ \bibnamefont {Pratt}}, \bibinfo {author}
  {\bibfnamefont {S.~J.}\ \bibnamefont {Blundell}}, \bibinfo {author}
  {\bibfnamefont {T.}~\bibnamefont {Lancaster}}, \bibinfo {author}
  {\bibfnamefont {R.~H.}\ \bibnamefont {Liu}}, \bibinfo {author} {\bibfnamefont
  {G.}~\bibnamefont {Wu}}, \bibinfo {author} {\bibfnamefont {X.~H.}\
  \bibnamefont {Chen}}, \bibinfo {author} {\bibfnamefont {I.}~\bibnamefont
  {Watanabe}}, \bibinfo {author} {\bibfnamefont {V.~K.}\ \bibnamefont {Malik}},
  \bibinfo {author} {\bibfnamefont {A.}~\bibnamefont {Dubroka}}, \bibinfo
  {author} {\bibfnamefont {M.}~\bibnamefont {Roessle}}, \bibinfo {author}
  {\bibfnamefont {K.~W.}\ \bibnamefont {Kim}}, \bibinfo {author} {\bibfnamefont
  {C.}~\bibnamefont {Baines}}, \ and\ \bibinfo {author} {\bibfnamefont
  {C.}~\bibnamefont {Bernhard}},\
  }\href{http://www.nature.com/nmat/journal/v8/n4/abs/nmat2396.html} {\bibfield
   {journal} {\bibinfo  {journal} {Nature Mater.}\ }\textbf {\bibinfo {volume}
  {8}},\ \bibinfo {pages} {310} (\bibinfo {year} {2009})}\BibitemShut {NoStop}%
\bibitem [{\citenamefont {Park}\ \emph {\textit{et~al.}}(2009)\citenamefont
  {Park}, \citenamefont {Inosov}, \citenamefont {Niedermayer}, \citenamefont
  {Sun}, \citenamefont {Haug}, \citenamefont {Christensen}, \citenamefont
  {Dinnebier}, \citenamefont {Boris}, \citenamefont {Drew}, \citenamefont
  {Schulz}, \citenamefont {Shapoval}, \citenamefont {Wolff}, \citenamefont
  {Neu}, \citenamefont {Yang}, \citenamefont {Lin}, \citenamefont {Keimer},\
  and\ \citenamefont {Hinkov}}]{ParkInosov09}%
  \BibitemOpen
  \bibfield  {author} {\bibinfo {author} {\bibfnamefont {J.~T.}\ \bibnamefont
  {Park}}, \bibinfo {author} {\bibfnamefont {D.~S.}\ \bibnamefont {Inosov}},
  \bibinfo {author} {\bibfnamefont {C.}~\bibnamefont {Niedermayer}}, \bibinfo
  {author} {\bibfnamefont {G.~L.}\ \bibnamefont {Sun}}, \bibinfo {author}
  {\bibfnamefont {D.}~\bibnamefont {Haug}}, \bibinfo {author} {\bibfnamefont
  {N.~B.}\ \bibnamefont {Christensen}}, \bibinfo {author} {\bibfnamefont
  {R.}~\bibnamefont {Dinnebier}}, \bibinfo {author} {\bibfnamefont {A.~V.}\
  \bibnamefont {Boris}}, \bibinfo {author} {\bibfnamefont {A.~J.}\ \bibnamefont
  {Drew}}, \bibinfo {author} {\bibfnamefont {L.}~\bibnamefont {Schulz}},
  \bibinfo {author} {\bibfnamefont {T.}~\bibnamefont {Shapoval}}, \bibinfo
  {author} {\bibfnamefont {U.}~\bibnamefont {Wolff}}, \bibinfo {author}
  {\bibfnamefont {V.}~\bibnamefont {Neu}}, \bibinfo {author} {\bibfnamefont
  {X.}~\bibnamefont {Yang}}, \bibinfo {author} {\bibfnamefont {C.~T.}\
  \bibnamefont {Lin}}, \bibinfo {author} {\bibfnamefont {B.}~\bibnamefont
  {Keimer}}, \ and\ \bibinfo {author} {\bibfnamefont {V.}~\bibnamefont
  {Hinkov}},\ }\href{http://link.aps.org/abstract/PRL/v102/e117006} {\bibfield
  {journal} {\bibinfo  {journal} {Phys. Rev. Lett.}\ }\textbf {\bibinfo
  {volume} {102}},\ \bibinfo {eid} {117006} (\bibinfo {year}
  {2009})}\BibitemShut {NoStop}%
\bibitem [{\citenamefont {Goko}\ \emph {\textit{et~al.}}(2009)\citenamefont
  {Goko}, \citenamefont {Aczel}, \citenamefont {Baggio-Saitovitch},
  \citenamefont {Bud'ko}, \citenamefont {Canfield}, \citenamefont {Carlo},
  \citenamefont {Chen}, \citenamefont {Dai}, \citenamefont {Hamann},
  \citenamefont {Hu}, \citenamefont {Kageyama}, \citenamefont {Luke},
  \citenamefont {Luo}, \citenamefont {Nachumi}, \citenamefont {Ni},
  \citenamefont {Reznik}, \citenamefont {Sanchez-Candela}, \citenamefont
  {Savici}, \citenamefont {Sikes}, \citenamefont {Wang}, \citenamefont {Wiebe},
  \citenamefont {Williams}, \citenamefont {Yamamoto}, \citenamefont {Yu},\ and\
  \citenamefont {Uemura}}]{GokoAczel09}%
  \BibitemOpen
  \bibfield  {author} {\bibinfo {author} {\bibfnamefont {T.}~\bibnamefont
  {Goko}}, \bibinfo {author} {\bibfnamefont {A.~A.}\ \bibnamefont {Aczel}},
  \bibinfo {author} {\bibfnamefont {E.}~\bibnamefont {Baggio-Saitovitch}},
  \bibinfo {author} {\bibfnamefont {S.~L.}\ \bibnamefont {Bud'ko}}, \bibinfo
  {author} {\bibfnamefont {P.~C.}\ \bibnamefont {Canfield}}, \bibinfo {author}
  {\bibfnamefont {J.~P.}\ \bibnamefont {Carlo}}, \bibinfo {author}
  {\bibfnamefont {G.~F.}\ \bibnamefont {Chen}}, \bibinfo {author}
  {\bibfnamefont {P.}~\bibnamefont {Dai}}, \bibinfo {author} {\bibfnamefont
  {A.~C.}\ \bibnamefont {Hamann}}, \bibinfo {author} {\bibfnamefont {W.~Z.}\
  \bibnamefont {Hu}}, \bibinfo {author} {\bibfnamefont {H.}~\bibnamefont
  {Kageyama}}, \bibinfo {author} {\bibfnamefont {G.~M.}\ \bibnamefont {Luke}},
  \bibinfo {author} {\bibfnamefont {J.~L.}\ \bibnamefont {Luo}}, \bibinfo
  {author} {\bibfnamefont {B.}~\bibnamefont {Nachumi}}, \bibinfo {author}
  {\bibfnamefont {N.}~\bibnamefont {Ni}}, \bibinfo {author} {\bibfnamefont
  {D.}~\bibnamefont {Reznik}}, \bibinfo {author} {\bibfnamefont {D.~R.}\
  \bibnamefont {Sanchez-Candela}}, \bibinfo {author} {\bibfnamefont {A.~T.}\
  \bibnamefont {Savici}}, \bibinfo {author} {\bibfnamefont {K.~J.}\
  \bibnamefont {Sikes}}, \bibinfo {author} {\bibfnamefont {N.~L.}\ \bibnamefont
  {Wang}}, \bibinfo {author} {\bibfnamefont {C.~R.}\ \bibnamefont {Wiebe}},
  \bibinfo {author} {\bibfnamefont {T.~J.}\ \bibnamefont {Williams}}, \bibinfo
  {author} {\bibfnamefont {T.}~\bibnamefont {Yamamoto}}, \bibinfo {author}
  {\bibfnamefont {W.}~\bibnamefont {Yu}}, \ and\ \bibinfo {author}
  {\bibfnamefont {Y.~J.}\ \bibnamefont {Uemura}},\
  }\href{http://link.aps.org/abstract/PRB/v80/e024508} {\bibfield  {journal}
  {\bibinfo  {journal} {Phys. Rev.~B}\ }\textbf {\bibinfo {volume} {80}},\
  \bibinfo {eid} {024508} (\bibinfo {year} {2009})}\BibitemShut {NoStop}%
\bibitem [{\citenamefont {Takeshita}\ \emph
  {\textit{et~al.}}(2009)\citenamefont {Takeshita}, \citenamefont {Kadono},
  \citenamefont {Hiraishi}, \citenamefont {Miyazaki}, \citenamefont {Koda},
  \citenamefont {Matsuishi},\ and\ \citenamefont {Hosono}}]{TakeshitaKadono09}%
  \BibitemOpen
  \bibfield  {author} {\bibinfo {author} {\bibfnamefont {S.}~\bibnamefont
  {Takeshita}}, \bibinfo {author} {\bibfnamefont {R.}~\bibnamefont {Kadono}},
  \bibinfo {author} {\bibfnamefont {M.}~\bibnamefont {Hiraishi}}, \bibinfo
  {author} {\bibfnamefont {M.}~\bibnamefont {Miyazaki}}, \bibinfo {author}
  {\bibfnamefont {A.}~\bibnamefont {Koda}}, \bibinfo {author} {\bibfnamefont
  {S.}~\bibnamefont {Matsuishi}}, \ and\ \bibinfo {author} {\bibfnamefont
  {H.}~\bibnamefont {Hosono}},\
  }\href{http://link.aps.org/doi/10.1103/PhysRevLett.103.027002} {\bibfield
  {journal} {\bibinfo  {journal} {Phys. Rev. Lett.}\ }\textbf {\bibinfo
  {volume} {103}},\ \bibinfo {pages} {027002} (\bibinfo {year}
  {2009})}\BibitemShut {NoStop}%
\bibitem [{\citenamefont {Wiesenmayer}\ \emph
  {\textit{et~al.}}(2011)\citenamefont {Wiesenmayer}, \citenamefont {Luetkens},
  \citenamefont {Pascua}, \citenamefont {Khasanov}, \citenamefont {Amato},
  \citenamefont {Potts}, \citenamefont {Banusch}, \citenamefont {Klauss},\ and\
  \citenamefont {Johrendt}}]{WiesenmayerLuetkens11}%
  \BibitemOpen
  \bibfield  {author} {\bibinfo {author} {\bibfnamefont {E.}~\bibnamefont
  {Wiesenmayer}}, \bibinfo {author} {\bibfnamefont {H.}~\bibnamefont
  {Luetkens}}, \bibinfo {author} {\bibfnamefont {G.}~\bibnamefont {Pascua}},
  \bibinfo {author} {\bibfnamefont {R.}~\bibnamefont {Khasanov}}, \bibinfo
  {author} {\bibfnamefont {A.}~\bibnamefont {Amato}}, \bibinfo {author}
  {\bibfnamefont {H.}~\bibnamefont {Potts}}, \bibinfo {author} {\bibfnamefont
  {B.}~\bibnamefont {Banusch}}, \bibinfo {author} {\bibfnamefont {H.-H.}\
  \bibnamefont {Klauss}}, \ and\ \bibinfo {author} {\bibfnamefont
  {D.}~\bibnamefont {Johrendt}},\
  }\href{http://link.aps.org/doi/10.1103/PhysRevLett.107.237001} {\bibfield
  {journal} {\bibinfo  {journal} {Phys. Rev. Lett.}\ }\textbf {\bibinfo
  {volume} {107}},\ \bibinfo {pages} {237001} (\bibinfo {year}
  {2011})}\BibitemShut {NoStop}%
\bibitem [{\citenamefont {Khasanov}\ \emph {\textit{et~al.}}(2011)\citenamefont
  {Khasanov}, \citenamefont {Sanna}, \citenamefont {Prando}, \citenamefont
  {Shermadini}, \citenamefont {Bendele}, \citenamefont {Amato}, \citenamefont
  {Carretta}, \citenamefont {De~Renzi}, \citenamefont {Karpinski},
  \citenamefont {Katrych}, \citenamefont {Luetkens},\ and\ \citenamefont
  {Zhigadlo}}]{KhasanovSanna11}%
  \BibitemOpen
  \bibfield  {author} {\bibinfo {author} {\bibfnamefont {R.}~\bibnamefont
  {Khasanov}}, \bibinfo {author} {\bibfnamefont {S.}~\bibnamefont {Sanna}},
  \bibinfo {author} {\bibfnamefont {G.}~\bibnamefont {Prando}}, \bibinfo
  {author} {\bibfnamefont {Z.}~\bibnamefont {Shermadini}}, \bibinfo {author}
  {\bibfnamefont {M.}~\bibnamefont {Bendele}}, \bibinfo {author} {\bibfnamefont
  {A.}~\bibnamefont {Amato}}, \bibinfo {author} {\bibfnamefont
  {P.}~\bibnamefont {Carretta}}, \bibinfo {author} {\bibfnamefont
  {R.}~\bibnamefont {De~Renzi}}, \bibinfo {author} {\bibfnamefont
  {J.}~\bibnamefont {Karpinski}}, \bibinfo {author} {\bibfnamefont
  {S.}~\bibnamefont {Katrych}}, \bibinfo {author} {\bibfnamefont
  {H.}~\bibnamefont {Luetkens}}, \ and\ \bibinfo {author} {\bibfnamefont
  {N.~D.}\ \bibnamefont {Zhigadlo}},\
  }\href{http://link.aps.org/doi/10.1103/PhysRevB.84.100501} {\bibfield
  {journal} {\bibinfo  {journal} {Phys. Rev. B}\ }\textbf {\bibinfo {volume}
  {84}},\ \bibinfo {pages} {100501} (\bibinfo {year} {2011})}\BibitemShut
  {NoStop}%
\bibitem [{\citenamefont {Shermadini}\ \emph
  {\textit{et~al.}}(2011)\citenamefont {Shermadini}, \citenamefont
  {Krzton-Maziopa}, \citenamefont {Bendele}, \citenamefont {Khasanov},
  \citenamefont {Luetkens}, \citenamefont {Conder}, \citenamefont
  {Pomjakushina}, \citenamefont {Weyeneth}, \citenamefont {Pomjakushin},
  \citenamefont {Bossen},\ and\ \citenamefont
  {Amato}}]{ShermadiniKrztonMaziopa11}%
  \BibitemOpen
  \bibfield  {author} {\bibinfo {author} {\bibfnamefont {Z.}~\bibnamefont
  {Shermadini}}, \bibinfo {author} {\bibfnamefont {A.}~\bibnamefont
  {Krzton-Maziopa}}, \bibinfo {author} {\bibfnamefont {M.}~\bibnamefont
  {Bendele}}, \bibinfo {author} {\bibfnamefont {R.}~\bibnamefont {Khasanov}},
  \bibinfo {author} {\bibfnamefont {H.}~\bibnamefont {Luetkens}}, \bibinfo
  {author} {\bibfnamefont {K.}~\bibnamefont {Conder}}, \bibinfo {author}
  {\bibfnamefont {E.}~\bibnamefont {Pomjakushina}}, \bibinfo {author}
  {\bibfnamefont {S.}~\bibnamefont {Weyeneth}}, \bibinfo {author}
  {\bibfnamefont {V.}~\bibnamefont {Pomjakushin}}, \bibinfo {author}
  {\bibfnamefont {O.}~\bibnamefont {Bossen}}, \ and\ \bibinfo {author}
  {\bibfnamefont {A.}~\bibnamefont {Amato}},\
  }\href{http://link.aps.org/doi/10.1103/PhysRevLett.106.117602} {\bibfield
  {journal} {\bibinfo  {journal} {Phys. Rev. Lett.}\ }\textbf {\bibinfo
  {volume} {106}},\ \bibinfo {pages} {117602} (\bibinfo {year}
  {2011})}\BibitemShut {NoStop}%
\bibitem [{\citenamefont {Charnukha}\ \emph
  {\textit{et~al.}}(2012)\citenamefont {Charnukha}, \citenamefont {Cvitkovic},
  \citenamefont {Prokscha}, \citenamefont {Pr\"opper}, \citenamefont {Ocelic},
  \citenamefont {Suter}, \citenamefont {Salman}, \citenamefont {Morenzoni},
  \citenamefont {Deisenhofer}, \citenamefont {Tsurkan}, \citenamefont {Loidl},
  \citenamefont {Keimer},\ and\ \citenamefont {Boris}}]{CharnukhaCvitkovic12}%
  \BibitemOpen
  \bibfield  {author} {\bibinfo {author} {\bibfnamefont {A.}~\bibnamefont
  {Charnukha}}, \bibinfo {author} {\bibfnamefont {A.}~\bibnamefont
  {Cvitkovic}}, \bibinfo {author} {\bibfnamefont {T.}~\bibnamefont {Prokscha}},
  \bibinfo {author} {\bibfnamefont {D.}~\bibnamefont {Pr\"opper}}, \bibinfo
  {author} {\bibfnamefont {N.}~\bibnamefont {Ocelic}}, \bibinfo {author}
  {\bibfnamefont {A.}~\bibnamefont {Suter}}, \bibinfo {author} {\bibfnamefont
  {Z.}~\bibnamefont {Salman}}, \bibinfo {author} {\bibfnamefont
  {E.}~\bibnamefont {Morenzoni}}, \bibinfo {author} {\bibfnamefont
  {J.}~\bibnamefont {Deisenhofer}}, \bibinfo {author} {\bibfnamefont
  {V.}~\bibnamefont {Tsurkan}}, \bibinfo {author} {\bibfnamefont
  {A.}~\bibnamefont {Loidl}}, \bibinfo {author} {\bibfnamefont
  {B.}~\bibnamefont {Keimer}}, \ and\ \bibinfo {author} {\bibfnamefont {A.~V.}\
  \bibnamefont {Boris}},\
  }\href{http://link.aps.org/doi/10.1103/PhysRevLett.109.017003} {\bibfield
  {journal} {\bibinfo  {journal} {Phys. Rev. Lett.}\ }\textbf {\bibinfo
  {volume} {109}},\ \bibinfo {pages} {017003} (\bibinfo {year}
  {2012})}\BibitemShut {NoStop}%
\bibitem [{\citenamefont {Shermadini}\ \emph
  {\textit{et~al.}}(2012)\citenamefont {Shermadini}, \citenamefont {Luetkens},
  \citenamefont {Khasanov}, \citenamefont {Krzton-Maziopa}, \citenamefont
  {Conder}, \citenamefont {Pomjakushina}, \citenamefont {Klauss},\ and\
  \citenamefont {Amato}}]{ShermadiniLuetkens12}%
  \BibitemOpen
  \bibfield  {author} {\bibinfo {author} {\bibfnamefont {Z.}~\bibnamefont
  {Shermadini}}, \bibinfo {author} {\bibfnamefont {H.}~\bibnamefont
  {Luetkens}}, \bibinfo {author} {\bibfnamefont {R.}~\bibnamefont {Khasanov}},
  \bibinfo {author} {\bibfnamefont {A.}~\bibnamefont {Krzton-Maziopa}},
  \bibinfo {author} {\bibfnamefont {K.}~\bibnamefont {Conder}}, \bibinfo
  {author} {\bibfnamefont {E.}~\bibnamefont {Pomjakushina}}, \bibinfo {author}
  {\bibfnamefont {H.-H.}\ \bibnamefont {Klauss}}, \ and\ \bibinfo {author}
  {\bibfnamefont {A.}~\bibnamefont {Amato}},\
  }\href{http://link.aps.org/doi/10.1103/PhysRevB.85.100501} {\bibfield
  {journal} {\bibinfo  {journal} {Phys. Rev. B}\ }\textbf {\bibinfo {volume}
  {85}},\ \bibinfo {pages} {100501} (\bibinfo {year} {2012})}\BibitemShut
  {NoStop}%
\bibitem [{\citenamefont {Bernhard}\ \emph {\textit{et~al.}}(2012)\citenamefont
  {Bernhard}, \citenamefont {Wang}, \citenamefont {Nuccio}, \citenamefont
  {Schulz}, \citenamefont {Zaharko}, \citenamefont {Larsen}, \citenamefont
  {Aristizabal}, \citenamefont {Willis}, \citenamefont {Drew}, \citenamefont
  {Varma}, \citenamefont {Wolf},\ and\ \citenamefont
  {Niedermayer}}]{BernhardWang12}%
  \BibitemOpen
  \bibfield  {author} {\bibinfo {author} {\bibfnamefont {C.}~\bibnamefont
  {Bernhard}}, \bibinfo {author} {\bibfnamefont {C.~N.}\ \bibnamefont {Wang}},
  \bibinfo {author} {\bibfnamefont {L.}~\bibnamefont {Nuccio}}, \bibinfo
  {author} {\bibfnamefont {L.}~\bibnamefont {Schulz}}, \bibinfo {author}
  {\bibfnamefont {O.}~\bibnamefont {Zaharko}}, \bibinfo {author} {\bibfnamefont
  {J.}~\bibnamefont {Larsen}}, \bibinfo {author} {\bibfnamefont
  {C.}~\bibnamefont {Aristizabal}}, \bibinfo {author} {\bibfnamefont
  {M.}~\bibnamefont {Willis}}, \bibinfo {author} {\bibfnamefont {A.~J.}\
  \bibnamefont {Drew}}, \bibinfo {author} {\bibfnamefont {G.~D.}\ \bibnamefont
  {Varma}}, \bibinfo {author} {\bibfnamefont {T.}~\bibnamefont {Wolf}}, \ and\
  \bibinfo {author} {\bibfnamefont {C.}~\bibnamefont {Niedermayer}},\
  }\href{http://link.aps.org/doi/10.1103/PhysRevB.86.184509} {\bibfield
  {journal} {\bibinfo  {journal} {Phys. Rev.~B}\ }\textbf {\bibinfo {volume}
  {86}},\ \bibinfo {pages} {184509} (\bibinfo {year} {2012})}\BibitemShut
  {NoStop}%
\bibitem [{\citenamefont {Aczel}\ \emph {\textit{et~al.}}(2008)\citenamefont
  {Aczel}, \citenamefont {Baggio-Saitovitch}, \citenamefont {Budko},
  \citenamefont {Canfield}, \citenamefont {Carlo}, \citenamefont {Chen},
  \citenamefont {Dai}, \citenamefont {Goko}, \citenamefont {Hu}, \citenamefont
  {Luke}, \citenamefont {Luo}, \citenamefont {Ni}, \citenamefont
  {Sanchez-Candela}, \citenamefont {Tafti}, \citenamefont {Wang}, \citenamefont
  {Williams}, \citenamefont {Yu},\ and\ \citenamefont
  {Uemura}}]{AczelBaggio08}%
  \BibitemOpen
  \bibfield  {author} {\bibinfo {author} {\bibfnamefont {A.~A.}\ \bibnamefont
  {Aczel}}, \bibinfo {author} {\bibfnamefont {E.}~\bibnamefont
  {Baggio-Saitovitch}}, \bibinfo {author} {\bibfnamefont {S.~L.}\ \bibnamefont
  {Budko}}, \bibinfo {author} {\bibfnamefont {P.~C.}\ \bibnamefont {Canfield}},
  \bibinfo {author} {\bibfnamefont {J.~P.}\ \bibnamefont {Carlo}}, \bibinfo
  {author} {\bibfnamefont {G.~F.}\ \bibnamefont {Chen}}, \bibinfo {author}
  {\bibfnamefont {P.}~\bibnamefont {Dai}}, \bibinfo {author} {\bibfnamefont
  {T.}~\bibnamefont {Goko}}, \bibinfo {author} {\bibfnamefont {W.~Z.}\
  \bibnamefont {Hu}}, \bibinfo {author} {\bibfnamefont {G.~M.}\ \bibnamefont
  {Luke}}, \bibinfo {author} {\bibfnamefont {J.~L.}\ \bibnamefont {Luo}},
  \bibinfo {author} {\bibfnamefont {N.}~\bibnamefont {Ni}}, \bibinfo {author}
  {\bibfnamefont {D.~R.}\ \bibnamefont {Sanchez-Candela}}, \bibinfo {author}
  {\bibfnamefont {F.~F.}\ \bibnamefont {Tafti}}, \bibinfo {author}
  {\bibfnamefont {N.~L.}\ \bibnamefont {Wang}}, \bibinfo {author}
  {\bibfnamefont {T.~J.}\ \bibnamefont {Williams}}, \bibinfo {author}
  {\bibfnamefont {W.}~\bibnamefont {Yu}}, \ and\ \bibinfo {author}
  {\bibfnamefont {Y.~J.}\ \bibnamefont {Uemura}},\
  }\href{http://link.aps.org/doi/10.1103/PhysRevB.78.214503} {\bibfield
  {journal} {\bibinfo  {journal} {Phys. Rev. B}\ }\textbf {\bibinfo {volume}
  {78}},\ \bibinfo {pages} {214503} (\bibinfo {year} {2008})}\BibitemShut
  {NoStop}%
\bibitem [{\citenamefont {Frankovsky}\ \emph
  {\textit{et~al.}}(2013)\citenamefont {Frankovsky}, \citenamefont {Luetkens},
  \citenamefont {Tambornino}, \citenamefont {Marchuk}, \citenamefont {Pascua},
  \citenamefont {Amato}, \citenamefont {Klauss},\ and\ \citenamefont
  {Johrendt}}]{FrankovskyLuetkens13}%
  \BibitemOpen
  \bibfield  {author} {\bibinfo {author} {\bibfnamefont {R.}~\bibnamefont
  {Frankovsky}}, \bibinfo {author} {\bibfnamefont {H.}~\bibnamefont
  {Luetkens}}, \bibinfo {author} {\bibfnamefont {F.}~\bibnamefont
  {Tambornino}}, \bibinfo {author} {\bibfnamefont {A.}~\bibnamefont {Marchuk}},
  \bibinfo {author} {\bibfnamefont {G.}~\bibnamefont {Pascua}}, \bibinfo
  {author} {\bibfnamefont {A.}~\bibnamefont {Amato}}, \bibinfo {author}
  {\bibfnamefont {H.-H.}\ \bibnamefont {Klauss}}, \ and\ \bibinfo {author}
  {\bibfnamefont {D.}~\bibnamefont {Johrendt}},\ }\bibinfo {howpublished}
  {\href{http://arxiv.org/abs/arXiv:1303.6833}{arXiv:1303.6833} (to be
  published in Phys. Rev.~B).}\BibitemShut {Stop}%
\bibitem [{\citenamefont {Yamazaki}(1982)}]{Yamazaki82}%
  \BibitemOpen
  \bibfield  {author} {\bibinfo {author} {\bibfnamefont {T.}~\bibnamefont
  {Yamazaki}},\ }\href{http://dx.doi.org/10.1016/0167-5087(82)90183-1}
  {\bibfield  {journal} {\bibinfo  {journal} {Nucl. Instr. Meth.}\ }\textbf
  {\bibinfo {volume} {199}},\ \bibinfo {pages} {133} (\bibinfo {year}
  {1982})}\BibitemShut {NoStop}%
\bibitem [{\citenamefont {Uemura}\ \emph {\textit{et~al.}}(1985)\citenamefont
  {Uemura}, \citenamefont {Yamazaki}, \citenamefont {Harshman}, \citenamefont
  {Senba},\ and\ \citenamefont {Ansaldo}}]{UemuraYamazaki85}%
  \BibitemOpen
  \bibfield  {author} {\bibinfo {author} {\bibfnamefont {Y.~J.}\ \bibnamefont
  {Uemura}}, \bibinfo {author} {\bibfnamefont {T.}~\bibnamefont {Yamazaki}},
  \bibinfo {author} {\bibfnamefont {D.~R.}\ \bibnamefont {Harshman}}, \bibinfo
  {author} {\bibfnamefont {M.}~\bibnamefont {Senba}}, \ and\ \bibinfo {author}
  {\bibfnamefont {E.~J.}\ \bibnamefont {Ansaldo}},\
  }\href{http://link.aps.org/doi/10.1103/PhysRevB.31.546} {\bibfield  {journal}
  {\bibinfo  {journal} {Phys. Rev.~B}\ }\textbf {\bibinfo {volume} {31}},\
  \bibinfo {pages} {546} (\bibinfo {year} {1985})}\BibitemShut {NoStop}%
\bibitem [{\citenamefont {Fischer}(1985)}]{Fischer85}%
  \BibitemOpen
  \bibfield  {author} {\bibinfo {author} {\bibfnamefont {K.~H.}\ \bibnamefont
  {Fischer}},\ }\href{\doibase/10.1002/pssb.2221300102} {\bibfield  {journal}
  {\bibinfo  {journal} {Phys. Stat. Sol. (B)}\ }\textbf {\bibinfo {volume}
  {130}},\ \bibinfo {pages} {13} (\bibinfo {year} {1985})}\BibitemShut
  {NoStop}%
\bibitem [{\citenamefont {van~der Marck}(1997)}]{VanDerMarck97}%
  \BibitemOpen
  \bibfield  {author} {\bibinfo {author} {\bibfnamefont {S.~C.}\ \bibnamefont
  {van~der Marck}},\ }\href{http://link.aps.org/doi/10.1103/PhysRevE.55.1514}
  {\bibfield  {journal} {\bibinfo  {journal} {Phys. Rev.~E}\ }\textbf {\bibinfo
  {volume} {55}},\ \bibinfo {pages} {1514} (\bibinfo {year}
  {1997})}\BibitemShut {NoStop}%
\bibitem [{\citenamefont {Harriger}\ \emph {\textit{et~al.}}(2011)\citenamefont
  {Harriger}, \citenamefont {Luo}, \citenamefont {Liu}, \citenamefont {Frost},
  \citenamefont {Hu}, \citenamefont {Norman},\ and\ \citenamefont
  {Dai}}]{HarrigerLuo11}%
  \BibitemOpen
  \bibfield  {author} {\bibinfo {author} {\bibfnamefont {L.~W.}\ \bibnamefont
  {Harriger}}, \bibinfo {author} {\bibfnamefont {H.~Q.}\ \bibnamefont {Luo}},
  \bibinfo {author} {\bibfnamefont {M.~S.}\ \bibnamefont {Liu}}, \bibinfo
  {author} {\bibfnamefont {C.}~\bibnamefont {Frost}}, \bibinfo {author}
  {\bibfnamefont {J.~P.}\ \bibnamefont {Hu}}, \bibinfo {author} {\bibfnamefont
  {M.~R.}\ \bibnamefont {Norman}}, \ and\ \bibinfo {author} {\bibfnamefont
  {P.}~\bibnamefont {Dai}},\
  }\href{http://link.aps.org/doi/10.1103/PhysRevB.84.054544} {\bibfield
  {journal} {\bibinfo  {journal} {Phys. Rev.~B}\ }\textbf {\bibinfo {volume}
  {84}},\ \bibinfo {pages} {054544} (\bibinfo {year} {2011})}\BibitemShut
  {NoStop}%
\bibitem [{\citenamefont {Zhang}\ \emph {\textit{et~al.}}(2011)\citenamefont
  {Zhang}, \citenamefont {Wang}, \citenamefont {Luo}, \citenamefont {Wang},
  \citenamefont {Liu}, \citenamefont {Zhao}, \citenamefont {Abernathy},
  \citenamefont {Marty}, \citenamefont {Lumsden}, \citenamefont {Chi},
  \citenamefont {Chang}, \citenamefont {Rodriguez-Rivera}, \citenamefont
  {Lynn}, \citenamefont {Xiang}, \citenamefont {Hu},\ and\ \citenamefont
  {Dai}}]{ZhangWang11}%
  \BibitemOpen
  \bibfield  {author} {\bibinfo {author} {\bibfnamefont {C.}~\bibnamefont
  {Zhang}}, \bibinfo {author} {\bibfnamefont {M.}~\bibnamefont {Wang}},
  \bibinfo {author} {\bibfnamefont {H.}~\bibnamefont {Luo}}, \bibinfo {author}
  {\bibfnamefont {M.}~\bibnamefont {Wang}}, \bibinfo {author} {\bibfnamefont
  {M.}~\bibnamefont {Liu}}, \bibinfo {author} {\bibfnamefont {J.}~\bibnamefont
  {Zhao}}, \bibinfo {author} {\bibfnamefont {D.~L.}\ \bibnamefont {Abernathy}},
  \bibinfo {author} {\bibfnamefont {K.}~\bibnamefont {Marty}}, \bibinfo
  {author} {\bibfnamefont {M.~D.}\ \bibnamefont {Lumsden}}, \bibinfo {author}
  {\bibfnamefont {S.}~\bibnamefont {Chi}}, \bibinfo {author} {\bibfnamefont
  {S.}~\bibnamefont {Chang}}, \bibinfo {author} {\bibfnamefont {J.~A.}\
  \bibnamefont {Rodriguez-Rivera}}, \bibinfo {author} {\bibfnamefont {J.~W.}\
  \bibnamefont {Lynn}}, \bibinfo {author} {\bibfnamefont {T.}~\bibnamefont
  {Xiang}}, \bibinfo {author} {\bibfnamefont {J.}~\bibnamefont {Hu}}, \ and\
  \bibinfo {author} {\bibfnamefont {P.}~\bibnamefont {Dai}},\
  }\href{http://www.nature.com/srep/2011/111019/srep00115/full/srep00115.html}
  {\bibfield  {journal} {\bibinfo  {journal} {Scientific Reports}\ }\textbf
  {\bibinfo {volume} {1}},\ \bibinfo {pages} {115} (\bibinfo {year}
  {2011})}\BibitemShut {NoStop}%
\bibitem [{\citenamefont {Wang}\ \emph {\textit{et~al.}}(2013)\citenamefont
  {Wang}, \citenamefont {Zhang}, \citenamefont {Lu}, \citenamefont {Tan},
  \citenamefont {Luo}, \citenamefont {Song}, \citenamefont {Wang},
  \citenamefont {Zhang}, \citenamefont {Goremychkin}, \citenamefont {Perring},
  \citenamefont {Maier}, \citenamefont {Yin}, \citenamefont {Haule},
  \citenamefont {Kotliar},\ and\ \citenamefont {Dai}}]{WangZhang13}%
  \BibitemOpen
  \bibfield  {author} {\bibinfo {author} {\bibfnamefont {M.}~\bibnamefont
  {Wang}}, \bibinfo {author} {\bibfnamefont {C.}~\bibnamefont {Zhang}},
  \bibinfo {author} {\bibfnamefont {X.}~\bibnamefont {Lu}}, \bibinfo {author}
  {\bibfnamefont {G.}~\bibnamefont {Tan}}, \bibinfo {author} {\bibfnamefont
  {H.}~\bibnamefont {Luo}}, \bibinfo {author} {\bibfnamefont {Y.}~\bibnamefont
  {Song}}, \bibinfo {author} {\bibfnamefont {M.}~\bibnamefont {Wang}}, \bibinfo
  {author} {\bibfnamefont {X.}~\bibnamefont {Zhang}}, \bibinfo {author}
  {\bibfnamefont {E.~A.}\ \bibnamefont {Goremychkin}}, \bibinfo {author}
  {\bibfnamefont {T.~G.}\ \bibnamefont {Perring}}, \bibinfo {author}
  {\bibfnamefont {T.~A.}\ \bibnamefont {Maier}}, \bibinfo {author}
  {\bibfnamefont {Z.}~\bibnamefont {Yin}}, \bibinfo {author} {\bibfnamefont
  {K.}~\bibnamefont {Haule}}, \bibinfo {author} {\bibfnamefont
  {G.}~\bibnamefont {Kotliar}}, \ and\ \bibinfo {author} {\bibfnamefont
  {P.}~\bibnamefont {Dai}},\ }\bibinfo {howpublished}
  {\href{http://arxiv.org/abs/arXiv:1303.7339}{arXiv:1303.7339}
  (unpublished).}\BibitemShut {Stop}%
\bibitem [{\citenamefont {Qureshi}\ \emph {\textit{et~al.}}(2012)\citenamefont
  {Qureshi}, \citenamefont {Steffens}, \citenamefont {Wurmehl}, \citenamefont
  {Aswartham}, \citenamefont {B\"uchner},\ and\ \citenamefont
  {Braden}}]{QureshiSteffens12}%
  \BibitemOpen
  \bibfield  {author} {\bibinfo {author} {\bibfnamefont {N.}~\bibnamefont
  {Qureshi}}, \bibinfo {author} {\bibfnamefont {P.}~\bibnamefont {Steffens}},
  \bibinfo {author} {\bibfnamefont {S.}~\bibnamefont {Wurmehl}}, \bibinfo
  {author} {\bibfnamefont {S.}~\bibnamefont {Aswartham}}, \bibinfo {author}
  {\bibfnamefont {B.}~\bibnamefont {B\"uchner}}, \ and\ \bibinfo {author}
  {\bibfnamefont {M.}~\bibnamefont {Braden}},\
  }\href{http://link.aps.org/doi/10.1103/PhysRevB.86.060410} {\bibfield
  {journal} {\bibinfo  {journal} {Phys. Rev.~B}\ }\textbf {\bibinfo {volume}
  {86}},\ \bibinfo {pages} {060410} (\bibinfo {year} {2012})}\BibitemShut
  {NoStop}%
\bibitem [{\citenamefont {Tranquada}\ \emph
  {\textit{et~al.}}(1989)\citenamefont {Tranquada}, \citenamefont {Shirane},
  \citenamefont {Keimer}, \citenamefont {Shamoto},\ and\ \citenamefont
  {Sato}}]{TranquadaShirane89}%
  \BibitemOpen
  \bibfield  {author} {\bibinfo {author} {\bibfnamefont {J.~M.}\ \bibnamefont
  {Tranquada}}, \bibinfo {author} {\bibfnamefont {G.}~\bibnamefont {Shirane}},
  \bibinfo {author} {\bibfnamefont {B.}~\bibnamefont {Keimer}}, \bibinfo
  {author} {\bibfnamefont {S.}~\bibnamefont {Shamoto}}, \ and\ \bibinfo
  {author} {\bibfnamefont {M.}~\bibnamefont {Sato}},\
  }\href{http://link.aps.org/doi/10.1103/PhysRevB.40.4503} {\bibfield
  {journal} {\bibinfo  {journal} {Phys. Rev.~B}\ }\textbf {\bibinfo {volume}
  {40}},\ \bibinfo {pages} {4503} (\bibinfo {year} {1989})}\BibitemShut
  {NoStop}%
\bibitem [{\citenamefont {Shamoto}\ \emph {\textit{et~al.}}(1993)\citenamefont
  {Shamoto}, \citenamefont {Sato}, \citenamefont {Tranquada}, \citenamefont
  {Sternlieb},\ and\ \citenamefont {Shirane}}]{ShamotoSato93}%
  \BibitemOpen
  \bibfield  {author} {\bibinfo {author} {\bibfnamefont {S.}~\bibnamefont
  {Shamoto}}, \bibinfo {author} {\bibfnamefont {M.}~\bibnamefont {Sato}},
  \bibinfo {author} {\bibfnamefont {J.~M.}\ \bibnamefont {Tranquada}}, \bibinfo
  {author} {\bibfnamefont {B.~J.}\ \bibnamefont {Sternlieb}}, \ and\ \bibinfo
  {author} {\bibfnamefont {G.}~\bibnamefont {Shirane}},\
  }\href{http://link.aps.org/doi/10.1103/PhysRevB.48.13817} {\bibfield
  {journal} {\bibinfo  {journal} {Phys. Rev.~B}\ }\textbf {\bibinfo {volume}
  {48}},\ \bibinfo {pages} {13817} (\bibinfo {year} {1993})}\BibitemShut
  {NoStop}%
\bibitem [{\citenamefont {Bourges}\ \emph {\textit{et~al.}}(1994)\citenamefont
  {Bourges}, \citenamefont {Sidis}, \citenamefont {Hennion}, \citenamefont
  {Villeneuve}, \citenamefont {Collin},\ and\ \citenamefont
  {Marucco}}]{BourgesSidis94}%
  \BibitemOpen
  \bibfield  {author} {\bibinfo {author} {\bibfnamefont {P.}~\bibnamefont
  {Bourges}}, \bibinfo {author} {\bibfnamefont {Y.}~\bibnamefont {Sidis}},
  \bibinfo {author} {\bibfnamefont {B.}~\bibnamefont {Hennion}}, \bibinfo
  {author} {\bibfnamefont {R.}~\bibnamefont {Villeneuve}}, \bibinfo {author}
  {\bibfnamefont {G.}~\bibnamefont {Collin}}, \ and\ \bibinfo {author}
  {\bibfnamefont {J.~F.}\ \bibnamefont {Marucco}},\
  }\href{http://dx.doi.org.ezproxy.aai.mpg.de/10.1016/0921-4534(94)92063-X}
  {\bibfield  {journal} {\bibinfo  {journal} {Physica~C: Superconductivity}\
  }\textbf {\bibinfo {volume} {235--240}},\ \bibinfo {pages} {1683} (\bibinfo
  {year} {1994})}\BibitemShut {NoStop}%
\bibitem [{\citenamefont {Petitgrand}\ \emph
  {\textit{et~al.}}(1999)\citenamefont {Petitgrand}, \citenamefont {Maleyev},
  \citenamefont {Bourges},\ and\ \citenamefont {Ivanov}}]{PetitgrandMaleyev99}%
  \BibitemOpen
  \bibfield  {author} {\bibinfo {author} {\bibfnamefont {D.}~\bibnamefont
  {Petitgrand}}, \bibinfo {author} {\bibfnamefont {S.~V.}\ \bibnamefont
  {Maleyev}}, \bibinfo {author} {\bibfnamefont {P.}~\bibnamefont {Bourges}}, \
  and\ \bibinfo {author} {\bibfnamefont {A.~S.}\ \bibnamefont {Ivanov}},\
  }\href{http://link.aps.org/doi/10.1103/PhysRevB.59.1079} {\bibfield
  {journal} {\bibinfo  {journal} {Phys. Rev. B}\ }\textbf {\bibinfo {volume}
  {59}},\ \bibinfo {pages} {1079} (\bibinfo {year} {1999})}\BibitemShut
  {NoStop}%
\bibitem [{\citenamefont {Diallo}\ \emph {\textit{et~al.}}(2010)\citenamefont
  {Diallo}, \citenamefont {Pratt}, \citenamefont {Fernandes}, \citenamefont
  {Tian}, \citenamefont {Zarestky}, \citenamefont {Lumsden}, \citenamefont
  {Perring}, \citenamefont {Broholm}, \citenamefont {Ni}, \citenamefont
  {Bud'ko}, \citenamefont {Canfield}, \citenamefont {Li}, \citenamefont
  {Vaknin}, \citenamefont {Kreyssig}, \citenamefont {Goldman},\ and\
  \citenamefont {McQueeney}}]{DialloPratt10}%
  \BibitemOpen
  \bibfield  {author} {\bibinfo {author} {\bibfnamefont {S.~O.}\ \bibnamefont
  {Diallo}}, \bibinfo {author} {\bibfnamefont {D.~K.}\ \bibnamefont {Pratt}},
  \bibinfo {author} {\bibfnamefont {R.~M.}\ \bibnamefont {Fernandes}}, \bibinfo
  {author} {\bibfnamefont {W.}~\bibnamefont {Tian}}, \bibinfo {author}
  {\bibfnamefont {J.~L.}\ \bibnamefont {Zarestky}}, \bibinfo {author}
  {\bibfnamefont {M.}~\bibnamefont {Lumsden}}, \bibinfo {author} {\bibfnamefont
  {T.~G.}\ \bibnamefont {Perring}}, \bibinfo {author} {\bibfnamefont {C.~L.}\
  \bibnamefont {Broholm}}, \bibinfo {author} {\bibfnamefont {N.}~\bibnamefont
  {Ni}}, \bibinfo {author} {\bibfnamefont {S.~L.}\ \bibnamefont {Bud'ko}},
  \bibinfo {author} {\bibfnamefont {P.~C.}\ \bibnamefont {Canfield}}, \bibinfo
  {author} {\bibfnamefont {H.-F.}\ \bibnamefont {Li}}, \bibinfo {author}
  {\bibfnamefont {D.}~\bibnamefont {Vaknin}}, \bibinfo {author} {\bibfnamefont
  {A.}~\bibnamefont {Kreyssig}}, \bibinfo {author} {\bibfnamefont {A.~I.}\
  \bibnamefont {Goldman}}, \ and\ \bibinfo {author} {\bibfnamefont {R.~J.}\
  \bibnamefont {McQueeney}},\
  }\href{http://link.aps.org/abstract/PRB/v81/e214407} {\bibfield  {journal}
  {\bibinfo  {journal} {Phys. Rev.~B}\ }\textbf {\bibinfo {volume} {81}},\
  \bibinfo {pages} {214407} (\bibinfo {year} {2010})}\BibitemShut {NoStop}%
\bibitem [{\citenamefont {Hossain}\ \emph {\textit{et~al.}}(2009)\citenamefont
  {Hossain}, \citenamefont {Bohnenbuck}, \citenamefont {Chuang}, \citenamefont
  {{Cruz Gonzalez}}, \citenamefont {Zegkinoglou}, \citenamefont {Haverkort},
  \citenamefont {Geck}, \citenamefont {Hawthorn}, \citenamefont {Wu},
  \citenamefont {Schussler-Langeheine}, \citenamefont {Mathieu}, \citenamefont
  {Tokura}, \citenamefont {Satow}, \citenamefont {Takagi}, \citenamefont
  {Yoshida}, \citenamefont {Denlinger}, \citenamefont {Elfimov}, \citenamefont
  {Hussain}, \citenamefont {Keimer}, \citenamefont {Sawatzky},\ and\
  \citenamefont {Damascelli}}]{HossainBohnenbuck09}%
  \BibitemOpen
  \bibfield  {author} {\bibinfo {author} {\bibfnamefont {M.~A.}\ \bibnamefont
  {Hossain}}, \bibinfo {author} {\bibfnamefont {B.}~\bibnamefont {Bohnenbuck}},
  \bibinfo {author} {\bibfnamefont {Y.-D.}\ \bibnamefont {Chuang}}, \bibinfo
  {author} {\bibfnamefont {A.~G.}\ \bibnamefont {{Cruz Gonzalez}}}, \bibinfo
  {author} {\bibfnamefont {I.}~\bibnamefont {Zegkinoglou}}, \bibinfo {author}
  {\bibfnamefont {M.~W.}\ \bibnamefont {Haverkort}}, \bibinfo {author}
  {\bibfnamefont {J.}~\bibnamefont {Geck}}, \bibinfo {author} {\bibfnamefont
  {D.~G.}\ \bibnamefont {Hawthorn}}, \bibinfo {author} {\bibfnamefont {H.-H.}\
  \bibnamefont {Wu}}, \bibinfo {author} {\bibfnamefont {C.}~\bibnamefont
  {Schussler-Langeheine}}, \bibinfo {author} {\bibfnamefont {R.}~\bibnamefont
  {Mathieu}}, \bibinfo {author} {\bibfnamefont {Y.}~\bibnamefont {Tokura}},
  \bibinfo {author} {\bibfnamefont {S.}~\bibnamefont {Satow}}, \bibinfo
  {author} {\bibfnamefont {H.}~\bibnamefont {Takagi}}, \bibinfo {author}
  {\bibfnamefont {Y.}~\bibnamefont {Yoshida}}, \bibinfo {author} {\bibfnamefont
  {J.~D.}\ \bibnamefont {Denlinger}}, \bibinfo {author} {\bibfnamefont {I.~S.}\
  \bibnamefont {Elfimov}}, \bibinfo {author} {\bibfnamefont {Z.}~\bibnamefont
  {Hussain}}, \bibinfo {author} {\bibfnamefont {B.}~\bibnamefont {Keimer}},
  \bibinfo {author} {\bibfnamefont {G.~A.}\ \bibnamefont {Sawatzky}}, \ and\
  \bibinfo {author} {\bibfnamefont {A.}~\bibnamefont {Damascelli}},\ }\bibinfo
  {howpublished} {\href{http://arxiv.org/abs/arXiv:0906.0035}{arXiv:0906.0035}
  (unpublished).}\BibitemShut {Stop}%
\bibitem [{\citenamefont {Mesa}\ \emph {\textit{et~al.}}(2012)\citenamefont
  {Mesa}, \citenamefont {Ye}, \citenamefont {Chi}, \citenamefont
  {Fernandez-Baca}, \citenamefont {Tian}, \citenamefont {Hu}, \citenamefont
  {Jin}, \citenamefont {Plummer},\ and\ \citenamefont {Zhang}}]{MesaYe12}%
  \BibitemOpen
  \bibfield  {author} {\bibinfo {author} {\bibfnamefont {D.}~\bibnamefont
  {Mesa}}, \bibinfo {author} {\bibfnamefont {F.}~\bibnamefont {Ye}}, \bibinfo
  {author} {\bibfnamefont {S.}~\bibnamefont {Chi}}, \bibinfo {author}
  {\bibfnamefont {J.~A.}\ \bibnamefont {Fernandez-Baca}}, \bibinfo {author}
  {\bibfnamefont {W.}~\bibnamefont {Tian}}, \bibinfo {author} {\bibfnamefont
  {B.}~\bibnamefont {Hu}}, \bibinfo {author} {\bibfnamefont {R.}~\bibnamefont
  {Jin}}, \bibinfo {author} {\bibfnamefont {E.~W.}\ \bibnamefont {Plummer}}, \
  and\ \bibinfo {author} {\bibfnamefont {J.}~\bibnamefont {Zhang}},\
  }\href{http://link.aps.org/doi/10.1103/PhysRevB.85.180410} {\bibfield
  {journal} {\bibinfo  {journal} {Phys. Rev.~B}\ }\textbf {\bibinfo {volume}
  {85}},\ \bibinfo {pages} {180410} (\bibinfo {year} {2012})}\BibitemShut
  {NoStop}%
\bibitem [{\citenamefont {Hossain}\ \emph {\textit{et~al.}}(2012)\citenamefont
  {Hossain}, \citenamefont {Bohnenbuck}, \citenamefont {Chuang}, \citenamefont
  {Haverkort}, \citenamefont {Elfimov}, \citenamefont {Tanaka}, \citenamefont
  {Cruz~Gonzalez}, \citenamefont {Hu}, \citenamefont {Lin}, \citenamefont
  {Chen}, \citenamefont {Mathieu}, \citenamefont {Tokura}, \citenamefont
  {Yoshida}, \citenamefont {Tjeng}, \citenamefont {Hussain}, \citenamefont
  {Keimer}, \citenamefont {Sawatzky},\ and\ \citenamefont
  {Damascelli}}]{HossainBohnenbuck12}%
  \BibitemOpen
  \bibfield  {author} {\bibinfo {author} {\bibfnamefont {M.~A.}\ \bibnamefont
  {Hossain}}, \bibinfo {author} {\bibfnamefont {B.}~\bibnamefont {Bohnenbuck}},
  \bibinfo {author} {\bibfnamefont {Y.~D.}\ \bibnamefont {Chuang}}, \bibinfo
  {author} {\bibfnamefont {M.~W.}\ \bibnamefont {Haverkort}}, \bibinfo {author}
  {\bibfnamefont {I.~S.}\ \bibnamefont {Elfimov}}, \bibinfo {author}
  {\bibfnamefont {A.}~\bibnamefont {Tanaka}}, \bibinfo {author} {\bibfnamefont
  {A.~G.}\ \bibnamefont {Cruz~Gonzalez}}, \bibinfo {author} {\bibfnamefont
  {Z.}~\bibnamefont {Hu}}, \bibinfo {author} {\bibfnamefont {H.-J.}\
  \bibnamefont {Lin}}, \bibinfo {author} {\bibfnamefont {C.~T.}\ \bibnamefont
  {Chen}}, \bibinfo {author} {\bibfnamefont {R.}~\bibnamefont {Mathieu}},
  \bibinfo {author} {\bibfnamefont {Y.}~\bibnamefont {Tokura}}, \bibinfo
  {author} {\bibfnamefont {Y.}~\bibnamefont {Yoshida}}, \bibinfo {author}
  {\bibfnamefont {L.~H.}\ \bibnamefont {Tjeng}}, \bibinfo {author}
  {\bibfnamefont {Z.}~\bibnamefont {Hussain}}, \bibinfo {author} {\bibfnamefont
  {B.}~\bibnamefont {Keimer}}, \bibinfo {author} {\bibfnamefont {G.~A.}\
  \bibnamefont {Sawatzky}}, \ and\ \bibinfo {author} {\bibfnamefont
  {A.}~\bibnamefont {Damascelli}},\
  }\href{http://link.aps.org/doi/10.1103/PhysRevB.86.041102} {\bibfield
  {journal} {\bibinfo  {journal} {Phys. Rev.~B}\ }\textbf {\bibinfo {volume}
  {86}},\ \bibinfo {pages} {041102} (\bibinfo {year} {2012})}\BibitemShut
  {NoStop}%
\bibitem [{\citenamefont {Weber}\ and\ \citenamefont
  {Mila}(2012)}]{WeberMila12}%
  \BibitemOpen
  \bibfield  {author} {\bibinfo {author} {\bibfnamefont {C.}~\bibnamefont
  {Weber}}\ and\ \bibinfo {author} {\bibfnamefont {F.}~\bibnamefont {Mila}},\
  }\href{http://link.aps.org/doi/10.1103/PhysRevB.86.184432} {\bibfield
  {journal} {\bibinfo  {journal} {Phys. Rev.~B}\ }\textbf {\bibinfo {volume}
  {86}},\ \bibinfo {pages} {184432} (\bibinfo {year} {2012})}\BibitemShut
  {NoStop}%
\bibitem [{\citenamefont {Fernandes}\ \emph
  {\textit{et~al.}}(2012)\citenamefont {Fernandes}, \citenamefont {Vavilov},\
  and\ \citenamefont {Chubukov}}]{FernandesVavilov12}%
  \BibitemOpen
  \bibfield  {author} {\bibinfo {author} {\bibfnamefont {R.~M.}\ \bibnamefont
  {Fernandes}}, \bibinfo {author} {\bibfnamefont {M.~G.}\ \bibnamefont
  {Vavilov}}, \ and\ \bibinfo {author} {\bibfnamefont {A.~V.}\ \bibnamefont
  {Chubukov}},\ }\href{http://link.aps.org/doi/10.1103/PhysRevB.85.140512}
  {\bibfield  {journal} {\bibinfo  {journal} {Phys. Rev.~B}\ }\textbf {\bibinfo
  {volume} {85}},\ \bibinfo {pages} {140512} (\bibinfo {year}
  {2012})}\BibitemShut {NoStop}%
\bibitem [{\citenamefont {Li}\ \emph {\textit{et~al.}}(2013)\citenamefont {Li},
  \citenamefont {Shen}, \citenamefont {Luo}, \citenamefont {Yang},
  \citenamefont {Tao}, \citenamefont {Cao},\ and\ \citenamefont
  {Xu}}]{LiShen13}%
  \BibitemOpen
  \bibfield  {author} {\bibinfo {author} {\bibfnamefont {Y.}~\bibnamefont
  {Li}}, \bibinfo {author} {\bibfnamefont {C.}~\bibnamefont {Shen}}, \bibinfo
  {author} {\bibfnamefont {Y.}~\bibnamefont {Luo}}, \bibinfo {author}
  {\bibfnamefont {X.}~\bibnamefont {Yang}}, \bibinfo {author} {\bibfnamefont
  {Q.}~\bibnamefont {Tao}}, \bibinfo {author} {\bibfnamefont {G.-H.}\
  \bibnamefont {Cao}}, \ and\ \bibinfo {author} {\bibfnamefont {Z.-A.}\
  \bibnamefont {Xu}},\ }\href{http://iopscience.iop.org/0295-5075/102/3/37003/}
  {\bibfield  {journal} {\bibinfo  {journal} {EPL}\ }\textbf {\bibinfo {volume}
  {102}},\ \bibinfo {pages} {37003} (\bibinfo {year} {2013})}\BibitemShut
  {NoStop}%
\end{thebibliography}
\end{document}